\documentclass[submission,Phys]{SciPost}
\usepackage[utf8]{inputenc}
\usepackage{color}
\usepackage{amsfonts}
\usepackage{amsmath}
\usepackage{amssymb}
\usepackage{amsthm}
\usepackage{amscd}
\usepackage{eucal}
\usepackage{bm}
\usepackage{booktabs}
\usepackage{graphicx}
\usepackage{subcaption}

\usepackage{hyperref} 
\hypersetup{
	colorlinks=true,       
	linkcolor=red,        
	citecolor=blue,        
	filecolor=magenta,     
	urlcolor=blue         
}

\numberwithin{equation}{section}
\numberwithin{figure}{section}
\numberwithin{table}{section}
\input{macro.lib}
\definecolor{dgreen}{rgb}{0.2 ,0.54, 0.2}

\title{Confinement in the three-state Potts quantum spin chain in extreme ferromagnetic limit}
\author{A. Krasznai$^{1,2}$\thanks{anna.krasznai@edu.bme.hu},  
S.~B.~Rutkevich$^{3}$\thanks{rutkevich@uni-wuppertal.de}, and G. Tak\'acs$^{1,4}$\thanks{takacs.gabor@ttk.bme.hu} \\
$^{1}${\small{}{}{}
Department of Theoretical Physics, Institute of
Physics,}\\
 {\small{}{}{}Budapest University of Technology and
Economics,}\\
 {\small{}{}{} M{\H u}egyetem rkp. 3., H-1111 Budapest,
Hungary}\\
$^{2}${\small{}{}{}
HUN-REN-BME-BCE Quantum Technology Research Group,}\\
 {\small{}{}{}Budapest University of Technology and
Economics,}\\
 {\small{}{}{} M{\H u}egyetem rkp. 3., H-1111 Budapest,
Hungary}\\
$^{3}${\small{}{}{} Fakult\"at f\"ur Mathematik und Naturwissenschaften,}\\
 {\small{}{}{} Bergische Universit\"at Wuppertal, 42097 Wuppertal, Germany}\\
$^{4}${\small{}{}{}
BME-MTA Statistical Field Theory ’Lend\"ulet’ Research
Group,}\\
 {\small{}{}{}Budapest University of Technology and
Economics,}\\
 {\small{}{}{} M{\H u}egyetem rkp. 3., H-1111 Budapest,
Hungary}
}

\date{April 10, 2026}

\begin{document}
\maketitle
\begin{abstract}
We investigate the dynamics of the three-state Potts quantum spin chain in the extreme ferromagnetic limit using a perturbative expansion in the transverse magnetic field. We demonstrate that a perturbative approach provides access to important features beyond the reach of previous studies, most notably the description of resonant excitations and the analytic prediction of post-quench time evolution. A central focus is the oblique quench regime — a feature unique to the Potts model with no Ising counterpart — in which unconfined kink excitations hybridise with the two-kink bound states. We provide a detailed examination of the analytic structure of the two-kink scattering amplitude, tracing the transformation of stable excitations into resonances near stability thresholds. Our analytical results for the excitation spectra and magnetisation dynamics show excellent agreement with numerical simulations of non-equilibrium dynamics.
\end{abstract}

\tableofcontents
\section{Introduction}

Confinement occurs when particles cannot exist in isolation, but instead are observed only within bound states, a phenomenon most famously exemplified by quarks in hadrons \cite{Wilson74,Nar04}. In systems with one spatial dimension, such as two-dimensional Quantum Field Theories (QFT) and quantum spin chains, the confinement of kink topological excitations is a common feature of models with discrete symmetry and order-disorder phase transitions \cite{McCoy78,1996NuPhB.473..469D,1998NuPhB.516..675D,FonZam2003,FZ06,Rut08a,RutP09,Rut18,Lagnese_2020,2020PhRvB.102a4426R,Mus22,Rut23s}. 

The most thoroughly investigated case is the Ising Field Theory (IFT) and its lattice counterpart, the mixed-field Ising spin chain. In the ferromagnetic phase, applying a longitudinal magnetic field induces a long-range attractive interaction between kinks, leading to their confinement into "meson" (two-kink) bound states.

Beyond the spectrum, it is also interesting to consider non-equilibrium dynamics initialised from a fully polarised state. In the Ising model in the ferromagnetic phase, it was shown to lead to dynamical confinement when the longitudinal magnetic field is parallel to the initial magnetisation \cite{Kor16}, and mesonic excitations dominate the dynamics. When the longitudinal magnetic field is anti-parallel to the initial magnetisation, it leads to Wannier-Stark localisation \cite{2019PhRvB..99r0302M,2020PhRvB.102d1118L}, with the dynamics dominated by bubbles nucleating over a false vacuum. These bubbles undergo Bloch oscillations, leading to the suppression of false vacuum decay \cite{Po22}. 

However, the $\mathbb{Z}_{2}$ symmetry of the Ising model significantly restricts its dynamics, limiting its "hadron" spectrum exclusively to mesons. In contrast, the three-state Potts model possesses an $S_{3}$ permutation symmetry that allows for both meson and "baryon" (three-kink) excitations in the ferromagnetic phase \cite{DG08,RutP09,Rut15B,LT2015,Roy26}. While recent studies of the Potts quantum spin chain have successfully used semiclassical quantisation to explain two-kink spectra \cite{Po25}, these methods face inherent limitations. Specifically, semiclassical approaches cannot account for the time evolution when the system is driven out of equilibrium. A particularly challenging non-equilibrium regime is the oblique quench, which has no Ising counterpart. In these cases, localisation is only partial, leading to unconfined kink excitations \cite{Po25}. The hybridisation between the unconfined two-particle continuum and the bound states leads to resonances in the spectrum that also require an analytical treatment beyond semiclassical approximations.

In this work, we address these challenges by examining the Potts model in the extreme ferromagnetic regime using a perturbative expansion in the transverse magnetic field. This approach provides several key advantages. It allows the calculation of the excitation spectrum and the post-quench time evolution of magnetisation up to second order, respectively. Specifically, the calculation successfully determines the spectrum of the resonance excitations via the analytic structure of scattering amplitudes, thus overcoming the limitations of the semiclassical and exact diagonalisation methods. Furthermore, it provides a predictive framework for post-quench dynamics that captures features such as high-frequency oscillations, well beyond the reach of previous methods.

The paper is organised as follows. Section \ref{sec:PSC} introduces the Potts Hamiltonian and discusses its spectrum. Section \ref{analytic_S} explores the analytic structure of scattering amplitudes and the transformation of stable bound states into resonances. Section \ref{sec:Pottsquench} presents our results on quench dynamics, comparing perturbative predictions with iTEBD numerical simulations. We conclude in Section \ref{sec:conclusions}. Appendices \ref{sec:pert_details}-\ref{sec:calc_potts_dyn_app} contain the details regarding the perturbative calculations of the excitation spectrum, vacuum structure and quench time evolution for the Potts spin chain, Appendix \ref{ApIs} demonstrates the perturbative approach to quench time evolution using the Ising case, while Appendix \ref{Numerical_details} gives the technical details of the iTEBD simulations.

\section{Spectrum of the three-state Potts quantum spin chain \label{sec:PSC}}
\subsection{A brief review of the Ising case}\label{subsec:Ising}
As a warm-up, we recall the essential features of confinement in the Ising spin chain. The mixed-field Ising quantum spin chain is defined by the Hamiltonian:
\begin{equation}
\label{HamIs}
     {H}_{IC} (h_x,h_z)=
        -\frac{1}{2}\sum_{j=-\infty}^\infty (\sigma_j^z \,\sigma_{j+1}^z+ h_x \sigma_j^x +h_z \sigma_j^z ) \,.
\end{equation}
Here $\sigma ^{x,z}$  are Pauli matrices, $j$ enumerates the chain sites, $h_x$ is the transverse and $h_z$ is the longitudinal magnetic field. 

At $h_z=0$, the Hamiltonian \eqref{HamIs} has the  $\mathbb{Z}_2$-symmetry 
\begin{equation}
[{H}_{IC} (h_x,0),U]=0,
\end{equation}
induced by  the spin-flip operator 
\begin{equation}
    U=\bigotimes_{j=-\infty}^{j=+\infty}\, \sigma_j^x.
\end{equation}
The model \eqref{HamIs}  is integrable at $h_z=0$,  and reduces to a free-fermion model by means of the
Jordan-Wigner transformation. The quantum phase transition takes place on the axis $h_z=0$ at $h_x=1$, which separates the paramagnetic
($h_x>1$) and ferromagnetic ($0<h_x<1$) phases. 

In the ferromagnetic phase $0<h_x<1$ at $h_z=0$, the Hamiltonian $\mathbb{Z}_2$-symmetry is spontaneously broken, and model \eqref{HamIs} has two degenerate ferromagnetic vacua $|\text{vac}\rangle^{(\mu)}$,
$\mu=1,2$, which are distinguished by the sign of the spontaneous magnetisation: 
\begin{equation}\label{sv}
\phantom{.}^{(1)}\!\langle\text{vac}|\sigma_j^z|\text{vac}\rangle^{(1)}=\bar{\sigma},\quad \phantom{.}^{(2)}\!\langle\text{vac}|\sigma_j^z|\text{vac}\rangle^{(2)}
=-\bar{\sigma},
\end{equation}
with $\bar{\sigma}=(1-h_x^2)^{1/8}>0$. Elementary excitations in this case are the kinks $K_{12}$ and $K_{21}$, which interpolate between the two vacua.  

Applying a longitudinal magnetic field $h_z>0$ breaks the model's integrability and explicitly breaks the $\mathbb{Z}_2$ symmetry of its   Hamiltonian. As a result, the vacuum $|\text{vac}\rangle^{(1)}$ decreases in energy and becomes the true ground state, while the vacuum $|\text{vac}\rangle^{(2)}$ increases in energy and transforms into a metastable false vacuum. The energy difference between the true and false vacua induces a long-range attractive interaction between kinks, which, in turn, leads to their confinement: isolated kinks do not exist anymore in the system, and the kinks are bound into compound particles called `mesons', the spectrum of which has been studied extensively \cite{Rut08a,Kor16,Po22,Ising_ak_gt}.  

The scaling limit of the spin chain \eqref{HamIs} to the Ising field theory (IFT) is defined in the vicinity of the second-order phase transition point $h_x=1$, i.e. in the limit $|h_x-1|\to 0$ with scaling $h_z\to 0$ appropriately \cite{2016NuPhB.911..805R}:
\begin{equation}
\label{HamIFT}
     {H}_\text{IFT} (\tau,h)={H}_*+\int dx \left(\tau \epsilon(x)+ h\sigma(x)\right)\,
\end{equation}
where ${H}_{*}$ is the fixed-point Hamiltonian (corresponding to a massless Majorana fermions), $\epsilon$ is the energy operator (fermion mass term), and $\sigma$ is the continuum field corresponding to the order parameter $\sigma^z_i$.  The physics of the IFT is determined by a single dimensionless scaling parameter $\eta$, defined as \cite{FZ06}:
\begin{equation}
\eta = 2 \pi \frac{\tau}{|h|^{8/15}}.
\end{equation}
Confinement in the IFT was first examined by McCoy and Wu \cite{McCoy78}, and the mass spectrum of mesons was determined by combining analytic and numerical methods \cite{McCoy78,FonZam2003,FZ06,Rut05,Rut08a,Rut09,LT2015}. An important feature of the spectrum is the existence of resonant excitations, first noted in \cite{McCoy78}.

In the limits $\eta \to \pm \infty$, corresponding to zero magnetic field $h = 0$, the IFT reduces to a relativistic QFT describing free spinless Majorana fermions with mass $m = 2\pi |\tau|$. These fermions represent ordinary particles in the paramagnetic phase ($\eta \to -\infty$), whereas in the ferromagnetic phase ($\eta \to +\infty$) they become topological excitations - kinks interpolating between two degenerate ferromagnetic vacua. To describe the evolution of the particle content and physical properties of the IFT as the scaling parameter $\eta$ varies between the limiting values $\pm\infty$, McCoy and Wu proposed the following scenario \cite{McCoy78}.

The case $0 < \eta < \infty$ corresponds to the ferromagnetic IFT in the presence of a magnetic field $h$, which lifts the degeneracy between the two ferromagnetic ground states and leads to confinement of kinks into the `meson' bound states. The mass $M_1(\eta)$ of the lightest meson approaches from above the two-kink mass value $2m$ as $\eta \to +\infty$, and increases as $\eta$ decreases. For a fixed positive value of $\eta$, there exists a finite number $N(\eta)$ of stable meson modes with masses $M_n(\eta)$ satisfying $M_n(\eta) < M_{n+1}(\eta)$ for $n = 1, \ldots, N(\eta)-1$. The number $N(\eta)$ diverges as $\eta \to +\infty$ and decreases as $\eta$ decreases.

The mass $M_{N(\eta)}(\eta)$ of the heaviest stable meson does not exceed the threshold value $2M_1(\eta)$. Thus, for generic $\eta>0$, the masses of the $N(\eta)$ stable mesons satisfy
\begin{equation}
2m < M_1(\eta) < M_2(\eta) < \ldots < M_{N(\eta)}(\eta) < 2M_1(\eta).
\end{equation}
When the parameter $\eta$ approaches a certain value $\eta_N$ from above, the mass of the $N$-th meson reaches the stability threshold $2M_1(\eta_N)$:
\begin{equation}
\lim_{\eta \to \eta_N + 0} M_N(\eta) = 2M_1(\eta_N).
\end{equation}
Upon decreasing $\eta$ below $\eta_N$, the $N$-th meson becomes unstable with respect to decay into two lightest ($n=1$) mesons and transforms into a resonance state.

Further decreasing $\eta$ to negative values leads to a continued reduction in the number of stable particles $N(\eta)$. Eventually, for $\eta < \eta_2$ (with some negative $\eta_2$), only a single stable particle remains in the theory. In the limit $\eta \to -\infty$, this particle becomes a free Majorana fermion, and
\begin{equation}
\lim_{\eta \to -\infty} M_1(\eta) = m.
\end{equation}
The McCoy-Wu scenario outlined above was subsequently confirmed and elaborated in numerous analytical and numerical investigations; see \cite{FZ06,Rut09} and references therein. Two particularly striking aspects of this scenario are the intrinsic presence of resonances in the IFT and the transformation of heavy stable mesons into resonances as the scaling parameter $\eta$ is tuned and the meson mass crosses the stability threshold $2M_1(\eta)$. The detailed mechanism of this transformation, which was not specified in \cite{McCoy78}, was later proposed by Fonseca and Zamolodchikov \cite{FZ06}. Unfortunately, a consistent analytical treatment remains elusive as it requires knowledge of the elastic two-particle scattering amplitude of the lightest ($n=1$) meson. Since these mesons are themselves composite particles, an analytic calculation of their mutual scattering amplitude remains a highly non-trivial and unresolved problem.

There are strong reasons to expect that the meson-resonance transformation scenario suggested in \cite{FZ06} is rather general and also occurs in other QFTs and spin-chain models exhibiting confinement. It turns out that resonance excitations also play an important role in the non-equilibrium dynamics of the Potts spin chain, and we present an analysis of their spectrum in Section \ref{analytic_S}. In the present section, we focus on the meson spectrum.   

\subsection{The three-state Potts model}
\subsubsection{The Potts spin chain Hamiltonian}
The three-state Potts quantum spin chain is  defined  by the Hamiltonian:
\begin{equation}\label{eq:potts_Hamiltonian}
H = -\sum_{j=-\infty}^{\infty} \sum_{\mu=1}^{3} ( P^{\mu}_j P^{\mu}_{j+1} + h_{\mu} P^{\mu}_j) - g\sum_{j=-\infty}^{\infty} \tilde{P}_j \, ,
\end{equation}
where the index $j\in \mathbb{Z}$ enumerates the sites of the infinite chain. It is convenient to modify the Hamiltonian \eqref{eq:potts_Hamiltonian} to the following form by adding a suitable constant:
\begin{subequations}
\label{eq:Ham1} 
\begin{align}
  H(\mathbf{v},g)&={H}_0+g \,V(\mathbf{v}),\\
\label{eq:H0}
& {H}_0=-\sum_{j=-\infty}^{\infty} \sum_{\mu=1}^{3} \left( -\frac{2}{3}+P^{\mu}_j P^{\mu}_{j+1} \right),\\
\label{eq:V}
&V(\mathbf{v})=- \sum_{j=-\infty}^{\infty} \left(\tilde{P}_j+
\sum_{\mu=1}^{3}v_\mu \left(P^{\mu}_j-c^\mu
\right)
\right),
\end{align}
\end{subequations}
where $\mathbf{v}=(v_1,v_2,v_3)$ is composed of the rescaled longitudinal magnetic fields $v_\mu=h_\mu/g$ and the constant shift is specified by $c^1=2/3$, and $c^2=c^3=-1/3$. The Hamiltonian acts on the infinite tensor product Hilbert space 
\begin{equation}
W_\infty =\bigotimes_{j=-\infty}^\infty [\mathbb{C}^3]_j.
\end{equation}
The three-dimensional local vector space $[\mathbb{C}^3]_j$, associated with the site $j$, has the basis $|\mu\rangle_j$, with $\mu=1,2,3$. 
In this basis, the action of operators  $P^{\mu}_j$, $\mu=1,2,3$, and $\tilde{P}_j$ in the space $[\mathbb{C}^3]_j$ is 
determined by the $3\times 3$ - matrices:
\begin{align}
     P_j^1=\begin{pmatrix}
     {2}/{3}&0&0\\
    0 & - {1}/{3}&0\\
     0&0&- {1}/{3}
     \end{pmatrix},
  \quad 
       P_j^2=\begin{pmatrix}
   - {1}/{3}&0&0\\
    0 &  {2}/{3}&0\\
     0&0&- {1}/{3}
     \end{pmatrix},   \\\nonumber
       P_j^3=\begin{pmatrix}
     -{1}/{3}&0&0\\
    0 & - {1}/{3}&0\\
     0&0& {2}/{3}
     \end{pmatrix},   \quad
     \tilde{P}_j= \begin{pmatrix}
    0&{1}/{3}&{1}/{3}\\
    {1}/{3} & 0&{1}/{3}\\
     {1}/{3}&{1}/{3}&0
     \end{pmatrix}.
\end{align}
The parameter $g>0$ is the transverse magnetic field, and parameters $h_\mu$, with $\mu=1,2,3$, constitute the three components of the longitudinal magnetic field.

The Hamiltonian \eqref{eq:Ham1} commutes with the unit-step translation operator $\hat{T}_1$:
\begin{equation}
[H,\hat{T}_1]=0.
\end{equation}
The latter maps the  three-dimension space $[\mathbb{C}^3]_j$ onto $[\mathbb{C}^3]_{j-1}$, acting on the basis vectors 
of the former as follows:
\begin{equation}\label{T1a}
\hat{T}_1|\mu\rangle_j=|\mu\rangle_{j-1},
\end{equation}
with $j\in\mathbb{Z}$, and $\mu=1,2,3$.

The Hamiltonian \eqref{eq:Ham1} corresponds to the thermodynamic limit of the Potts spin chain, which is the 
main subject of our interest. However, in certain intermediate calculations, we also address the finite Potts spin chain with an even number of sites $N$. The Hamiltonian $H_{N}$ of the latter is obtained from the right-hand side of \eqref{eq:Ham1}, in which 
the index $j$ runs from $-N/2+1$ to $N/2$. The Hamiltonian $H_{N}$ acts on  the $3^N$-dimensional vector space 
\begin{equation}
W_N=\bigotimes_{j=-N/2+1}^{N/2} [\mathbb{C}^3]_j.
\end{equation}
The spectrum of the model \eqref{eq:Ham1} at $h_\mu=0$ can be constructed perturbatively in the extreme ferromagnetic limit $g\ll1$ \cite{Rap06}. It turns out that the spectrum of this model can also be computed in the extreme ferromagnetic limit  $g\ll1$ at any fixed values of the components  $v_\mu={h_\mu}/{g}\ne0$ of the rescaled longitudinal magnetic field. A similar technique has been used previously for studying the kink confinement in the extreme anisotropic limit of the mixed-field Ising spin chain \cite{Rut08a,Rut10C}, and of the antiferromagnetic XXZ spin chain \cite{Shiba_80,Bera17,Rut18,Rut22, Rut24}. Perturbative expansion in the transverse field was also found to be useful for the study of transport in the confining Ising spin chain \cite{2019PhRvB..99r0302M}.

\subsubsection{The purely transverse case}

The Hamiltonian $H(\mathbf{v},g)$ at $\mathbf{v}=0$ is known as the transverse field Potts quantum spin chain. It is invariant under the group $S_3$  of permutations of three `colours', and displays a continuous order-disorder phase transition. In the ferromagnetic phase $0<g<1$, it has three energetically degenerate polarised vacua $|\text{vac}\rangle^{(\mu)}$, which for $g=0$ take the form:
\begin{align}
&H_0 |0\rangle^{(\mu)}=0\,,\\
&|0\rangle^{(\mu)}=\bigotimes_{j\in \mathbb{Z}}|\mu\rangle_j\,.
\end{align}
enumerated by the index $\mu=1,2,3$. The quasi-particle spectrum is made up from six types of kinks $K_{\mu\nu}$  ('quarks') interpolating between $|\text{vac}\rangle^{(\mu)}$ and $|\text{vac}\rangle^{(\nu)}$, with $\mu,\nu=1,2,3$. 

The scaling limit of the purely transverse spin chain, called the Potts field theory (PFT), is integrable, and the kink scattering matrix \cite{CZ92} and form factors of physically relevant operators \cite{KS88} in this model are exactly known. Note, however, that the Potts quantum spin chain with a finite $g$ is not integrable, even in the purely transverse case.

\subsubsection{Confinement in the Potts model}

Similarly to the IFT, application of a longitudinal magnetic field breaks the integrability of the three-state Potts field theory (PFT) and leads to confinement of kinks, as shown by Delfino and Grinza \cite{DG08}. These authors performed the symmetry analysis of the kink bound states in the $q$-state PFT and showed that in contrast to the $q=2$ Ising model, both mesons and baryons are allowed at $q=3$. The meson masses in the $q$-state PFT in the weak confinement regime were analytically calculated to leading order in $h$ in \cite{RutP09}, while the masses of several lightest baryons in the three-state PFT in the leading order in $h$ have been calculated in \cite{Rut15B}. These predictions were later confirmed by direct numerical calculations \cite{LT2015}.

Recently, confinement of kinks was studied away from the scaling limit, i.e. in the three-state Potts spin chain defined by the Hamiltonian \eqref{eq:Ham1} \cite{Po25}.  Similarly to the field theory version, switching on a (weak) longitudinal field in the ferromagnetic phase leads to confinement, with the spectrum containing meson and baryon excitations.\footnote{We note that the realisation of baryon excitations in condensed matter systems has attracted recent interest \cite{2007PhRvL..98p0405R,2020arXiv200707258L,2023PhRvR...5d3020W}.} It was found that the spectra of two-kink excitations can be explained well using semiclassical quantisation \cite{Po25}.

In the Potts model, the non-equilibrium dynamics following quantum quenches is largely analogous to the Ising case when the longitudinal field is either parallel or antiparallel to the initial magnetisation. The main difference from the Ising case is the presence of baryonic oscillation signatures. However, for the Potts model, it is also possible to realise oblique quenches in which the longitudinal field is in a direction different from the initial magnetisation, and such quenches have no Ising counterparts \cite{Po25}. Unlike the Ising model, where the longitudinal field confines all excitations, the $S_3$ symmetry of the Potts model allows specific field orientations to retain a residual degeneracy. In these cases, confinement is only partially effective, which leads to the presence of unconfined kink excitations. Additionally, the hybridisation of the corresponding unconfined two-kink continuum with the bound states leads to the appearance of resonances in the spectrum, which cannot be treated using semiclassical methods. 

Due to the appearance of false vacuum states, the excitation spectrum also contains bubbles which correspond to the nucleation of the true vacuum from the false vacuum according to the Coleman scenario \cite{1977PhRvD..15.2929C}. Contrary to field theory, these excitations can be localised in a lattice setting (Wannier-Stark localisation) \cite{2019PhRvB..99r0302M, 2020PhRvB.102d1118L, 2024PhRvL.133x0402L, Ising_ak_gt}. Similarly to mesons, the bubble excitations can also hybridise with an unconfined two-kink continuum, leading to the appearance of resonances \cite{Po25}. Once again, while the bubble spectrum admits a semiclassical treatment, the resonances cannot be incorporated into this description.  

In the present work, we use a perturbative approach to overcome the aforementioned limitations of the methods used in \cite{Po25} by determining the excitation spectrum of the resonances and describing the post-quench dynamics of the model.

\subsection{Structure of the Hilbert space}

The structure of the vector space of the low-energy excitations of the unperturbed  Hamiltonian $H_0$ is, to a great extent, similar to the infinite antiferromagnetic  XXZ spin-1/2 chain in the anisotropic limit \cite{Jimbo94}.

Denote by $\mathcal{L}$ the {\it space of states} for the Hamiltonian $H_0$, i.e. the Hilbert space 
spanned by the eigenvectors $\otimes_{j=-\infty}^\infty |\mu_j\rangle_j$ of the Hamiltonian
$H_0$ that have finite energies (eigenvalues). The Hilbert space $\mathcal{L}$
splits into the direct sum of nine subspaces (sectors):
\begin{equation}
\mathcal{L}=\bigoplus_{\mu,\mu'=1}^3 \mathcal{L}_{\mu\mu'}.
\end{equation}
The subspace $\mathcal{L}_{\mu\mu'}$ is spanned by the basis vectors $\otimes_{j=-\infty}^\infty |\mu_j\rangle_j$, which satisfy the requirement (i) that
there exist two integers $j_1,j_2\in \mathbb{Z}$, such that $j_1<j_2$; and
$\mu_j=\mu$ for all $j<j_1$, while $\mu_j=\mu'$, for $j>j_2$. This means that the relevant configurations interpolate between the spin polarisations $\mu$ and $\mu'$. The subspaces 
$\mathcal{L}_{\mu\mu'}$ with $\mu=\mu'$, and with $\mu\ne \mu'$ are called the {\it  topologically neutral}, and 
{\it  topologically charged sectors}, respectively. 
Since the local spin operators $P_j^\mu$ map each sector $\mathcal{L}_{\mu\mu'}$ onto itself, different sectors are separate from each other. 

In our studies of non-equilibrium dynamics in Section \ref{sec:Pottsquench}, we start from a state polarised in direction $1$. As a result, we are interested in the topologically neutral sector $\mathcal{L}_{11}$, whose structure we describe below in detail.

The Hilbert space $\mathcal{L}_{11}$ is the direct sum of the subspaces $\mathcal{L}_{11}^{(n)}$, with $n=0,2,3,4,\ldots$:
\begin{equation}\label{LL}
\mathcal{L}_{11}=\mathcal{L}_{11}^{(0)}\oplus\mathcal{L}_{11}^{(2)}\oplus\mathcal{L}_{11}^{(3)}\oplus\ldots.
\end{equation}
The basis vector  $|\Phi\rangle =\otimes_{j=-\infty}^\infty |\mu_j\rangle_j$  belongs to the subspace $\mathcal{L}_{11}^{(n)}$, if it satisfies the requirement (i) with $\mu=\mu'=1$, and 
\begin{equation}\label{H0A}
H_0\, |\Phi\rangle= n\, |\Phi\rangle,
\end{equation}
i.e., it contains $n$ domain walls. The space $\mathcal{L}_{11}^{(0)}$ is one-dimensional spanned by the fully polarised state $|0\rangle^{(1)}\in \mathcal{L}_{11}^{(0)}$, while all subspaces $\mathcal{L}_{11}^{(n)}$ with $n>0$ have infinite dimensions. 

The subject of our particular interest is the two-kink subspace $\mathcal{L}_{11}^{(2)}$. 
The basis in this subspace is formed by the localised two-kink states $|\mathbf{K}_{1,\nu}(j_1)\mathbf{K}_{\nu,1}(j_2)\rangle$, with $\nu=2,3$ and $-\infty<j_1<j_2<\infty$. These states correspond to fully polarised domains joined at single sites and are defined as follows:
\begin{equation}\label{2ki}
|\mathbf{K}_{1,\nu}(j_1)\mathbf{K}_{\nu,1}(j_2)\rangle=\left(\bigotimes_{j=-\infty}^{j_1}|1\rangle_j \right)\left(
\bigotimes_{j=j_1+1}^{j_2}|\nu\rangle_j\right)\left(\bigotimes_{j=j_2+1}^\infty|1 \rangle_j\right)\,,
\end{equation}
which satisfy the normalisation condition: 
\begin{equation}
\langle \mathbf{K}_{1,\nu}(j_2)\mathbf{K}_{\nu,1}(j_1)| \mathbf{K}_{1,\nu'}(j_1')\mathbf{K}_{\nu',1}(j_2')\rangle=\delta_{\nu\nu'}\delta_{j_1,j_1'}\delta_{j_2,j_2'},
\end{equation}
where $\langle \mathbf{K}_{1,\nu}(j_2)\mathbf{K}_{\nu,1}(j_1)|$ is the 'bra'-vector corresponding to the `ket'-vector 
$| \mathbf{K}_{1,\nu}(j_1)\mathbf{K}_{\nu,1}(j_2)\rangle$. The unit-step translation operator $\hat{T}_1$ defined by equation \eqref{T1a}  acts  on the states \eqref{2ki} in the following way:
\begin{equation}\label{Tr1}
\hat{T}_1| \mathbf{K}_{1,\nu}(j_1)\mathbf{K}_{\nu,1}(j_2)\rangle = | \mathbf{K}_{1,\nu}(j_1-1)\mathbf{K}_{\nu,1}(j_2-1)\rangle.
\end{equation}
We also use the parity eigenbasis in $\mathcal{L}_{11}^{(2)}$  formed by the states 
\begin{equation}\label{Kiot}
|\mathbf{K}(j_1)\mathbf{K}(j_2)\rangle_\iota=\frac{1}{\sqrt{2}}\Big(
| \mathbf{K}_{1,2}(j_1)\mathbf{K}_{2,1}(j_2)\rangle+\iota | \mathbf{K}_{1,3}(j_1)\mathbf{K}_{3,1}(j_2)\rangle
\Big),
\end{equation}
with $\iota=\pm$, denoting the even and odd states respectively.

For the subsequent analysis, it is also essential to introduce another basis in $\mathcal{L}_{11}^{(2)}$ given by the momentum eigenstates (‘Bethe states’) 
\begin{align}\nonumber
&|{K}_{1,\nu}(p_1){K}_{\nu,1}(p_2)\rangle=\sum_{j_1=-\infty}^\infty\sum_{j_2=j_1+1}^\infty \sum_{\nu'=2}^3\Big[e^{i(p_1j_1+p_2 j_2)}\delta_{\nu,\nu'}+
S_{\nu,\nu'}(p_1,p_2) e^{i(p_1j_2+p_2 j_1)}\Big]\\
\label{eq:bethe_states}
&\times| \mathbf{K}_{1,\nu'}(j_1)\mathbf{K}_{\nu',1}(j_2)\rangle,
\end{align}
where $-\pi<p_1,p_2\leq\pi$ are the quasimomenta of the two kinks, $\nu=2,3$ and $S_{\nu,\nu'}(p_1,p_2)$ are the two-kink scattering amplitudes defined in Eq. \eqref{scm_app}. We also use the parity-even/odd two-kink Bethe  states
\begin{align}\label{Bl2}
&|{K}(p_1){K}(p_2)\rangle_\iota=\frac{1}{\sqrt{2}}\Big(|{K}_{1,2}(p_1){K}_{2,1}(p_2)\rangle+\iota |{K}_{1,3}(p_1){K}_{3,1}(p_2)\rangle\Big), \quad \text{with }\iota=\pm
\end{align}
which diagonalise the two-kink scattering matrix \eqref{eq:scat_mx_diag_app}. Yet another basis is obtained by transforming to the centre of mass frame:
\begin{equation}\label{jP23}
|j,P\rangle_\nu
=\sum_{j_1=-\infty}^\infty \exp\left[i P\left(j_1+\frac{j}{2}\right)\right]|\mathbf{K}_{1,\nu}(j_1)\mathbf{K}_{\nu,1}(j_1+j)\rangle,
\end{equation}
with $\nu=2,3$, $j=1,2,\ldots$, which satisfy 
\begin{equation}
\hat{T}_1 |j,P\rangle_\nu=e^{i P}|j,P\rangle_\nu,
\end{equation}
where $-\pi<P\leq\pi$ is the overall quasimomentum. We can also introduce  even/odd combinations of these states by :
\begin{equation}\label{jP_iota}
|j,P\rangle_+=\frac{1}{\sqrt{2}}(|j,P\rangle_2+|j,P\rangle_3),\quad 
|j,P\rangle_-=\frac{1}{\sqrt{2}}(|j,P\rangle_2-|j,P\rangle_3).
\end{equation}
Due to equation \eqref{H0A}, all states in $|\Phi\rangle\in\mathcal{L}_{11}^{(n)}$ are the eigenstates of the unperturbed Hamiltonian $H_0$ with the same eigenvalue $n$. Therefore, to the first order in the transverse magnetic field $g$, it is only necessary to 
diagonalise the perturbed Hamiltonian \eqref{eq:Ham1} inside the subspace $\mathcal{L}_{11}^{(2)}$. We note that while higher-order corrections generally mix the two-kink subspace with states of higher kink numbers, previous results show that this mixing is small and a two-kink approximation \cite{Rut08a,2016NuPhB.911..805R,2024PhRvL.133x0402L,Ising_ak_gt} provides a very accurate description of the spectrum as shown in \cite{Po25}, similarly to the Ising case \cite{Rut08a,2016NuPhB.911..805R,2024PhRvL.133x0402L,Ising_ak_gt}.

Denoting the projection operator onto the two-kink subspace $\mathcal{L}_{11}^{(2)}$ as $\mathcal{P}_{11}^{(2)}$, the restriction of the Hamiltonian $H(\mathbf{v},g)$ to the two-kink subspace can be defined as 
\begin{equation}\label{prH}
H^{(2)}(\mathbf{v},g)=\mathcal{P}_{11}^{(2)} H(\mathbf{v},g) \mathcal{P}_{11}^{(2)}.
\end{equation}
Now we consider the eigenvalue problem 
\begin{equation}\label{RS}
H(\mathbf{v},g)|\Psi(\mathbf{v},g)\rangle =E(\mathbf{v},g)\, |\Psi(\mathbf{v},g)\rangle,
\end{equation}
with $|\Psi(\mathbf{v},g=0)\rangle\in \mathcal{L}_{11}$. Since the Hamiltonian $H(\mathbf{v},g)$ commutes with the  translation operator $\hat{T}_1$, we can require that
\begin{equation}
\hat{T}_1|\Psi(\mathbf{v},g)\rangle=e^{i P}\, |\Psi(\mathbf{v},g)\rangle,
\end{equation}
with quasimomentum $-\pi<P\leq \pi$. The formal expansion of both sides of equation \eqref{RS} in integer  powers of $g$ yields:
\begin{subequations}\label{prt}
\begin{align}
&|\Psi(\mathbf{v},g)\rangle=|\Psi_0(\mathbf{v})\rangle+g\, |\Psi_1(\mathbf{v})\rangle+g^2 |\Psi_2(\mathbf{v})\rangle+\ldots,\\
&E(\mathbf{v},g)=E_0(\mathbf{v})+g\, E_1(\mathbf{v})+g^2 E_2(\mathbf{v})+\ldots\,.
\end{align}
\end{subequations}
Subsequently equating the coefficients in the resulting power series leads to the Rayleigh-Schrödinger perturbation theory \cite{LL3,Gr18}. For the deformed vacuum state $|\text{vac}(\mathbf{v},g)\rangle^{(1)}$, the zero-order terms in the perturbation expansions \eqref{prt} read as follows:
\begin{equation}\label{zo}
|\text{vac}(\mathbf{v},g)\rangle^{(1)}=|0\rangle^{(1)}+O(g), \quad E_\text{vac}(\mathbf{v},g)=O(g^2).
\end{equation}
The perturbative calculation of the deformed vacuum states \eqref{zo} for $g\neq0$ is described in Appendix \ref{sec:vac}.

Turning to  the two-kink excitations, one finds that any two-kink state $|\Psi\rangle \in \mathcal{L}_{11}^{(2)}$ is the 
eigenstate of the unperturbed Hamiltonian $H_0$ corresponding to the same energy $E_0=2$: 
\begin{equation}\label{2kH}
H_0 |\Psi\rangle = 2  |\Psi\rangle,\quad  \text{for any } |\Psi\rangle\in \mathcal{L}_{11}^{(2)}.
\end{equation}
Switching on $g\neq 0$ in the Hamiltonian \eqref{eq:Ham1} lifts this degeneracy at the first
order in $g$, and the `good' zero-order eigenstates \cite{LL3,Gr18}  $|\Psi_0(\mathbf{v})\rangle$ can be obtained by diagonalising the Hamiltonian $H^{(2)}(\mathbf{v},g)$ defined by \eqref{prH}. In the following sections, we diagonalise the Hamiltonian  $H^{(2)}(\mathbf{v},g)$ at three different choices of the vector $\mathbf{v}$.
 \begin{itemize}
 \item
In Subsection \ref{diH0}, we put  $\mathbf{v}=\mathbf{0}$, corresponding to the purely transverse case. In this case, the excitations are two-kink states of the form \eqref{eq:bethe_states}.
 \item
The case $\mathbf{v}=(v_1,0,0)$ is studied in Section \ref{diav1}. In this case, the longitudinal magnetic field $\mathbf{h}=g\,  \mathbf{v}$ is either parallel (at $v_1>0$) 
  or antiparallel (at $v_1<0$) to the orientation of the magnetisation of the first vacuum $|0\rangle^{(1)}$. Following \cite{Po25}, we refer to these two cases as {\it positively} and {\it negatively aligned}, and the two-kink eigenstates are called mesons/bubbles, respectively. For $v_1>0$, the interaction between the two-kinks is attractive, and the excitations are true bound states (a.k.a. mesons), while for $v_1<0$, the interaction is repulsive and the excitations correspond to bubbles of the true vacuum forming inside a bulk false vacuum, which are stabilised by Wannier-Stark localisation. 
  \item  
 In Subsection \ref{diav2}, we address the {\it oblique} regime  $\mathbf{v}=(0,v_2,0)$, in which the field $\mathbf{v}$ is neither parallel nor anti-parallel to the magnetisation of the state $|0\rangle^{(1)}$. As in  \cite{Po25}, we distinguish positive and negative {oblique} regimes, according to the sign of $v_2$. Besides the presence of mesonic excitations for $v_2<0$ and bubbles for $v_2>0$, we also find unconfined kink excitations in this case. Hybridisation of these states with the meson/bubble states leads to the appearance of resonances in the spectrum. Resonance excitations were observed in \cite{Po25} using quench spectroscopy; however, they cannot be described by the semiclassical methods used in that paper. We determine the two-kink scattering amplitudes, which we use in Section \ref{analytic_S} to calculate the resonance spectrum exploiting their analytic structure following the Fonseca-Zamolodchikov approach \cite{FZ06}.
 \end{itemize}
\subsection{The spectrum of two-kink states}
\subsubsection{Pure transverse chain: kink excitations \label{diH0}}
The Hamiltonian $H(\mathbf{v},g)$ (defined in Eq. \eqref{eq:Ham1}) with the parameters $\mathbf{v}=\mathbf{0}$ corresponds to the infinite three-state Potts spin chain with purely transverse magnetic field. Here, we discuss the results of the perturbative analysis of this model in the extreme ferromagnetic limit 
$g\ll1$ in the two-kink sector. Additional details of the calculation are provided in Appendix \ref{diH0_tech}. 

We start from the eigenvalue problem \eqref{RS} with $\mathbf{v}=\mathbf{0}$. To find the `good' zero-order wave function $|\Psi_0(\mathbf{0})\rangle\in \mathcal{L}_{11}^{(2)}$ that stands in the expansion \eqref{prt}, one has to diagonalise the Hamiltonian  
$H^{(2)}(\mathbf{0},g)$. The solution to the latter problem is given by the two-kink `Bethe states' $|{K}_{1,\nu}(p_1){K}_{\nu,1}(p_2)\rangle$ defined by \eqref{eq:bethe_states} or 
$|{K}(p_1){K}(p_2)\rangle_\iota$ defined by \eqref{Bl2}. Both of these two-kink states diagonalise simultaneously the operators $H^{(2)}(\mathbf{0},g)$, and $\hat{T}_1$; for example
\begin{align}
&  \hat{T}_1 \,  |{K}(p_1){K}(p_2)\rangle_\iota=\exp[i(p_1+p_2)]\, |{K}(p_1){K}(p_2)\rangle_\iota, \quad \text{with }\iota=\pm,\\
&H^{(2)}(\mathbf{0},g)|{K}(p_1){K}(p_2)\rangle_\iota=[\omega(p_1)+\omega(p_2)]|{K}(p_1){K}(p_2)\rangle_\iota,
\end{align}
where
\begin{equation}\label{eq:perturbative_disprel}
\omega(p)=1-\frac{2 g}{3}  \cos p
\end{equation}
is the kink dispersion law to the linear order in the small parameter $g$. 

We note that the dynamical properties of the $q$-state Potts spin chain in the presence of the pure transverse magnetic field were previously studied perturbatively by Rapp and Zaránd \cite{Rap06}. Here we go beyond their results by presenting the explicit formulas \eqref{scm_app} for the two-kink scattering matrix in the case $q=3$, which is essential for the analysis of the resonance spectrum performed in Section \ref{analytic_S}.

\subsubsection{Two-kink states obtained from $\mathcal{L}_{11}^{(2)}$: the aligned case \label{diav1}}
\begin{figure}[htb]
\centering
\includegraphics[width=1\linewidth, angle=00]{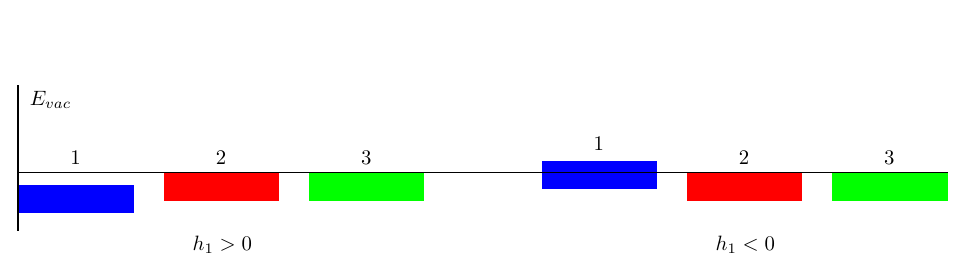}
\caption{Vacuum Energy Levels in Aligned Quenches.\\
Schematic representation of the energy levels for the three vacua $|vac(v_{1},g)\rangle^{(1)}$ (blue), $|vac(v_{1},g)\rangle^{(2)}$ (red), and $|vac(v_{1},g)\rangle^{(3)}$ (green), with $v_1=h_1/g$. In the positively aligned case ($h_{1} > 0$), vacuum 1 becomes the unique true ground state, leading to kink confinement into mesons. In the negatively aligned case ($h_{1} < 0$), vacuum 1 becomes a metastable false vacuum, where dynamics is dominated by the nucleation of "bubbles" of the true vacua 2 and 3.
\label{fig:align} } 
\end{figure}

Application of the longitudinal magnetic field $\mathbf{v}=(v_1,0,0)$, with $v_1=h_1/g$,  explicitly breaks the $S_3$ symmetry of the Hamiltonian \eqref{eq:Ham1}, with the effect of this field on the energies of the vacua of the model shown schematically in Fig. \ref{fig:align}. For $h_1>0$, the energy of the first vacuum $|\text{vac}(\mathbf{v},g)\rangle^{(1)}$ decreases 
with respect to the energies of the second and third ones $|\text{vac}(\mathbf{v},g)\rangle^{(\nu)}$, $\nu=2,3$. As a result, the state  $|\text{vac}(\mathbf{v},g)\rangle^{(1)}$ becomes the true ground state of the spin chain, while for $h_1<0$ it becomes the metastable (false) vacuum, with an energy larger than those of the two degenerate ground states  $|\text{vac}(\mathbf{v},g)\rangle^{(2,3)}$.
In both cases, the application of the weak longitudinal magnetic field leads to the formation of localised two-kink states a.k.a. 'mesons' and 'bubbles'. In \cite{Po25}, these excitations were described in the semiclassical approximation. For $h_1>0$, the mesons correspond to the confinement of the kinks, while the bubbles appearing for $h_1<0$ correspond to the nucleation of the true vacuum. These states are fully analogous to the mesons and bubbles obtained in the Ising spin chain \cite{Rut08a,2024PhRvL.133x0402L,Ising_ak_gt}.

Here we briefly discuss the spectrum of these two-kink bound states to the leading order in the weak transverse magnetic field $g$, with the details of the calculation relegated to Appendix \ref{diav1_tech}. We use the short-cut notation $H^{(2)}(v_1,g)$ for the 
Hamiltonian  \eqref{prH}. The eigenstates can be expressed in terms of the basis $|j,P\rangle_\iota$ defined by \eqref{jP_iota} as follows:
\begin{equation}
    H^{(2)}(v_1,g) |\pi_\iota\rangle= E_\iota |\pi_\iota\rangle, \quad \text{where} \quad|\pi_\iota\rangle=\sum_{j=1}^\infty \psi_\iota(j;E_\iota,P) |j,P\rangle_\iota.
\end{equation}
Together with the boundary condition that the wave-function vanishes as $j\rightarrow \infty$, the above equation determines the energy eigenvalues. The resulting
energy spectrum is discrete, and we obtain in Appendix  \ref{diav1_tech} the dispersion laws $E_{\iota,n}(P,h_1)$ of mesons at $h_1>0$, and bubbles at $h_1<0$ to the first order in the small parameter $g$.
\begin{figure}[htb]
\centering
\includegraphics[width=.8\linewidth, angle=00]{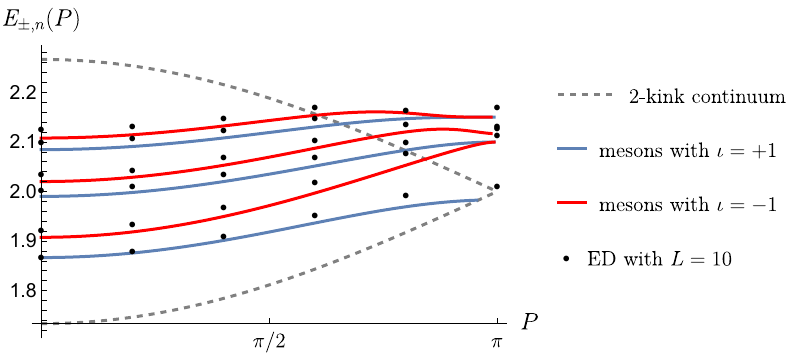}
\caption{Meson Dispersion Laws in the Aligned Case.\\
The dispersion laws $E_{\iota,n}(P)$  of the lightest mesons with $n=1,2,3$, and $\iota=+$ (solid blue), 
and $\iota=-$ (solid red). The dashed grey curves represent the boundaries of the two-kink continuous spectrum at $h_1=h_2=0$, calculated via \eqref{bas}. Numerical results from exact diagonalization (ED) for a chain length $L=10$ with periodic boundary conditions (black dots) demonstrate good agreement with the perturbative analytical curves. Parameters: $g=0.2$, $h_1=0.05$. 
\label{fig:DL} } 
\end{figure}

The energy spectra $E_{\pm,n}(P,h_1)$ of the three lightest mesons at $g=0.2$, $h_1=0.05$ are shown in Fig. \ref{fig:DL}. The  boundaries of the  two-kink continuum at $h_1=0$ are given by
\begin{equation}\label{bas}
 E_\text{min}(P)=2-\frac{4g}{3}\cos(P/2), \quad 
 E_\text{max}(P)=2+\frac{4g}{3}\cos(P/2), 
 \end{equation} 
using Eq. \eqref{eq:perturbative_disprel}.

A critical feature of the two-kink spectrum is the distinction between collisional and collisionless states \cite{Ising_ak_gt,Po25}. This distinction is most intuitively understood through a semiclassical lens. 
\begin{itemize}
    \item Mesonic excitations ($h_1 > 0$): For the lightest (lowest-energy) mesons, the constituent kinks follow trajectories that cross, resulting in periodic mutual collisions. As one moves to higher-energy mesonic states, the attractive longitudinal field induces Wannier-Stark localisation \cite{2019PhRvB..99r0302M,2020PhRvB.102d1118L}. This effect confines the constituent kinks to separate regions, preventing them from colliding. In this collisionless regime, the kinks undergo independent Bloch oscillations \cite{Po22}. Consequently, the energy levels become uniformly spaced, with the gap determined by the Bloch oscillation frequency.
    \item Bubble excitations ($h_1 < 0$): The above scenario is mirrored for bubbles, with the primary difference being that the bubble spectrum is bounded from above \cite{2024PhRvL.133x0402L}. This results in an inverted energy hierarchy in the bubble sector, where the highest-energy states are collisional, whereas the lower-energy states correspond to collisionless configurations stabilised by Wannier-Stark localisation.
\end{itemize}
The collisional particles lie inside the interval of the two-kink continuum at $h_1=0$ \cite{Po25}, i.e., between $E_\text{min}(P)$ and $E_\text{max}(P)$.
\subsubsection{Two-kink states obtained from $\mathcal{L}_{11}^{(2)}$: the oblique case \label{diav2}}

\begin{figure}[htb]
\centering
\includegraphics[width=1\linewidth, angle=00]{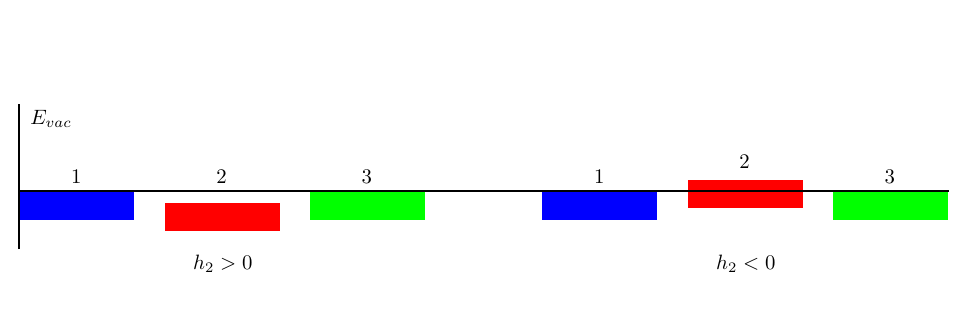}
\caption{Vacuum Energy Levels in the Oblique Case.\\
Energy levels for the vacua $|vac(v_{2},g)\rangle^{(1)}$ (blue), $|vac(v_{2},g)\rangle^{(2)}$ (red), and $|vac(v_{2},g)\rangle^{(3)}$ (green) under an oblique field, with $v_2=h_2/g$. For $h_{2} > 0$ (left), vacuum 2 is the true ground state while vacua 1 and 3 remain the degenerate false vacua. For $h_{2} < 0$ (right), vacua 1 and 3 are degenerate true ground states, leaving vacuum 2 as the false vacuum. This degeneracy allows for the existence of unconfined kink excitations.
\label{fig:align2} } 
\end{figure}

This section addresses the oblique regime, defined by the rescaled longitudinal field $\mathbf{v} = (0, v_2, 0)$. This configuration is uniquely characteristic of the three-state Potts model and has no counterpart in the Ising spin chain. The presence of a non-zero $v_2$ deforms the three ferromagnetic vacua and shifts their energies, as illustrated in Fig. \ref{fig:align2}.  A critical feature of this regime is that it allows for the coexistence of localised two-kink states (mesons for $v_2 < 0$ and bubbles for $v_2 > 0$) with unconfined kink excitations \cite{Po25},  which interpolate between the degenerate vacuum states. The hybridisation of these bound states with the two-particle continuum leads to the formation of resonances, which are considered in detail in Section~\ref{analytic_S}.

Throughout this Section, the  notation $H^{(2)}(v_2,g)$ 
for the Hamiltonian \eqref{prH} is used instead of $H^{(2)}(\mathbf{v},g)$. The details of the perturbative calculation are relegated to Appendix \ref{diav2_tech}.  

To determine the spectrum, we solve the eigenvalue problem restricted to the two-kink subspace $\mathcal{L}_{11}^{(2)}$. At total quasimomentum $P$, we expand the eigenstate in terms of the basis $|j,P\rangle_\nu$ determined by \eqref{jP23}:
\begin{equation}\label{eq:eig_eq_oblique}
H^{(2)}(v_2,g) |\Psi\rangle= E |\Psi\rangle, \quad \text{where} \quad |\Psi\rangle=\sum_{j=1}^\infty \sum_{\nu=2}^3\psi_\nu(j;E,P) |j,P\rangle_\nu.
\end{equation}
As shown in Appendix \ref{diav2_tech}, the above equation determines that the wavefunction coefficients have the form
\begin{align}\label{ps2}
\psi_2(j;E,P)&=[\sign(- h_2 )]^{j+1}\,J_{j-c_0}(Z),
\nonumber\\
Z=\frac{4g}{3|h_2|} &\cos(P/2),\qquad
c_0=\frac{2-E}{h_2}
\end{align}
and
\begin{equation}\label{ps3}
\psi_3(j;E,P)=\begin{cases}
A \sin[p(j-1)]+B\cos[p(j-1)], & \text{for }E_\text{min}(P)<E<E_\text{max}(P),\\
C \exp[-\lambda (j-1)],& \text{for } E<E_\text{min}(P),\\
C (-1)^{j+1}\exp[-\lambda (j-1)],& \text{for } E>E_\text{max}(P)\,,
\end{cases}
\end{equation}
where
\begin{align}
&p=\arccos\frac{3(2-E)}{4g \cos(P/2)},\qquad\lambda=
\mathrm{arccosh}\,\frac{3|E-2|}{4g \cos(P/2)}.
\end{align}
The excitations below the two-kink continuum ($E<E_\text{min}(P)$) correspond to collisionless bubbles, while those above the two-kink continuum ($E>E_\text{max}(P)$) are collisionless mesons. 

Eq. \eqref{eq:eig_eq_oblique} furthermore implies the boundary conditions 
\begin{subequations}\label{bc2}
\begin{align}
&\left(2-h_2 -E \right) \psi_2(1)-\frac{2 g }{3 } \cos(P/2)\, \psi_2(2)-\frac{ g }{3 } \psi_3(1) =0,\\
&\left(2 -E \right) \psi_3(1)-\frac{2 g }{3 } \cos(P/2)\, \psi_3(2)-\frac{ g }{3 } \psi_2(1) =0,
\end{align}
\end{subequations}
where for simplicity we suppressed $E$ and $P$ from the arguments of the function 
$\psi_\nu(j;E,P)$. At $j\to\infty$, the wave-function $\psi_2(j)$ must vanish, while the absolute value of $\psi_3(j)$ must remain bounded.

In the interval $E_\text{min}(P)<E<E_\text{max}(P)$, 
substitution of \eqref{ps2}, \eqref{ps3} into the equations of the boundary conditions \eqref{bc2} leads to two linear equations, determining the coefficients $A$ and $B$. For $P=0$, the wavefunction $\psi_3(j;E,P)$, given by Eq.$\,$\eqref{ps3}, can be represented in the form:
\begin{equation}\label{eq:psi3_P0}
\psi_3(j;E,0)=B_{in} \,e^{-i p j}+B_{out} \,e^{i p j}.
\end{equation}
Then Eq.$\,$\eqref{bc2} yields
\begin{align}\label{Bin1}
&B_{in}(z,v_2)=\frac{1 }{z-z^{-1}}\left[ 2 \,\mathrm{sign}\,v_2\,\,J_{-c_0(z,v_2)}\left(\frac{4}{3|v_2|}\right)+\frac{z}{2}
\, J_{1-c_0(z,v_2)}\left(\frac{4}{3|v_2|}\right)
\right],\\
&B_{out}(z,v_2)=B_{in}(z^{-1},v_2),
\end{align}
where $z=e^{i p}$ with $p$ taking real values, $v_2=\frac{h_2}{g}$, and $c_0(z,v_2)=\frac{2(z+z^{-1})}{3v_2}$. We remark that the symmetry \eqref{sim2} implies
\begin{equation}\label{symB}
B_{in}(z,v_2)=B_{in}(-z,-v_2), \quad B_{out}(z,v_2)=B_{out}(-z,-v_2),
\end{equation}
using that $v_2$ is real. Finally, we note that the wave-function $\psi_2(j;E,P=0)$ can be simply written as
\begin{equation}
    \psi_2(j;E,0)=[\sign(- h_2 )]^{j+1}\,J_{j-\frac{4 g \cos p}{3 h_2}}\left(\frac{4 g}{3|h_2|}\right).
    \label{eq:psi2_P0}
\end{equation}
Note that the states in this range are two-kink scattering states and therefore their energy spectrum is continuous. While standard semiclassical methods fail in this regime, our perturbative approach allows us to identify resonance peaks appearing in the quench spectroscopy analysis performed in Section \ref{sec:Pottsquench}, by analysing the analytic structure of the two-kink scattering amplitude, as detailed in Section \ref{analytic_S}.  

When the energy $E$ is outside the two-kink continuum, substituting Eqs.$\,$(\ref{ps2},\ref{ps3}) into the boundary condition equation \eqref{bc2} yields two linear secular equations that determine the coefficient $C$ and the discrete energy spectrum $\{E_n(P,h_2)\}_{n=1}^\infty$.
In the negative oblique regime  $h_2<0$, this energy spectrum corresponds to collisionless mesons,
 \begin{align}\label{eq:collisionless_mesons}
E_1(P,h_2)<E_2(P,h_2)<E_3(P,h_2)<\ldots,
\end{align}
with $E_1(P,h_2)>E_\text{max}(P)$, 
while in the positive oblique  regime  $h_2>0$, it corresponds to collisionless bubbles,
 \begin{align}\label{eq:collisionless_bubbles}
E_1(P,h_2)>E_2(P,h_2)>E_3(P,h_2)>\ldots,
\end{align}
with $E_1(P,h_2)<E_\text{min}(P)$.

\section{Resonance excitations in the oblique regime \label{analytic_S}}

As discussed above, a remarkable feature of the three-state Potts quantum spin chain is the coexistence of localised two-kink bound states—mesons and bubbles—with a continuum of stable, unconfined kinks in the oblique regime. 

In this section, we provide a detailed analysis of the two-kink scattering amplitude in the small-$g$ limit to describe the interactions within this mixed particle content. We calculate the energies of stable bound states and the positions of resonances by analysing the poles of the scattering amplitude in the complex plane. We then demonstrate how resonances arise from the hybridisation of the unconfined two-particle continuum with bound states that would otherwise be stable in a purely confining regime. By tracking the trajectories of these poles as the longitudinal field parameter $v_2$ is varied, we describe the physical transformation of stable mesons and bubbles into unstable resonances.
This analysis confirms that the three-state Potts model follows the general scenario proposed by Fonseca and Zamolodchikov for the Ising field theory, where heavy stable mesons cross a stability threshold and decay into the lightest available particles \cite{FZ06}. These results provide the theoretical basis for the quench spectroscopy features observed in Section \ref{sec:Pottsquench}, which remain beyond the reach of semiclassical approximations.

\subsection{The bound state and resonance spectrum}

In the oblique regime — defined by $h_2 \neq 0$ and $h_1 = h_3 = 0$ — the system's dynamics is governed by localised two-kink states and a continuum of unconfined kinks.  At zero total quasimomentum ($P=0$), we consider the wave-function $\psi_3(j; E, P=0)$ as defined in \eqref{eq:psi3_P0}. The scattering amplitude $S(z,v_2)=\frac{B_{in}(z^{-1},v_2)}{B_{in}(z,v_2)}$ of two  unconfined kinks is given by:
\begin{equation}\label{Szv}
S(z, v_2) = \frac{B_{in}(z^{-1}, v_2)}{B_{in}(z, v_2)} =-\frac{
z^{-1}\, J_{1-c_0(z,v_2)}\left(\frac{4}{3|v_2|}\right)+4\,\sign v_2\,\,J_{-c_0(z,v_2)}\left(\frac{4}{3|v_2|}\right)}{z\, J_{1-c_0(z,v_2)}\left(\frac{4}{3|v_2|}\right)+4\,\sign v_2\,\, J_{-c_0(z,v_2)}\left(\frac{4}{3|v_2|}\right)}\,.
\end{equation}
The scattering states correspond to $z=e^{ip}$ with the relative momentum variable $p$ taking a real value, i.e. $|z|=1$. 

\begin{figure}{t}
\centering
\begin{subfigure}{0.46\textwidth}
    \includegraphics[width=\textwidth]{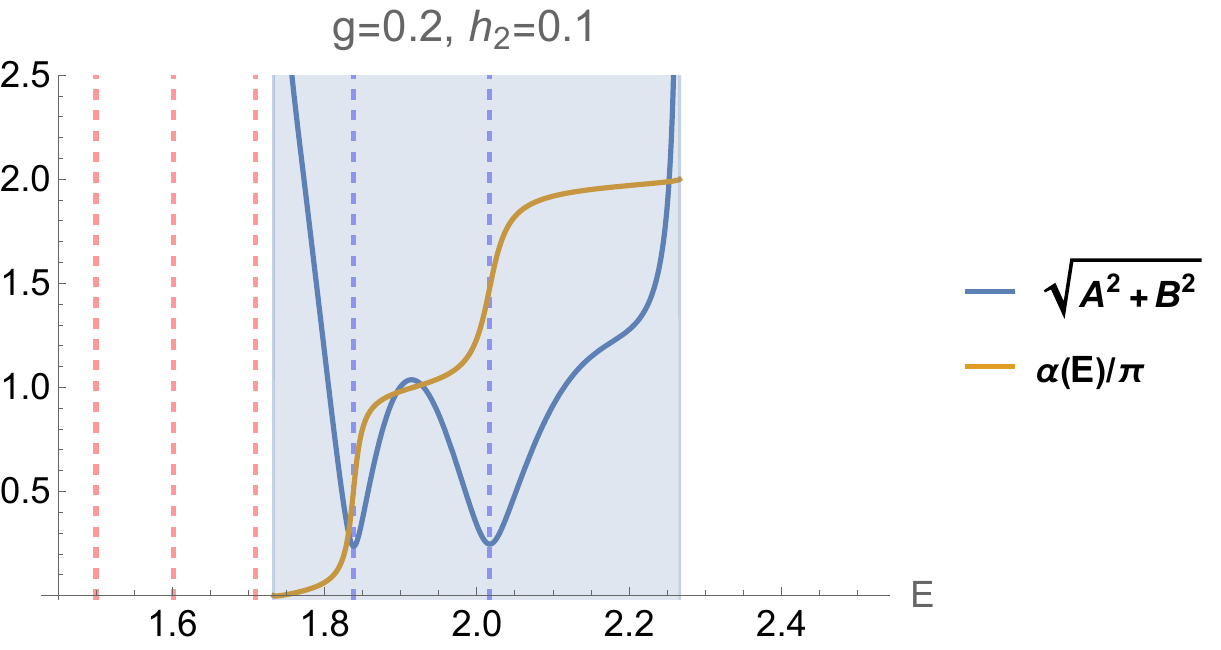}
   \caption{}
    \label{fig:1a}
\end{subfigure}
\hfill
\begin{subfigure}{0.46\textwidth}
    \includegraphics[width=\textwidth]{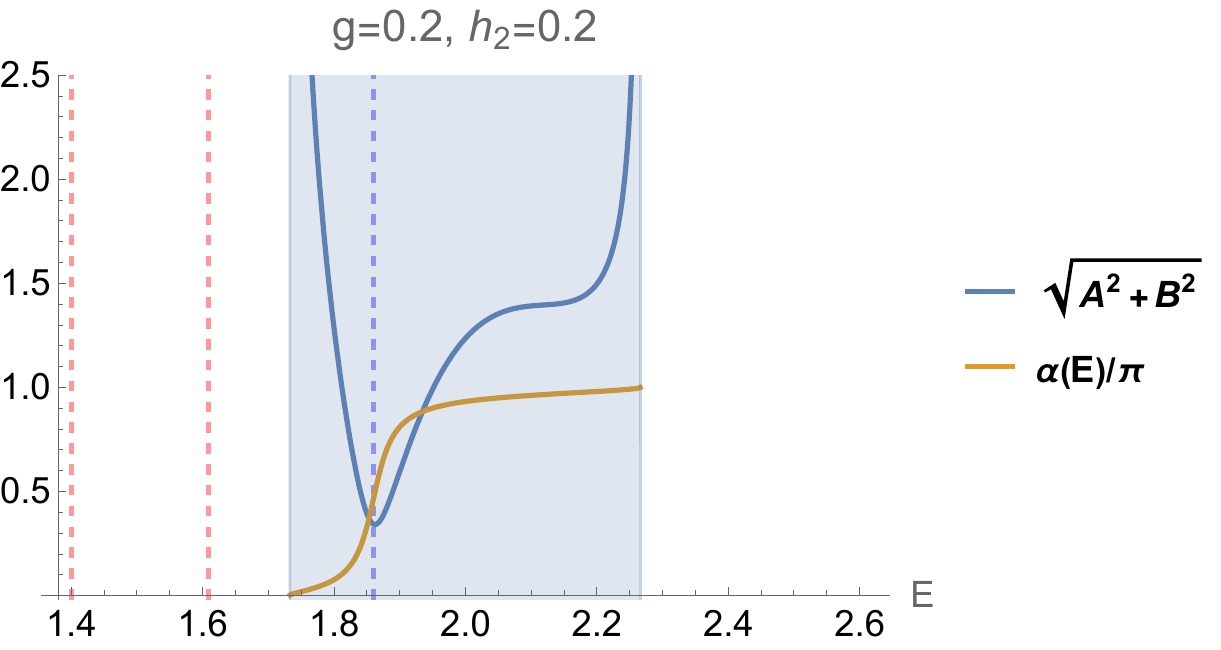}
   \caption{}
    \label{fig:2a}
\end{subfigure}
\hfill
\begin{subfigure}{0.46\textwidth}
    \includegraphics[width=\textwidth]{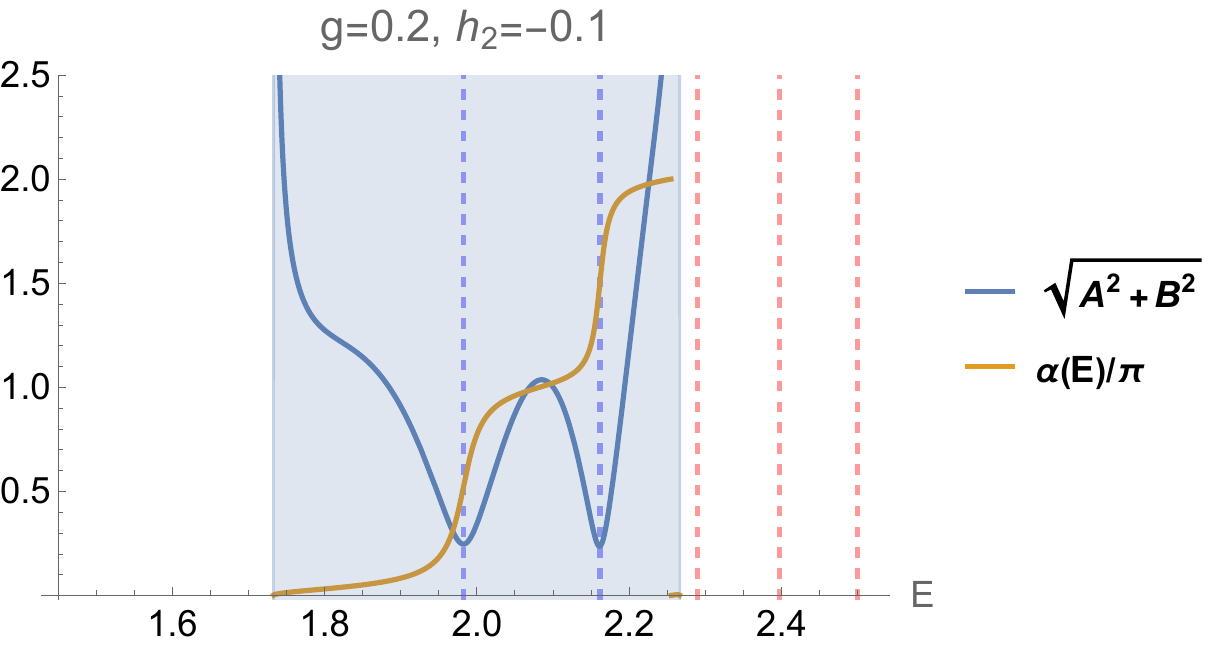}
   \caption{}
    \label{fig:-1a}
\end{subfigure}
\hfill
\begin{subfigure}{0.46\textwidth}
    \includegraphics[width=\textwidth]{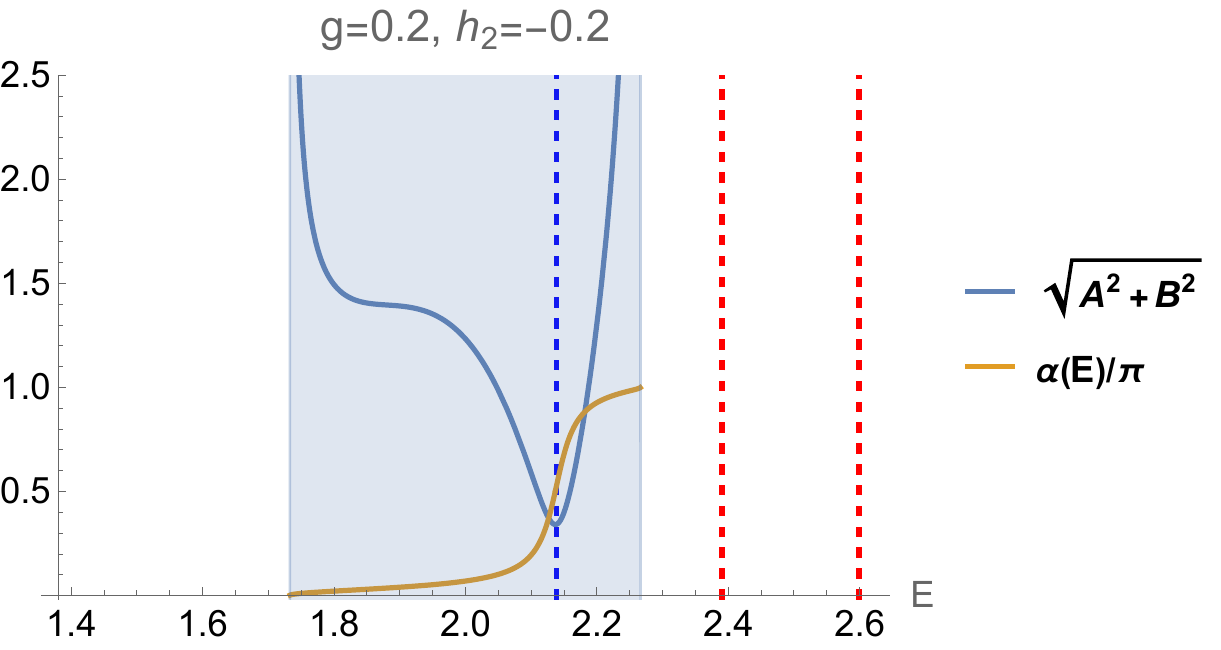}
   \caption{}
    \label{fig:-2a}
\end{subfigure}
\caption{Scattering Phase and Amplitude in the Oblique Regime. \\
Two-kink wave-function amplitude $\sqrt{A^{2}+B^{2}}$ (blue) and normalized scattering phase $\alpha/\pi=(-i\log S)/\pi$ (orange) as a function of total energy $E$ at $P=0$. Vertical red gridlines mark the energies of stable collisionless bubbles (a, b) and mesons (c, d), while vertical blue gridlines indicate resonance positions. The shaded blue region denotes the continuous two-kink spectrum $(E_{min}, E_{max})$. Subfigures correspond to $h_{2} = 0.1$ (a), $0.2$ (b), $-0.1$ (c), and $-0.2$ (d) at $g=0.2$.
 \label{fig:Mob}}
\end{figure}

To find the spectrum of bound states, we analytically continue the scattering amplitude to the complex $z$-plane, where it is a meromorphic function for a fixed real value of $v_2\ne 0$. It has two classes of poles:
\begin{itemize}
    \item Stable bound states: poles located on the real interval $z_n \in (-1, 1)$ correspond to stable two-kink bound states, with energies:
    \begin{equation}
    E_{n}=2 -\frac{2 g}{3}\, (z_n+z_n^{-1})\,.
    \end{equation} 
\begin{itemize}
    \item In the positive oblique regime ($v_2 > 0$), these poles are positive ($z_n \in (0, 1)$), representing collisionless bubbles.
    \item In the negative oblique regime ($v_2 < 0$), these poles are negative ($z_n \in (-1, 0)$), representing collisionless mesons.
\end{itemize}
\item Resonance states: poles $z_l$ that lie in the upper half-plane ($\text{Im } z_l > 0$) outside the unit circle ($|z_l| > 1$) determine the positions and widths of resonances
    \begin{equation}
    E_{\text{res},l}=2 -\frac{2 g}{3}\, (z_l+z_l^{-1})\,,
    \end{equation} 
where the real/imaginary parts give the energy/width, respectively.
\end{itemize}
Numerical values for these energies at $g = 0.2$ and $P = 0$ are detailed in Table \ref{tab:Energies}. The energy dependences of the wave-function amplitude $\sqrt{A^2 + B^2}$ and the scattering phase $\alpha=-i\log S$ are illustrated in Figure \ref{fig:Mob}. As shown in Fig. \ref{fig:Mob}, the stable collisionless particles appear as discrete energy levels (red gridlines) outside the continuous two-kink spectrum (shaded blue region). Conversely, resonance positions (blue gridlines) coincide with broader peaks within the continuum, where hybridisation between bound states and unconfined kinks occurs.
\begin{table}[!h]
\begin{center}
\label{Energies}
\begin{tabular}{ |c||c|c|c|c|c|c| }
\hline
$h_2$&$E_1$&$E_2$&$E_3$&$E_4$&$E_{res,1}$&$E_{res,2}$\\
\hline
\hline
0.1&1.70975&1.60194&1.50022&1.40001&1.8382 - 0.00898 i&2.0169 - 0.01813 i\\
\hline
0.2&1.60932&1.40079&1.20003&1.00000&1.8608 - 0.02246 i&\\
\hline
-0.1&2.29025&2.39806&2.49978&2.59999&1.9831 - 0.01813 i&2.1618 - 0.00898 i\\
\hline
-0.2&2.39071&2.5992&2.79997&3.00000&2.1392 - 0.02246 i&\\
\hline
\end{tabular}
\end{center}
\caption{Energy Spectra of Bound States and Resonances ($g=0.2, P=0$).\\
This table lists the discrete energies ($E_n$) of stable collisionless bubbles ($h_2 > 0$) and mesons ($h_2 < 0$), alongside the complex energies of resonance states ($E_{\text{res},l}$). The real part of the resonance energy indicates the peak position in the Fourier spectrum, while the imaginary part represents the decay width resulting from hybridisation with the unconfined two-kink continuum.}
\label{tab:Energies}
\end{table}
The complex energies $E_{\text{res},l}$ presented in Table \ref{tab:Energies} provide a quantitative basis for the quench spectroscopy features observed in the iTEBD simulations as we discuss later in Section \ref{sec:Pottsquench}. While the stable bound states ($E_n$) appear as sharp, delta-like peaks in quench spectroscopy, the resonance states are characterised by a finite decay width, determined by the imaginary part of the pole position in the complex energy plane.

\subsection{Trajectory of poles and zeroes of the scattering amplitude}

To understand the physical transformation of stable excitations into resonances, it is instructive to trace the evolution of the poles $p_n(v_2)$ and zeroes $k_n(v_2)$ of the two-kink scattering amplitude $\mathcal{S}(p,v_2)=S(z,v_2)|_{z=\exp(i p)}$ as the oblique parameter $v_2$ is tuned. Since the 
locations of poles $p_n(v_2)$ of the scattering amplitude determine the energies $E_n(v_2,g)$ of the two-kink bound states, 
\begin{equation}\label{Env}
E_n(v_2,g)=2-\frac{4 g}{3} \cos[ p_n(v_2)], \quad \text{if  }\, \mathrm{Im}\, e^{i p_n(v_2)}=0, \text{ and }  \mathrm{Re}\, e^{i p_n(v_2)}\in (-1,1),
\end{equation}
this gives us insight into the  behaviour of these energies near the stability threshold, and elucidates the transformation of the stable mesons (at $v_2<0$) and  bubbles (at $v_2>0$)  into the resonances. 

We represent the scattering amplitude in terms of the Jost function $F(p, v_2)$:
\begin{equation}\label{sca}
\mathcal{S}(p,v_2)=-\frac{F(-p,v_2)}{F(p,v_2)}\,,
\end{equation}
where the Jost function is defined as:
\begin{equation}\label{Jo}
F(p,v_2)=e^{i p}\,  J_{1-\frac{4 \cos p}{3 v_2}}\left(\frac{4}{3 |v_2|}
\right)+4\,\sign v_2\,\, J_{-\frac{4 \cos p}{3 v_2}}\left(\frac{4}{3 |v_2|}
\right)\,.
\end{equation}
This function is $2\pi$-periodic in $p$, and also analytic in the complex $p$-plane at any fixed real $v_2\ne0$.
The poles of the scattering amplitude \eqref{sca} are located at zeroes of the Jost function \eqref{Jo}:
\begin{equation}
F(p,v_2)|_{p=p_n(v_2)}=0.
\end{equation}
In the following, we focus on the positive oblique regime $v_2>0$. The analysis of the negative oblique regime $v_2<0$ is entirely parallel, since the location of poles $p_n(v_2)$  in the negative  oblique regime $v_2<0$  can be found from the equality
\begin{equation}
p_n(-v_2)=p_n(v_2)+\pi,
\end{equation}
that follows from  the symmetry relation $F(p+\pi,-v_2)=-F(p,v_2)$.

\begin{figure}
\centering
\begin{subfigure}{0.45\textwidth}
    \includegraphics[width=\textwidth]{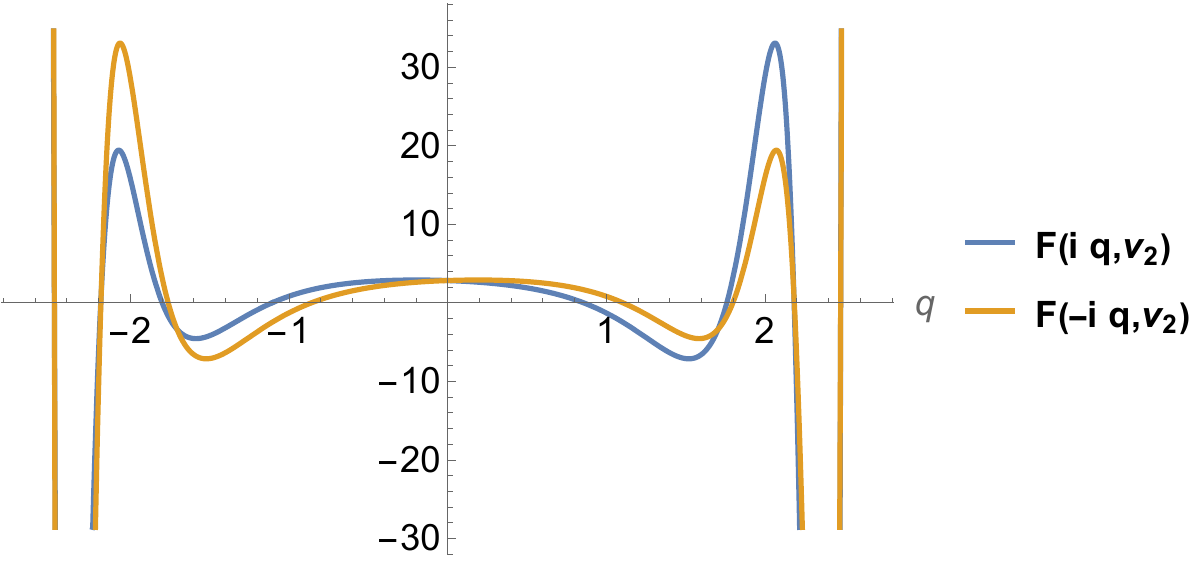}
   \caption{$v_2=2.0$}
    \label{fig:Jost2}
\end{subfigure}
\hfill
\begin{subfigure}{0.45\textwidth}
    \includegraphics[width=\textwidth]{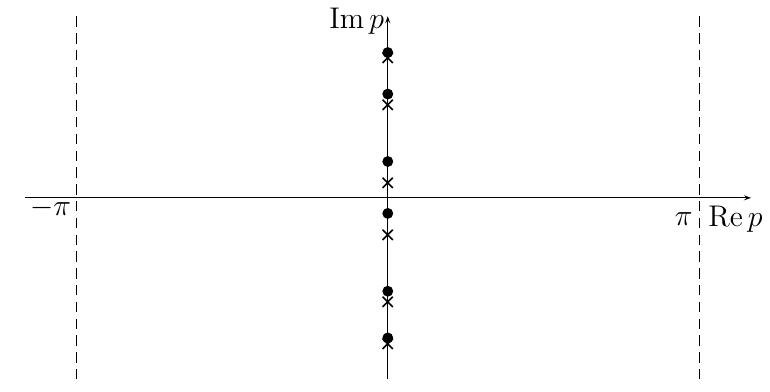}
   \caption{$v_2=2.0$}
    \label{fig:sc2}
\end{subfigure}
\hfill
\begin{subfigure}{0.45\textwidth}
    \includegraphics[width=\textwidth]{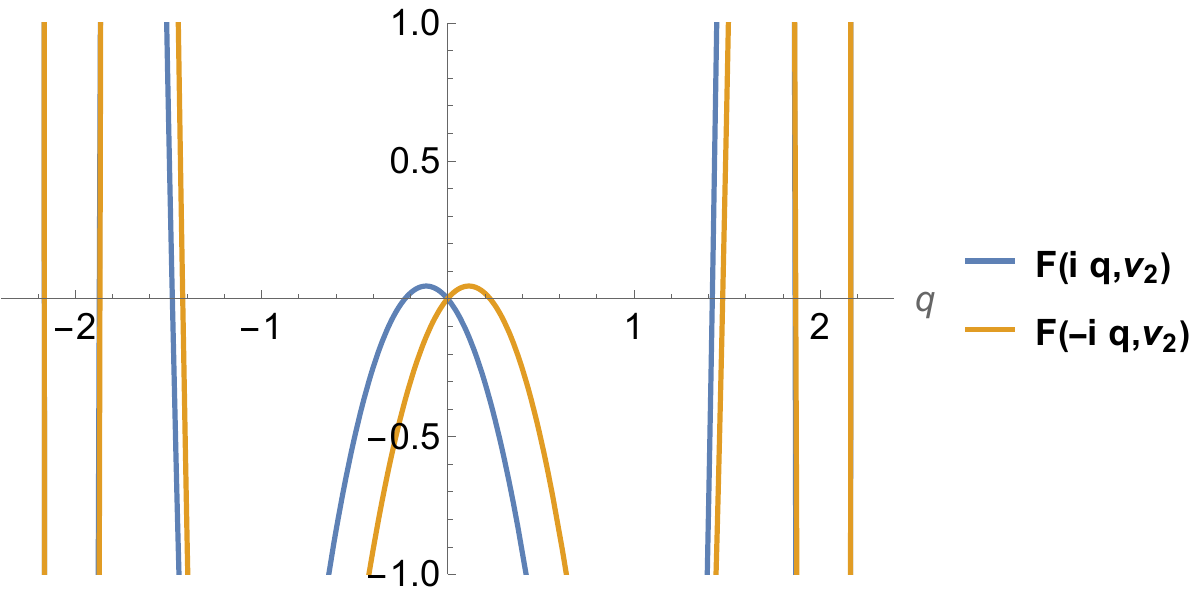}
    \caption{$v_2=1.47112\ldots$}
    \label{fig:Jost1a}
\end{subfigure}
\hfill
\begin{subfigure}{0.45\textwidth}
    \includegraphics[width=\textwidth]{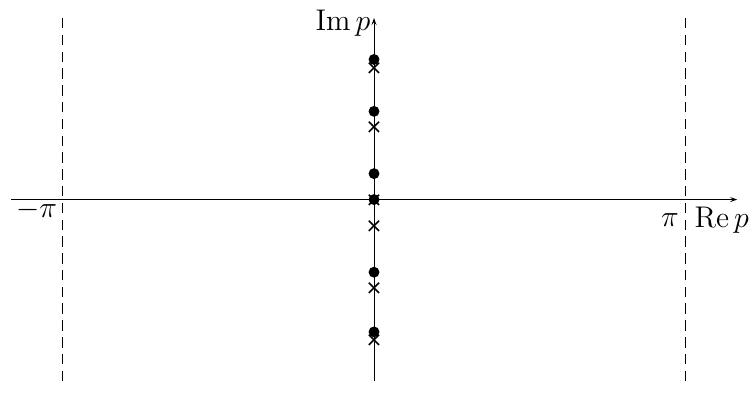}
    \caption{$v_2=1.47112\ldots$}
    \label{fig:sc1a}
\end{subfigure}
\hfill
\begin{subfigure}{0.45\textwidth}
    \includegraphics[width=\textwidth]{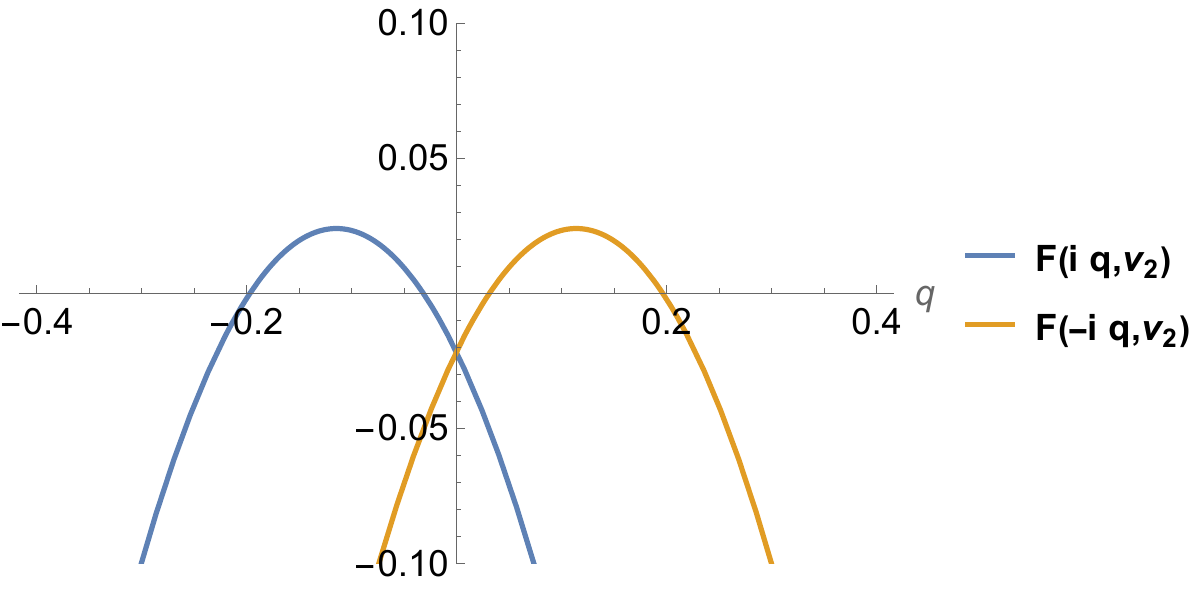}
    \caption{$v_2=1.468$}
    \label{fig:Jost3}
\end{subfigure}
\hfill
\begin{subfigure}{0.45\textwidth}
    \includegraphics[width=\textwidth]{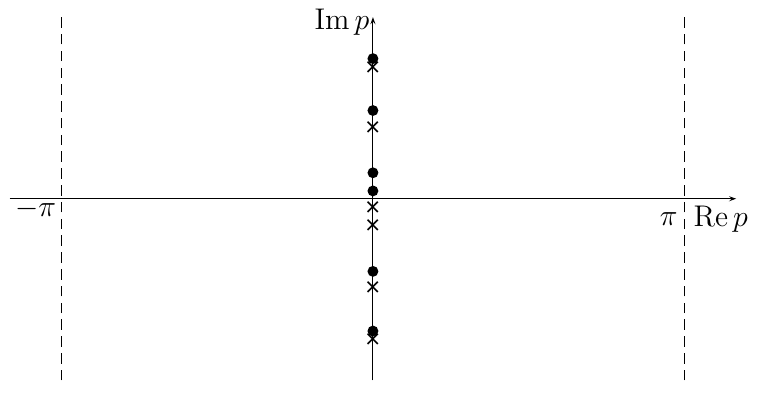}
    \caption{$v_2=1.468$}
    \label{fig:sc3a}
\end{subfigure}
\hfill
\begin{subfigure}{0.45\textwidth}
    \includegraphics[width=\textwidth]{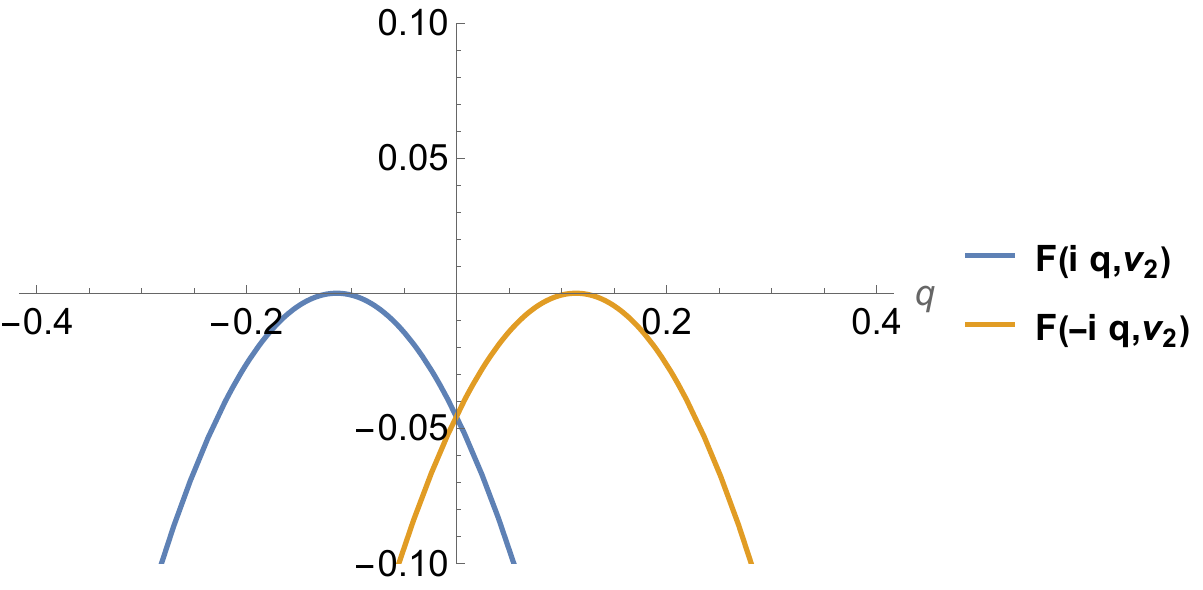}
      \caption{$v_2=1.4646\ldots$}
    \label{fig:third1}
\end{subfigure}
\hfill
\begin{subfigure}{0.45\textwidth}
    \includegraphics[width=\textwidth]{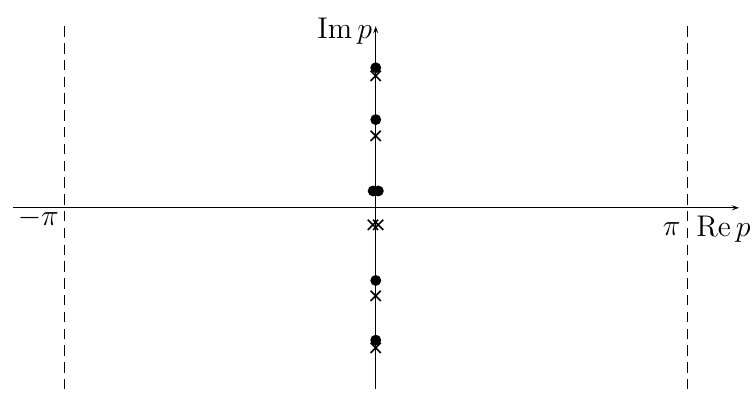}
    \caption{$v_2=1.4646\ldots$}
    \label{fig:sc4a}
\end{subfigure}
\hfill
\begin{subfigure}{0.45\textwidth}
    \includegraphics[width=\textwidth]{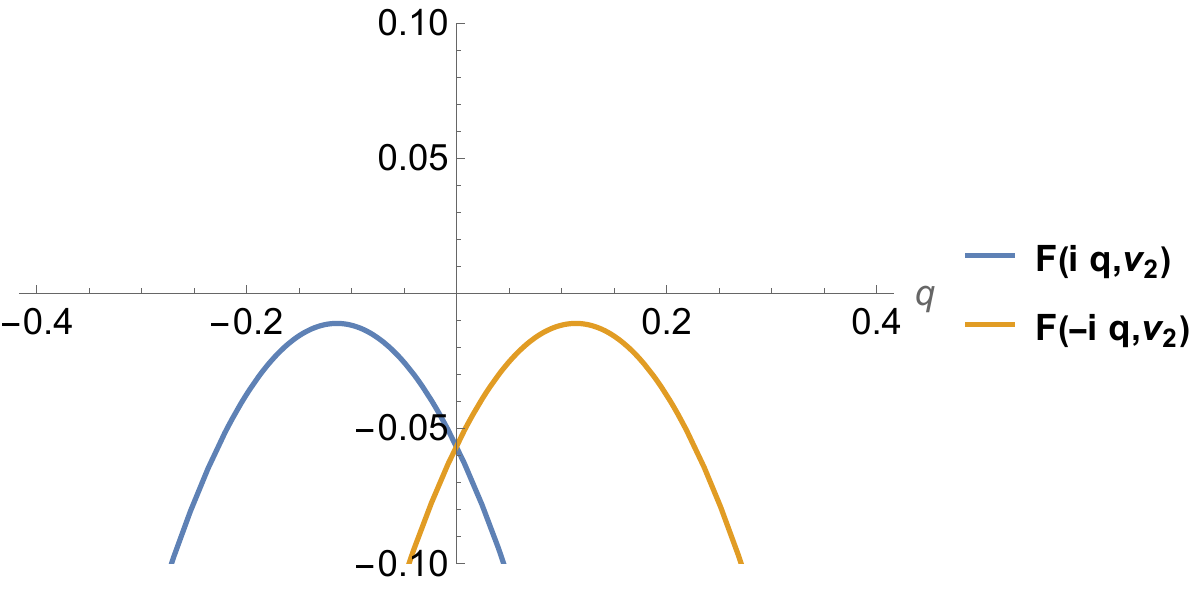}
      \caption{$v_2=1.463$}
    \label{fig:third}
\end{subfigure}
\hfill
\begin{subfigure}{0.45\textwidth}
    \includegraphics[width=\textwidth]{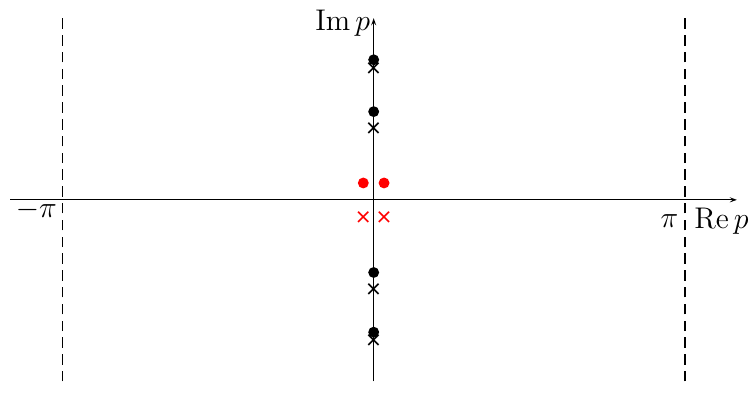}
    \caption{$v_2=1.463$}
    \label{fig:sc5a}
\end{subfigure}       
\caption{Evolution of the Jost functions $F(\pm iq, v_{2})$ (shown by blue/orange curves in the left panel) and the corresponding poles (crosses) and zeroes (dots) of the two-kink scattering amplitude $\mathcal{S}(p, v_{2})$ in the complex $p$-plane (right panel) as the oblique parameter $v_{2}$ is decreased. The sequence (a–j) illustrates the "Fonseca-Zamolodchikov" scenario \cite{FZ06}: a stable bound-state pole moves toward the origin $p=0$, crosses into the lower half-plane as a virtual state, and eventually gains a real component to transform into a resonance.
}
\label{fig:figures}
\end{figure}

At sufficiently large values of $v_2$, the zeroes of the Jost function $F(p,v_2)$ in the strip $-\pi<\mathrm{Re}\, p\le \pi$ are exclusively on the imaginary $p$-axis, as illustrated  in Figures \ref{fig:Jost2}, \ref{fig:sc2} for $v_2=2$. The left subfigure \ref{fig:Jost2} shows the momentum dependences of the Jost functions $F(p,v_2)$, and $F(-p,v_2)$ at purely imaginary momenta $p=i q$. According to \eqref{sca}, the zeroes of these functions determine the locations of the poles and the zeroes of the scattering amplitude $\mathcal{S}(p,v_2)$, which are schematically displayed in Fig. \ref{fig:sc2}. These figures demonstrate that the poles and zeroes of the function $\mathcal{S}(p,v_2)$ alternate on the imaginary $p$-axis forming pairs.  In the `physical' upper half-plane  $\mathrm{Im}\, p>0$, the poles  $p_n(v_2)$  
and zeroes  $k_n(v_2)$ are ordered as:
\begin{equation}
0<\mathrm{Im}\, p_1(v_2)<\mathrm{Im}\, k_1(v_2)<\mathrm{Im}\, p_2(v_2)<\mathrm{Im}\, k_2(v_2)<\ldots.
\end{equation}
In the lower half-plane  $\mathrm{Im}\, p<0$, poles and zeroes of the scattering amplitude $\mathcal{S}(p,v_2)$ 
are located at $\{-k_n(v_2)\}_{n=1}^\infty$, and at $\{-p_n(v_2)\}_{n=1}^\infty$, respectively, as required by the unitarity
relation: 
\begin{equation}
\mathcal{S}(p,v_2)\mathcal{S}(-p,v_2)=1.
\end{equation}
Upon decreasing the parameter $v_2$, all poles and zeroes of the scattering amplitude move along the imaginary $p$-axis
towards the origin $p=0$. When the parameter $v_2$ reaches its first critical value $a_1=1.47112\ldots$, the pole $p_1(v_2)$
arrives at the origin $p_1(a_1)=0$, where it merges with the zero of the function $\mathcal{S}(p,v_2)$ arriving from the lower half-plane. This configuration of poles and zeroes of the function $\mathcal{S}(p,a_1)$ is shown in Fig. \ref{fig:sc1a}.
Fig. \ref{fig:Jost1a} illustrates the $q$-dependencies of the Jost functions $F(i q,v_2)$ and $F(-i q,v_2)$ at $v_2=a_1$. 

Note, that at $v_2=a_1$, the energy of the first bubble  $E_1(v_2,g)$ reaches its stability threshold, which 
coincides with the lower bound of the two-kink continuum spectrum:
\begin{equation}
E_1(a_1,g)=2-\frac{4g}{3}=E_\text{min}(g).
\end{equation}
Since the function $p_1(v_2)$ remains analytic in $v_2$ in the vicinity of the critical point $a_1$, it varies at $v_2\to a_1$ as:
\begin{equation}\label{p1v2}
p_1(v_2)=i c_1(v_2-a_1)+ O\left((v_2-a_1)^2\right),
\end{equation}
with some  $c_1>0$. Due to \eqref{Env}, the difference $E_1(v_2,g)-E_\text{min}(g)$ vanishes quadratically in $(v_2-a_1)$, when $v_2$ approaches the 
critical value $a_1$ from above:
\begin{equation}\label{E1v2}
E_1(v_2,g)=E_\text{min}(g)-\frac{2 g}{3}c_1^2\, (v_2-a_1)^2+O\left((v_2-a_1)^3\right).
\end{equation}
\begin{figure}
\centering
\includegraphics[width=0.7\linewidth]{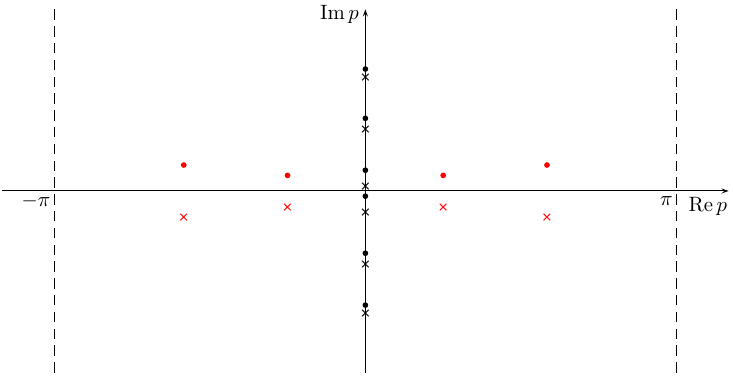}
\caption{ Pole Configuration in the Deep Oblique Regime.\\
Schematic distribution of poles (crosses) and zeroes (dots) of the scattering amplitude $\mathcal{S}(p, v_{2})$ in the complex $p$-plane for $v_{2} \simeq 0.5$. Black crosses on the positive imaginary axis represent stable collisionless bubbles. Red crosses shifted away from the axis represent resonance states characterized by a finite decay width.
 \label{fig:sc}}
\end{figure}
Fig. \ref{fig:Jost3} displays the profile of the Jost functions at $v_2=1.468$, which is slightly below the first critical value $a_1$, with the corresponding configuration of poles and zeroes shown in Fig. \ref{fig:sc3a}. The pole $p_1(v_2)$ drops into the lower half-plane $\mathrm {Im}\, p<0$ remaining purely imaginary. Formulas \eqref{p1v2}, and \eqref{E1v2} still can be used in this case, but no stable excitation exists at $v_2<a_1$, with the energy given by Eq. \eqref{E1v2} corresponding to the so-called `virtual level', see e.g \cite{taylor2012scattering}.

As the parameter $v_2$ is decreased further to the second critical value $v_2=b_1=1.4646\ldots$, two poles $p_1(v_2)$ and $-k_1(v_2)$ 
of the scattering amplitude $\mathcal{S}(p,v_2)$ merge at the negative $p$-half-axes, as it is shown in Fig. \ref{fig:sc4a}, with
the functions  $p_1(v_2)$, $k_1(v_2)$,  and $E_1(v_2,g)$ having a square-root branching point at $v_2=b_1$.

Decreasing $v_2$ further leads to the situation shown in Figure
\ref{fig:sc5a}, where the two poles shown by red crosses, which are located in the lower half-plane, are shifted away from the imaginary $p$-axis by gaining non-zero real parts $\sim \pm \sqrt{b_1-v_2}$. The corresponding energy $E_1(v_2,g)$, in turn, gains a negative imaginary part $ \sim - \sqrt{b_1-v_2}$, indicating that the virtual state transformed into a resonance. 

As the parameter $v_2$ decreases further, the above scenario repeats and the bubbles enumerated by the indices $n=2,3,\ldots$ transfer one by one into resonances. Fig.~\ref{fig:sc} shows the resulting configuration of poles and zeroes of the scattering amplitude $\mathcal{S}(p,v_2)$ at $v_2\simeq 0.5$ schematically.

\section{Evolution of magnetisation after a quantum quench \label{sec:Pottsquench}}
In this section, we apply a perturbative analysis to the post-quench dynamics of the three-state Potts spin chain. Our goal is to calculate the time-dependent local magnetisation, $M_\mu(t)$, up to the second order in the transverse field $g$.

Before delving into the calculation itself, we note that perturbative approaches to quantum quenches have already been explored in the context of massive quantum field theories \cite{2014JPhA...47N2001D,2017JPhA...50h4004D,2018ScPP....5...27H,2019JHEP...08..047H,Rut80,Rut12}. This approach works for small quenches, where the quench magnitude can be defined as the post-quench energy density relative to the gap scale. One of the main drawbacks of this method is that it is based on the pre-quench Hilbert space, whereas the time evolution of the quench is determined by the post-quench spectrum. 

An alternative approach is to expand in the post-quench Hilbert space using a form-factor approach \cite{2012JSMTE..04..017S,2014JSMTE..10..035B}, which has two main advantages. First, the time evolution has the correct frequency spectrum of the post-quench energy system \cite{2018ScPP....5...27H}. Second, resummation of secular contributions to the time evolution reproduces the long-time asymptotic behaviour of observables, most notably their exponential relaxation, and also gives a systematic perturbative expansion for the relaxation time. The relaxation time obtained this way for the Ising field theory was found to agree well with numerical simulations \cite{2016NuPhB.911..805R}. The drawback of the post-quench form factor expansion is that it requires knowledge of the energies of eigenstates of the post-quench Hamiltonian, their overlaps with the initial states, as well as the matrix elements of observables in the eigenstate basis. This information is available when both the post-quench Hamiltonian is integrable, and the initial state corresponds to a so-called integrable quench \cite{2017NuPhB.925..362P,2019ScPP....6...62P}, 
which is the case considered in \cite{2012JSMTE..04..017S,2014JSMTE..10..035B}.

In order to demonstrate the efficiency of our perturbation technique, we apply it in Appendix \ref{ApIs} to describe 
the evolution of the magnetisation in the Ising spin chain in the limit of the weak transverse magnetic field in the post-quench Hamiltonian. The comparison of the perturbative results with the well-known form-factor result clarifies the way in which our perturbative expansion works. Essentially, it can only capture the time evolution on sufficiently short time scales, provided the quench itself is suitably small. 

Since the Potts model is non-integrable, we can only resort to a perturbative approach to construct its time evolution. Despite its limitations, we demonstrate that it agrees reasonably well with the numerically computed time evolution in Ref. \cite{Po25}. In addition, we also demonstrate that the spectrum in the two-kink sector, constructed above in Sections \ref{sec:PSC} and \ref{analytic_S}, shows an excellent agreement with the Fourier spectrum obtained by quench spectroscopy. 

\subsection{Perturbation theory for Potts quench dynamics}
We explore essentially the same quench protocol as that studied in \cite{Po25}. We start from the finite Potts spin chain having an even number of sites $N$, whose Hamiltonian is given by a simple modification of \eqref{eq:Ham1}:
\begin{equation}\label{HamN}
H_N(\mathbf{v},g)=-\sum_{j=-N/2+1}^{N/2} \sum_{\mu=1}^{3} \left( -\frac{2}{3}+P^{\mu}_j P^{\mu}_{j+1} +gv_\mu(P^{\mu}_j-c^\mu)\right)-g 
\sum_{j=-N/2+1}^{N/2} \tilde{P}_j\,,
\end{equation}
where the constant shift is specified by $c^1=2/3$, and $c^2=c^3=-1/3$.  The initial state is purely ferromagnetic along the direction $1$:
\begin{equation}
|0\rangle^{(1)}=\bigotimes_{j=-N/2+1}^{N/2}|1\rangle_j.
\label{eq:init_state}
\end{equation}
For further calculations, let us denote by $ |\text{vac}(\mathbf{v},g|N)\rangle$ the deformation of the vacuum $|0\rangle^{(1)}$ of the Hamiltonian $H_N(\mathbf{0},0)$:
\begin{equation}
    H_N(\mathbf{v},g) |\text{vac}(\mathbf{v},g|N)\rangle=E_\text{vac}(\mathbf{v},g|N)|\text{vac}(\mathbf{v},g|N)\rangle.
\end{equation}
It is convenient to represent the deformed vacuum  $|\text{vac}(\mathbf{v},g|N)\rangle$  in the form:
\begin{equation}
|\text{vac}(\mathbf{v},g|N)\rangle=\frac{|\Omega(\mathbf{v},g|N)\rangle}{[\langle\Omega(\mathbf{v},g|N))|\Omega(\mathbf{v},g|N))\rangle]^{1/2}}, 
\end{equation} 
where the  vector $|\Omega(\mathbf{v},g|N)\rangle$ admits the  following asymptotic expansion at $g\to0$:
\begin{equation}\label{dvac}
|\Omega(\mathbf{v},g|N)\rangle=|0\rangle^{(1)}+g |\Omega_1(\mathbf{v}|N)\rangle+g^2 |\Omega_2(\mathbf{v}|N)\rangle+\ldots, 
\end{equation} 
such that
\[
\phantom{.}^{(1)}\langle 0| \Omega_n(\mathbf{v}|N)\rangle=0, \quad\text{for all } n=1,2,\ldots.
\]
The details of the perturbative analysis of the true/false vacuum states are relegated to the Appendix \ref{sec:vac}.

At $t>0$, the average value $A_\alpha(t|N)$ of the observable corresponding to the  operator $\hat{A}_\alpha$ reads:
\begin{equation}\label{timAv}
A_\alpha(t|N)=\phantom{.}^{(1)}\langle 0|\exp[i H_N(\mathbf{v},g)t]\, \hat{A}_\alpha\,\exp[-i H_N(\mathbf{v},g)t]|0\rangle^{(1)}.
\end{equation}
We consider the following three operators, which are basically the three magnetisations shifted by a constant:
\begin{subequations}\label{A123}
\begin{align}
& \hat{A}_1=\frac{1}{N}\sum_{j=-N/2+1}^{N/2} \left(P_j^1-\frac{2}{3}\right)=\hat{M}_1-\frac{2}{3},\\
 & \hat{A}_2=\frac{1}{N}\sum_{j=-N/2+1}^{N/2} \left(P_j^2+\frac{1}{3}\right)=\hat{M}_2+\frac{1}{3},\\
  & \hat{A}_3=\frac{1}{N}\sum_{j=-N/2+1}^{N/2} \left(P_j^3+\frac{1}{3}\right)=\hat{M}_3+\frac{1}{3}.
\end{align}
\end{subequations}
where $\hat{M}_\alpha$ are the three magnetisation operators  with $\alpha=1,2,3$. Since 
\begin{equation}
  \hat{A}_1+  \hat{A}_2+  \hat{A}_3=0,
\end{equation}
only two of the operators are independent. The operators $\hat{A}_\alpha$ are preferred over the magnetisations $\hat{M}_\alpha$ because they satisfy the relation 
\begin{equation}
\hat{A}_\alpha|0\rangle^{(1)}=0,
\end{equation}
which leads to a substantial simplification of the calculation of the expectation value \eqref{timAv}.
The perturbative calculation of the aforementioned expectation value is discussed in detail in Appendix \ref{sec:calc_potts_dyn_app}. Here, we only summarise the key ideas and the results of the computation.

To calculate the $N\rightarrow\infty$ limit, necessary for the subsequent comparison with the numerical results, we introduce the following notations as well \begin{subequations}\label{ThLim_maintxt}
\begin{align}
&A_\alpha(t):=\lim_{N\to\infty} A_\alpha(t|N),\qquad|\Omega_1(\mathbf{v}) \rangle:=\lim_{N\to\infty}  |\Omega_1(\mathbf{v}|N) \rangle,\\\label{dHam}
&\Delta H(\mathbf{v},g):=\lim_{N\to\infty}\left[
H_N(\mathbf{v},g)-E_\text{vac}(\mathbf{v},g|N)
\right].
\end{align}
\end{subequations}
After taking the thermodynamic limit, we apply a further key simplification, provided by the two-kink approximation, which essentially corresponds to replacing the Hamiltonian $\Delta H(\mathbf{v},g)$ by its restriction to the subspace $\mathcal{L}_{11}^{(2)}$:
\begin{align}\label{DHam}
&\Delta H(\mathbf{v},g) \to  H^{(2)}(\mathbf{v},g)={\mathcal{P}}_{11}^{(2)}\,   \Delta H(\mathbf{v},g)\,  {\mathcal{P}}_{11}^{(2)}\,.
\end{align}
Using the results of Appendix \ref{sec:calc_potts_dyn_app}, this approximation implies that
\begin{align}\label{2k}
&A_\alpha(t)\to A_\alpha^{(2)}(t)
=g^2\,\langle   \Omega_1(\mathbf{v})|\hat{A}_\alpha | \Omega_1(\mathbf{v})\rangle-
2 g^2 \, 
\mathrm{Re}\, \langle  \Omega_1(\mathbf{v})|
\hat{A}_\alpha\, e^{-i t\,H^{(2)}(\mathbf{v},g)}|\Omega_1(\mathbf{v}) \rangle
\\\nonumber
&+g^2 \, \langle  \Omega_1(\mathbf{v})|e^{i t\,H^{(2)}(\mathbf{v},g)}
\hat{A}_\alpha e^{-i t\,H^{(2)}(\mathbf{v},g) }|\Omega_1(\mathbf{v}) \rangle+O(g^3). 
\end{align}
The first term on the right-hand side of \eqref{2k} is time independent and can be expressed as shown in Eq.$\,$\eqref{eq:cst_term}. The second term describes the resonant
oscillations of the magnetisations $A_\alpha^{(2)}(t)$ with high frequencies $\omega\simeq 2+O(g)$,
while the last one  represents the low-frequency contribution to the quantities 
$A_\alpha^{(2)}(t)$.

We expect that at $g\ll1$, the difference between $A_\alpha(t)$ and $A_\alpha^{(2)}(t)$ remains small
during the first stage of the evolution, i.e. for not too large $t$. However, at sufficiently large $t$, the multi-kink contributions
to the state $|\Psi(t)\rangle=e^{- i t\, \Delta H(\mathbf{v},g)}|\Omega_1(\mathbf{v})\rangle $ should become considerable, as the full Hamiltonian does not conserve the number of kinks. This leads to a substantial deviation of $A_\alpha(t)$ from 
its two-kink approximation $A_\alpha^{(2)}(t)$. We return to this issue later. 

\begin{figure}
\centering
\begin{subfigure}{0.45\textwidth}
    \includegraphics[width=\textwidth]{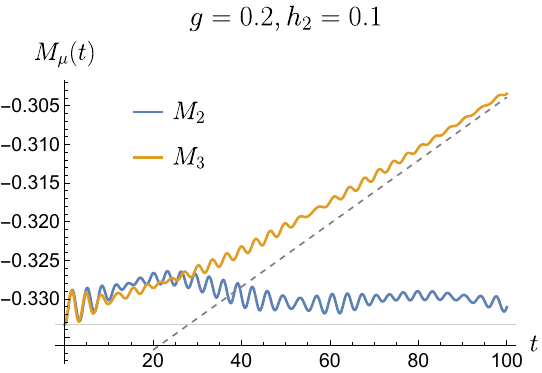}
   \caption{}
\label{fig:M2t}
\end{subfigure}
\hfill
\begin{subfigure}{0.45\textwidth}
    \includegraphics[width=\textwidth]{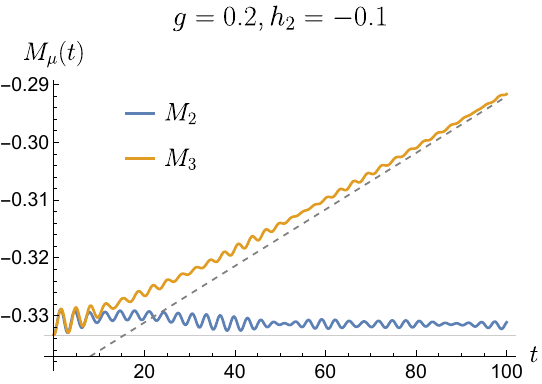}
    \caption{}
 \label{fig:Mt}
\end{subfigure}
\caption{Analytical Magnetisation Dynamics at $g=0.2$.\\
Predicted time evolution for magnetizations $M_{2}(t)$ (blue) and $M_{3}(t)$ (orange) in the positive ($h_{2}=0.1$) and negative ($h_{2}=-0.1$) oblique regimes. The dynamics is calculated using the two-kink approximation in equation \eqref{2k}. Dashed grey lines represent the predicted large-$t$ linear slope for $M_{3}(t)$ as derived in equations \eqref{eq:inf_time_behav} and\eqref{eq:cs_numvals}.
}
 \label{fig:M}
\end{figure}

In the rest of this Section, we concentrate on the oblique regimes and put $\mathbf{v}=(0,h_2/g,0)$.  We use the notation $H^{(2)}(h_2,g)$ for the Hamiltonian, instead of $H^{(2)}(\mathbf{v},g)$. The calculation of the second and third terms of the right-hand side of the formula \eqref{2k} in this regime is explained in detail in Appendix \ref{sec:calc_potts_dyn_app}. The resulting post-quench evolution of the magnetisations $M_2(t)$, and $M_3(t)$ is displayed in Fig. \ref{fig:M} at $g=0.2$ in the positive (a),
and negative (b) oblique regimes with $h_2=0.1$, and $h_2=-0.1$, respectively. As one can see from Fig. \ref{fig:M}, the magnetisation $M_3(t)$ linearly increases at large $t$ in  the initial stage
of the post-quench evolution, which is described by our perturbative approach:
\begin{equation}
M_3(t)+\frac{1}{3}= c\,  t+O(1), \quad \text{at }t\to\infty.
\label{eq:inf_time_behav}
\end{equation}
The coefficient $c$ that stands in this formula depends, of course, on the Hamiltonian parameters 
$h_2$ and $g$. By the means of Appendix \ref{sec:calc_potts_dyn_app}, the explicit form of $c$ reads as
\begin{equation}\label{ch2g}
c=\frac{g^3}{54 \pi}\int_0^\pi dp\, \sin p \, \frac{ \left[\psi_2(1;E,0)+\psi_3(1;E,0)\right]^2}{|\mathrm{B}(p)|^2}
\end{equation}
where the wave-functions $\psi_\nu(j;E,P=0)$ are determined by the equations \eqref{eq:psi3_P0} and \eqref{eq:psi2_P0}, $E(p)=2-\frac{4 g}{3}\cos p$ and $B(p)$ is defined as
\begin{equation}
    \mathrm{B}(p):=B_{in} (e^{ip},h_2/g)
\end{equation}
where the function $B_{in}(z,v_2)$ is given by \eqref{Bin1}. For the values of the parameters of the Hamiltonian chosen in Fig.  \ref{fig:M}, one finds:
\begin{equation}
c=\begin{cases}
4.091\cdot 10^{-4},& \text{at }g=0.2, \, h_2=0.1, \\ 4.895 \cdot 10^{-4},& \text{at }g=0.2, \, h_2=-0.1.
\label{eq:cs_numvals}
\end{cases}
\end{equation}
For later use, we also include the Fourier transform of the expression \eqref{2k}. Here, we only focus on the second term on the right-hand side of \eqref{2k}, which is the only contributing term in the relevant frequency domain, as explained in the next section.

Using the results of Appendix \ref{sec:calc_potts_dyn_app}, the real part of the Fourier transform
\begin{equation}
    F_\alpha^{(1)}(\omega):= \int_0^\infty dt \,e^{-i \omega t} \, \left[-2 g^2 \, 
\mathrm{Re}\,  \langle  \Omega_1(h_2)|\hat{A}_\alpha\, e^{-i tH^{(2)}(h_2,g)}|\Omega_1(h_2) \rangle\right]\,,\label{eq:FT_analitical_def}
\end{equation}
for the relevant frequency domain $\omega\in (2-\frac{4g}{3},2+\frac{4g}{3}) $, takes the form:
\begin{subequations}\label{Ftrao_F1F2}
\begin{align}
&\mathrm{Re}\, F_1^{(1)}(\omega)=\frac{3 g}{8 \sin p}\,\frac{[\psi_2(1,\omega, 0)+\psi_3(1,\omega, 0)]^2}{36|\mathrm{B}(p)|^2 }\,,\\
&\mathrm{Re}\, F_2^{(1)}(\omega)=-\frac{3 g}{8 \sin p}\,\frac{\psi_2(1,\omega, 0)\, [\psi_2(1,\omega, 0)+\psi_3(1,\omega, 0)]}{36|\mathrm{B}(p)|^2 }\,,
\end{align}
\end{subequations}
where 
\begin{equation}
p=\arccos \frac{3( 2-\omega)}{4 g}.
\end{equation}

\subsection{Comparison to numerical simulations \label{numerics}}

\subsubsection{Time evolution of the magnetisation}
To validate the perturbative results, we realised the quench protocol described at the beginning of Section \ref{sec:Pottsquench} using the infinite Time Evolving Block Decimation (iTEBD) method following Ref. \cite{Po25}. The details of the simulations are provided in Appendix \ref{Numerical_details}. 

The initial state is chosen as $|0\rangle^{(1)}$ given by the $N\rightarrow\infty$ limit of (\ref{eq:init_state}), which is evolved in time by the Hamiltonian (\ref{eq:Ham1}) with exactly one non-zero finite longitudinal field in each case, as discussed previously. We examine four distinct scenarios: $h_1>0$ (positively aligned), $h_1<0$ (negatively aligned), $h_2>0$ (positive oblique) and $ h_2<0$ (negative oblique). During the time evolution, we calculated the local magnetisations 
\begin{equation}
     M_\mu(t)  = \langle\psi(t)|P^\mu|\psi(t)\rangle.
\end{equation}
In aligned quenches, symmetry reduces the system to have a single independent magnetisation (chosen as $M_1$), whereas oblique quenches require two (typically chosen as $M_2$ and $M_3$). The underlying reason is that the three magnetisations sum up to zero, and in the aligned cases, the transformation swapping the spin directions 2 and 3 (leaving 1 intact) is a symmetry of the quench.  

\begin{figure}
\centering
\begin{subfigure}{0.45\textwidth}
    \includegraphics[width=\textwidth]{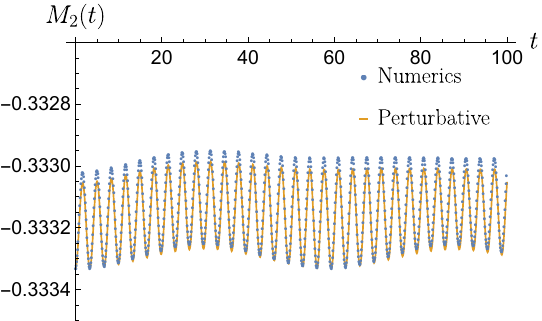}
    \caption{$g=0.05$, $h_2=0.1$}
    
\end{subfigure}
\hfill
\begin{subfigure}{0.45\textwidth}
    \includegraphics[width=\textwidth]{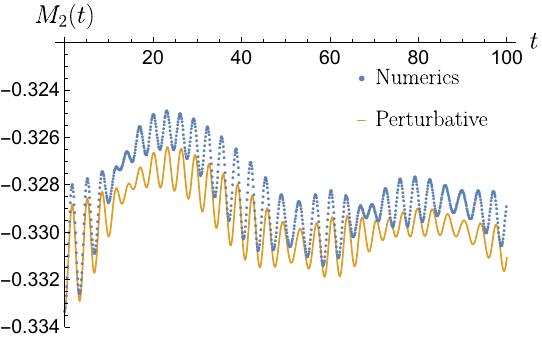}
    \caption{$g=0.2$, $h_2=0.1$}
    
\end{subfigure}
\hfill
\begin{subfigure}{0.45\textwidth}
    \includegraphics[width=\textwidth]{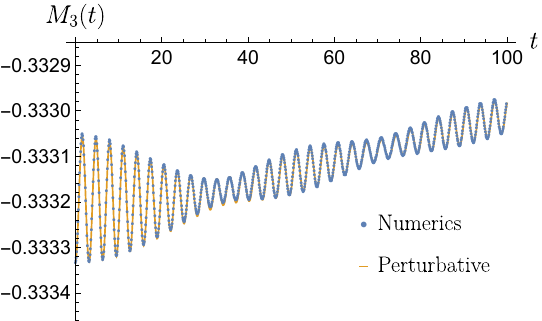}
    \caption{$g=0.05$, $h_2=0.1$}
    
\end{subfigure}
\hfill
\begin{subfigure}{0.45\textwidth}
    \includegraphics[width=\textwidth]{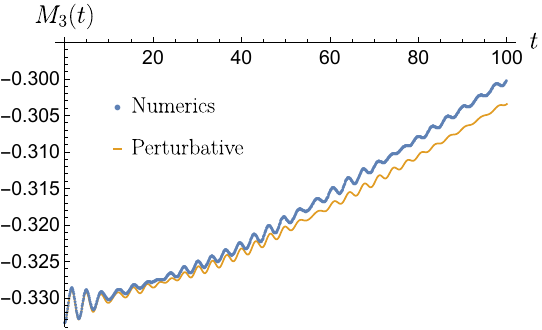}
    \caption{$g=0.2$, $h_2=0.1$}
    
\end{subfigure}
\hfill
\begin{subfigure}{0.45\textwidth}
    \includegraphics[width=\textwidth]{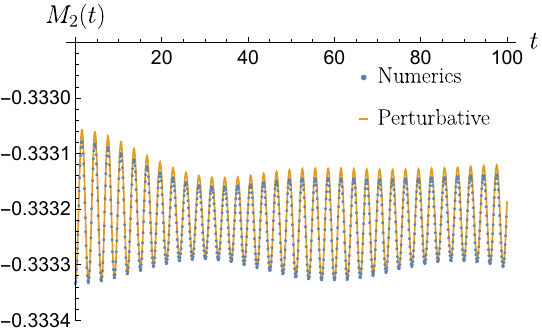}
    \caption{$g=0.05$, $h_2=-0.1$}
    
\end{subfigure}
\hfill
\begin{subfigure}{0.45\textwidth}
    \includegraphics[width=\textwidth]{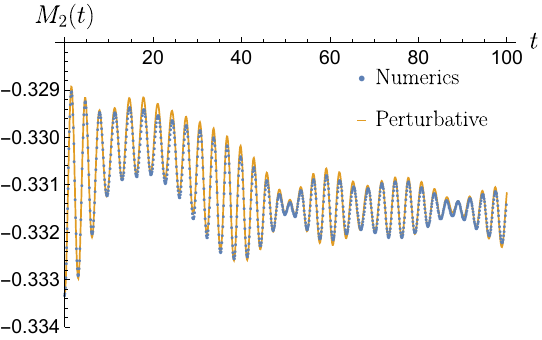}
    \caption{$g=0.2$, $h_2=-0.1$}
    
\end{subfigure}
\hfill
\begin{subfigure}{0.45\textwidth}
    \includegraphics[width=\textwidth]{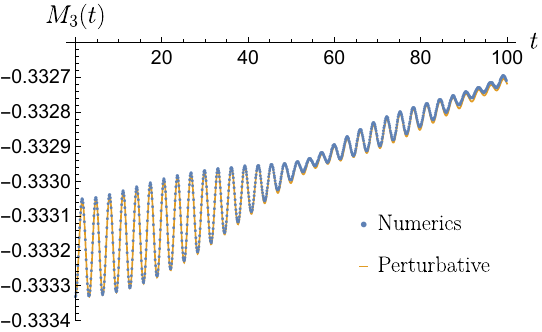}
      \caption{$g=0.05$, $h_2=-0.1$}
    
\end{subfigure}
\hfill
\begin{subfigure}{0.45\textwidth}
    \includegraphics[width=\textwidth]{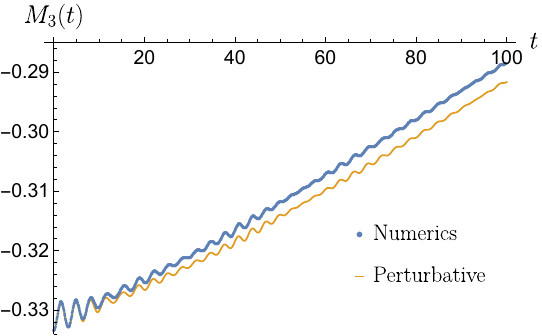}
    \caption{$g=0.2$, $h_2=-0.1$}
    
\end{subfigure}
\caption{Comparison of Perturbative Results and iTEBD Simulations.\\
Validation of the perturbative analytical results (orange) against numerical iTEBD data (blue) for the oblique quench. The plots illustrate the convergence of the second-order expansion to the full numerical simulation as the transverse field $g$ is reduced from $0.2$ (right column) to $0.05$ (left column).}
\label{fig:magnetisation_t_num}
\end{figure}

In Fig. \ref{fig:magnetisation_t_num}, we show the independent magnetisations $M_2(t)$ and $M_3(t)$ in the case of oblique quenches calculated numerically with the iTEBD method for a specific parameter setting in each case, together with the analytical prediction for the magnetisations in the two-kink approximation that was determined by \eqref{2k}. The perturbative analytical results show clear convergence toward the full iTEBD numerical data as the transverse field $g$ is reduced from $0.2$ to $0.05$. However, this convergence is notably slower in the Potts model than in the Ising spin chain. (see \ref{fig:MIs}). The accuracy of the perturbative results for the time evolution of $M_\mu(t)$ is highly sensitive to the magnitude of $g$ (as illustrated in Fig. \ref{fig:magnetisation_t_num}), while the spectral features (quench spectroscopy) remain significantly more robust, as demonstrated later in this section. 

\subsubsection{Quench spectroscopy in aligned quenches}
The quasi-particle content relevant for the quench dynamics can be obtained from the Fourier transform of the magnetisations \cite{Kor16,Po25}: 
\begin{equation}
    M_\mu(\omega)\equiv M_\mu\left(\frac{2\pi k}{N_s\delta t}\right):=\frac{1}{\sqrt{N_s}}\sum_{n=0}^{N-1}e^{-2\pi i \frac{kn}{N_s}}M(n\delta t) 
    \label{eq:FT_numerical_def}
\end{equation}
where $N_s$ is the total number of the discrete time steps of the simulation, $\delta t$ is the time step, and $k$ is an integer such that $k\in[0,\ldots N_s-1]$. The specific values used in the simulations are provided in Appendix \ref{Numerical_details}. 

In the following, we plot the numerically calculated $|M_\mu(\omega)|^2$ power spectra. Due to translational invariance, only eigenstates with total momentum $P=0$ contribute to the Fourier transform. In the following plots, the spectrum of the $P=0$ eigenstates obtained by perturbation theory is shown by vertical lines. Besides comparing the peak locations with the analytical results, we also explain the selection of the eigenstates expected to contribute in the different cases. A more detailed discussion, together with exact diagonalisation results for the spectrum, can be found in Ref.\cite{Po25}.

\begin{figure}
\centering
\begin{subfigure}{0.495\textwidth}
    \includegraphics[width=\textwidth]{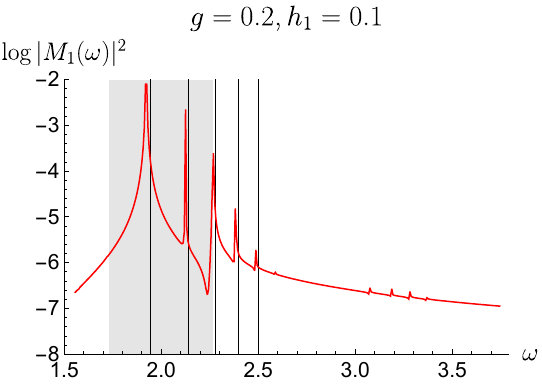}
   \caption{}
\end{subfigure}
\hfill
\begin{subfigure}{0.495\textwidth}
    \includegraphics[width=\textwidth]{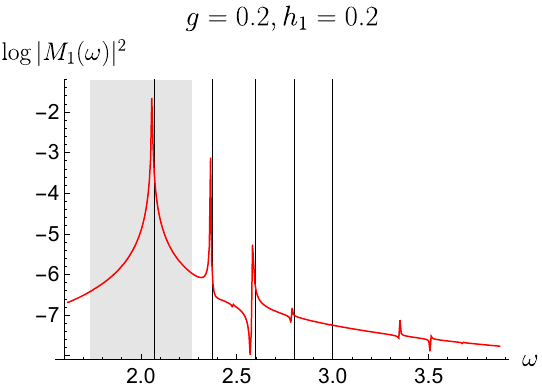}
    \caption{}
\end{subfigure}
\caption{Quench Spectroscopy of Positively Aligned Quenches.\\
Fourier power spectrum log$\,|M_{1}(\omega)|^{2}$ from iTEBD simulations for $h_{1}=0.1$ (a) and $h_{1}=0.2$ (b). Black vertical lines represent the masses of the first few $\iota=+$ mesons calculated via the perturbative secular equation \eqref{eq:secular_eq_of_meson_bubble_energies_app} and show precise alignment with the primary peaks. The grey region indicates the two-kink continuum; peaks within this region correspond to collisional mesons. The higher-energy secondary peaks indicate the presence of baryonic (three-kink) excitations.}
\label{fig:M1_om_pos_aligned}
\end{figure}

\paragraph{Positive aligned quenches -} Fig. \ref{fig:M1_om_pos_aligned} shows the typical Fourier spectrum of $M_1(\omega)$ for the case of a positively aligned quench.
Turning on a positive $h_1$ results in a doubly degenerate false and a single true vacuum state. As the initial state $|0\rangle^{(1)}$ is favoured by the longitudinal field, we expect excitations built upon the true vacuum to contribute to the quench with the largest overlaps. These are predominantly meson states, described in Section \ref{diav1}, with their energy determined by the solutions of the perturbative secular equation (\ref{eq:secular_eq_of_meson_bubble_energies_app}). Their overlap with the initial state is expected to decrease with their energy \cite{2024PhRvL.133x0402L,Ising_ak_gt}. Furthermore, because of the $\mathbb{Z}_2$ symmetry of the quench, only the $\iota=+$ (even) meson states contribute, and therefore we show the first few $\iota=+$ meson masses ($P=0$ energy eigenvalues), calculated from (\ref{eq:secular_eq_of_meson_bubble_energies_app}) using black vertical lines in Fig.~\ref{fig:M1_om_pos_aligned}. The locations of the peaks in the Fourier spectrum match the meson masses with reasonable precision. The grey region indicates the range of the continuous spectrum of two-kink states at zero longitudinal field. The boundaries of the region are determined by relation (\ref{bas}) at $P=0$:
\begin{equation}
 E_\text{min}(P=0)=2-\frac{4g}{3}, \quad 
 E_\text{max}(P=0)=2+\frac{4g}{3}. 
\label{eq:grey_bound}
 \end{equation} 
As already mentioned before, the meson excitations falling in this interval are collisional, i.e., the constituent kinks scatter on each other, while the ones outside are non-collisional, which means they consist of a pair of kinks localised by Bloch oscillations \cite{Ising_ak_gt,Po25}. We note that the exact diagonalisation (ED) results for the meson masses (with $L=10$ and periodic boundary conditions) match better with the peaks in $M_1(\omega)$, and they even allow for the identification of the additional small, high-energy peaks in the Fourier transform as baryons (three-kink bound states) \cite{Po25}. However, the exact diagonalisation method suffers from other limitations: it is not sufficient for a complete classification of the eigenstates; moreover, it cannot be applied to locate resonances in oblique cases, as discussed later. 

\begin{figure}
\centering
\begin{subfigure}{0.495\textwidth}
    \includegraphics[width=\textwidth]{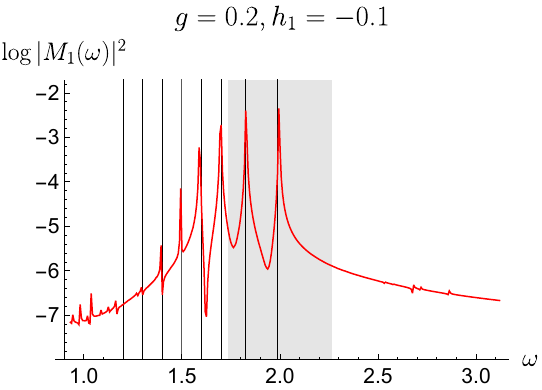}
   \caption{}
\end{subfigure}
\hfill
\begin{subfigure}{0.495\textwidth}
    \includegraphics[width=\textwidth]{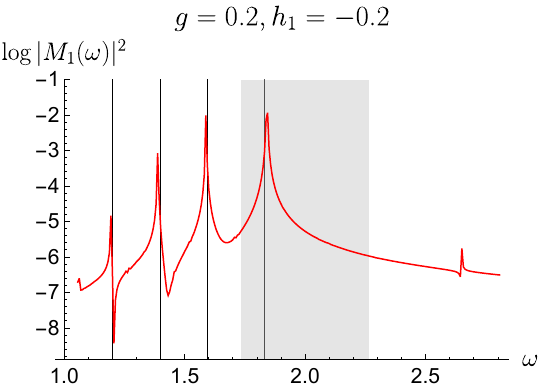}
    \caption{}
\end{subfigure}
\caption{Quench Spectroscopy of Negatively Aligned Quenches.\\
Fourier power spectrum log$\,|M_{1}(\omega)|^{2}$ for $h_{1}=-0.1$ (a) and $h_{1}=-0.2$ (b). Black vertical lines mark the analytically predicted masses for $\iota=+$ bubbles. The spectrum is dominated by these bubble states nucleated over the false vacuum. As in the aligned case, the grey region denotes the collisional regime, and high-energy signatures of baryonic bubbles are visible.}
\label{fig:M1_om_neg_aligned}
\end{figure}

\paragraph{Negative aligned quenches -} In the negatively aligned case, the typical Fourier spectrum of $M_1(\omega)$ resulting from the simulations is illustrated in Fig. \ref{fig:M1_om_neg_aligned}. 
In this case, there are two degenerate true vacua and a single false vacuum. The initial state $|0\rangle^{(1)}$ is disfavoured by the longitudinal field; therefore, the excitations built upon the false vacuum are expected to dominate the quench. These are bubble states that correspond to the nucleation of a true vacuum domain, with their energies determined by the solutions of Eq.$\,$(\ref{eq:secular_eq_of_meson_bubble_energies_app}). Again, due to the aforementioned symmetry of the initial state, only the $\iota=+$ (even) bubbles contribute to the time evolution. The energy of the bubbles decreases with their size, i.e., with their species label; therefore, their overlap with the initial state increases with their energy \cite{2024PhRvL.133x0402L,Ising_ak_gt}. The first few (highest energy) $\iota=+$ bubble masses, calculated from (\ref{eq:secular_eq_of_meson_bubble_energies_app}), are plotted in Fig. \ref{fig:M1_om_neg_aligned} by black vertical lines which again coincide well with the locations of the peaks in $M_1(\omega)$. The grey-coloured region in the figure again denotes the two-kink continuous spectrum at $h_1 = h_2 = 0$ with its boundaries determined by (\ref{eq:grey_bound}). Just as before, the excitations inside this interval are collisional, while the ones outside are non-collisional bubbles. Again, we remark that the bubble masses calculated via ED (with $L = 10$ and periodic boundary conditions) match better with the peaks in the Fourier transform. Nevertheless, we repeatedly emphasise that the bubble energy eigenstates cannot be identified from the ED states without the help of the perturbative results or semiclassical quantisation, as done in Ref.\cite{Po25}. The ED energy spectrum additionally enables us to identify the additional small peaks in the Fourier spectrum as baryonic bubbles (3-kink localised states built upon the false vacuum) \cite{Po25}.

\subsubsection{Quench spectroscopy in oblique quenches}

In Ref. \cite{Po25}, the positively and negatively aligned quenches considered above were also referred to as (standard) confining and anticonfining quenches, respectively, based on the main mechanism of localisation. In contrast, the oblique regimes are characterised as "partial confining" or "partial anticonfining" due to the coexistence of localised bound states and unconfined kink excitations. This mixed particle content leads to a complex spectral structure that cannot be captured by semiclassical methods but is precisely described by our perturbative framework. As mentioned, aligned quenches can also be realised in the Ising case; however, the oblique quenches have no Ising analogues.

\begin{figure}
\centering
\begin{subfigure}{0.495\textwidth}
    \includegraphics[width=\textwidth]{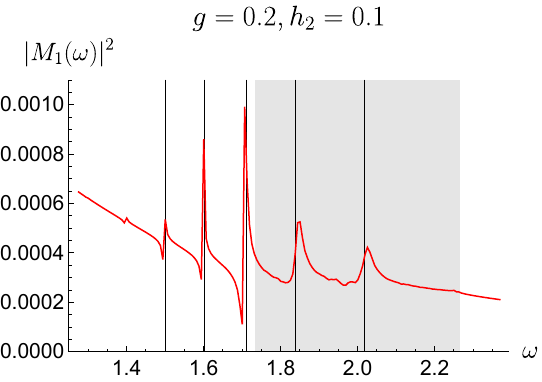}
   \caption{}\label{fig:oblique_quench_pos_a}
\end{subfigure}
\hfill
\begin{subfigure}{0.495\textwidth}
    \includegraphics[width=\textwidth]{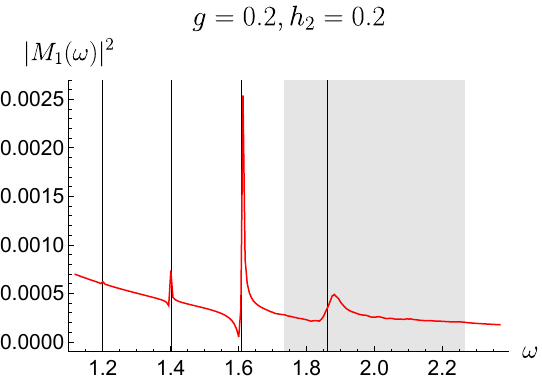}
    \caption{}
\end{subfigure}
\caption{Quench Spectroscopy in the Positive Oblique Case.\\
Fourier power spectrum of the numerically calculated magnetisation $M_{1}(t)$ for parameters $g=0.2, h_{2}=0.1$ (a) and $h_{2}=0.2$ (b). The black vertical gridlines represent analytically determined masses for the first few collisionless bubbles (sharp peaks outside the grey region) and resonance frequencies (broader peaks within the grey region). The grey shaded area denotes the two-kink continuous spectrum at $h_{1}=h_{2}=0$, within which the bubbles become unstable due to hybridization with the unconfined kink continuum. The numerically observed small cusps at the boundaries of this region align with the predicted spectral thresholds.}
\label{fig:M1_om_pos_obl}
\end{figure}

\paragraph{Positive oblique quenches -} In the positive oblique case, we chose the two independent magnetisations as $M_1(t)$ and $M_2(t)$. The typical Fourier spectrum for $M_1(t)$ is visible in Fig. \ref{fig:M1_om_pos_obl}. $M_2(\omega)$ is qualitatively very similar and exhibits peaks at the same locations. 
The presence of $h_2>0$ leads to a single true vacuum and doubly degenerate false vacua. The initial state $|0\rangle^{(1)}$ is disfavoured by the longitudinal field, and therefore, the elementary excitations built upon the false vacua have the largest overlaps with the initial state. The calculation of the energy spectrum of such excitations is discussed in Sections \ref{diav2} and \ref{analytic_S}. As discussed previously in detail, outside the interval $(E_\text{min}(0),E_\text{max}(0))$ (where $E_\text{min}(0)$ and  $E_\text{max}(0)$ are given by (\ref{eq:grey_bound})) we get a discrete energy spectrum (\ref{eq:collisionless_bubbles}) describing collisionless bubbles. These excitations appear in the Fourier spectrum as sharp peaks. In the interval $(E_\text{min}(0),E_\text{max}(0))$, the resulting spectrum is continuous and admits resonances (unstable collisional bubbles) as explained in Section \ref{analytic_S}.  Nevertheless, at the resonance locations, we still expect wide resonant peaks. Note that the resonance peaks are notably wider and less symmetric than their collisionless counterparts. As the resonance energy moves deeper into the two-kink continuum, the increasing imaginary part causes the peak to blend into the background spectral density, eventually becoming a broad hump or a cusp at the stability threshold. For instance, at $h_2 = 0.1$, the results in Table \ref{tab:Energies} show that the first resonance $E_{\text{res},1}$ has a smaller imaginary part compared to the second resonance $E_{\text{res},2}$. This accurately predicts why the first resonance peak in Fig. \ref{fig:oblique_quench_pos_a} appears sharper than the second one.

In Fig. \ref{fig:M1_om_pos_obl}, we show the analytically calculated collisionless bubble masses and the locations of resonances by black gridlines. The grey region again represents the two-kink continuous spectrum at $h_1 = h_2 = 0$, lying within the interval ($E_\text{min}(0),E_\text{max}(0)$). The black lines within this region denote the resonances (unstable collisional bubbles), while the lines outside this interval stand for the collisionless bubbles. The positions of the lines match with the locations of the peaks in $M_1(\omega)$ (and $M_2(\omega)$) with good precision. The boundary (threshold) of the continuous energy spectrum corresponds to the appearance of a cusp in the Fourier spectrum, which is observed numerically, and its location approximately agrees with the analytical prediction. We remark that the collisionless bubble masses and the location of the cusp can be determined more precisely by ED; however, the ED method cannot locate the resonances. 

\begin{figure}
\centering
\begin{subfigure}{0.495\textwidth}
    \includegraphics[width=\textwidth]{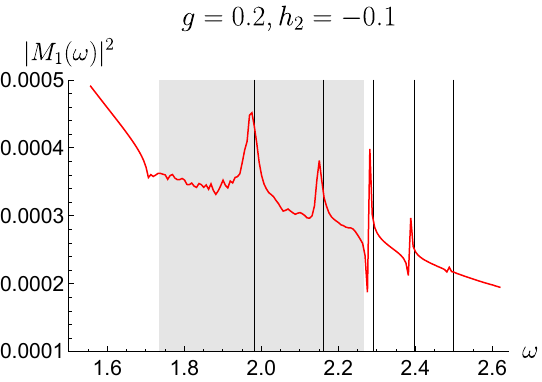}
   \caption{}\label{fig:oblique_quench_neg_a}
\end{subfigure}
\hfill
\begin{subfigure}{0.495\textwidth}
    \includegraphics[width=\textwidth]{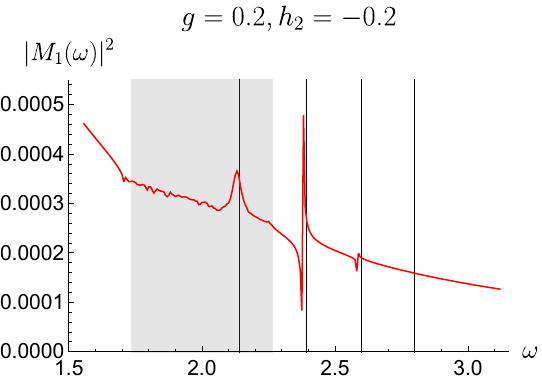}
    \caption{}
\end{subfigure}
\caption{Quench Spectroscopy in the Negative Oblique Case.\\
Fourier power spectrum of the numerically calculated magnetization $M_{1}(t)$ for parameters $g=0.2, h_{2}=-0.1$ (a) and $h_{2}=-0.2$ (b). Here, the dynamics is dominated by meson excitations built upon the true vacua. The black vertical gridlines represent analytically determined masses for the first few collisionless mesons (sharp peaks outside the grey region) and resonance frequencies (broader peaks within the grey region). The grey shaded area denotes the two-kink continuous spectrum at $h_{1}=h_{2}=0$. The position of the small cusp at the edge of the grey region matches the analytically calculated boundary of the continuous two-particle spectrum.}
\label{fig:M1_om_neg_obl}
\end{figure}

\paragraph{Negative oblique quenches -} In the negative oblique case, the dynamics and its explanation is similar to the positive oblique case, with the bubbles replaced by mesons. The typical Fourier spectrum of $M_1(t)$ is shown in Fig. \ref{fig:M1_om_neg_obl}, while the other independent magnetisation, chosen as $M_2(t)$, has the same qualitative behaviour \cite{Po25}. 
Setting $h_2<0$ results in a single false vacuum and doubly degenerate true vacua, meaning that $|0\rangle^{(1)}$ is favoured by the longitudinal field, thus the quasi-particles built upon the true vacua
contribute dominantly to the quench. The corresponding energy spectrum was calculated in Sections \ref{diav2} and \ref{analytic_S}. Outside the interval $(E_\text{min}(0),E_\text{max}(0))$ (marked by a grey region in the figure), the energy spectrum is discrete (\ref{eq:collisionless_mesons}) and describes collisionless mesons which are present in the quench spectroscopy as sharp peaks. In the interval $(E_\text{min}(0),E_\text{max}(0))$, the spectrum is continuous and exhibits resonances (unstable collisional mesons). Similarly to our analysis in the positive oblique case, for $h_2 =- 0.1$, the results in Table \ref{tab:Energies} show that the first resonance $E_{\text{res},1}$ has a larger imaginary part compared to the second resonance $E_{\text{res},2}$ which explains why the second resonance peak in Fig. \ref{fig:oblique_quench_neg_a} appears sharper than the first one.

The edge of the continuous spectrum again corresponds to a cusp in the Fourier spectrum. Once again, the ED meson masses (and the location of the cusp computed by ED) are more precise, but the method fails to determine the positions of the resonances for reasons analogous to those encountered in the positively aligned case. 

\begin{figure}
\centering
\begin{subfigure}{0.495\textwidth}
    \includegraphics[width=\textwidth]{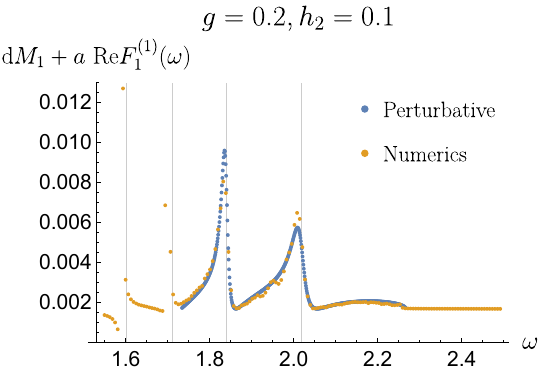}
   \caption{}
\end{subfigure}
\hfill
\begin{subfigure}{0.495\textwidth}
    \includegraphics[width=\textwidth]{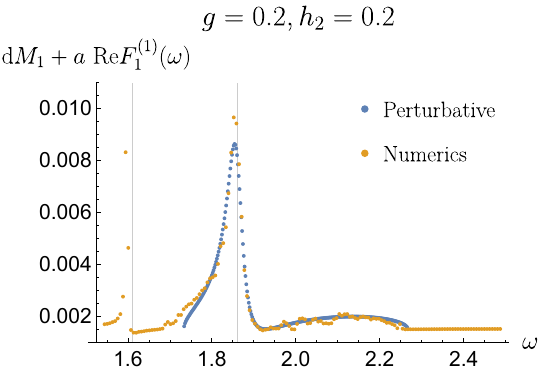}
    \caption{}
\end{subfigure}
\caption{Quantitative Comparison of Spectroscopy in Positive Oblique Quenches.\\
Comparison between the rescaled analytical Fourier transform $a\,\mathrm{Re}\,F_{1}^{(1)}(\omega) + dM_{1}$ (blue dots) and numerical iTEBD results (orange dots) for the parameters $g=0.2$, $h_2=0.1$ (a) and $h_2=0.2$ (b). The analytical model precisely captures the line shapes of the broad, asymmetric resonance peaks. Grey vertical lines denote the quasi-particle masses and resonance frequencies determined via perturbation theory.
}
\label{fig:reM1_om_pos_ob}
\end{figure}

\begin{figure}
\centering
\begin{subfigure}{0.495\textwidth}
    \includegraphics[width=\textwidth]{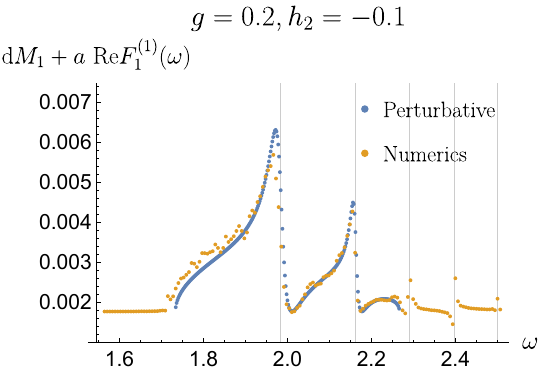}
   \caption{}
\end{subfigure}
\hfill
\begin{subfigure}{0.495\textwidth}
    \includegraphics[width=\textwidth]{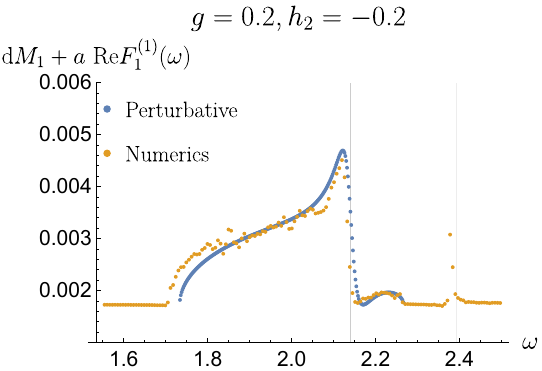}
    \caption{}
\end{subfigure}
\caption{Quantitative Comparison of Spectroscopy in Negative Oblique Quenches\\
Analytical vs. numerical Fourier spectra for negative oblique quenches ($g=0.2, h_{2}=-0.1, -0.2$). The analytical model precisely captures the line shapes of the broad, asymmetric resonance peaks. Grey vertical lines denote the quasi-particle masses and resonance frequencies determined via perturbation theory.}
\label{fig:reM1_om_neg_ob}
\end{figure}

In the following, we compare the Fourier transform of the magnetisation $M_1(\omega)$ calculated numerically in the simulations for the oblique cases with the analytical results in Eq. (\ref{Ftrao_F1F2}). We expect the Fourier transform of $A_\alpha(t)$ given by (\ref{2k}) to match the corresponding numerically calculated magnetisation $M_\alpha(\omega)$ in the relevant frequency domain of the low-energy excitation spectrum up to a constant shift $\text{d}M_\alpha$ and a constant multiplicative factor $a$. The constant shift was set manually, while the multiplicative factor originates from the different definitions of the Fourier transforms as discussed in Appendix \ref{Numerical_details}. The difference between the definition of $A_\alpha(t)$ and $M_\alpha(t)$ \eqref{A123} becomes irrelevant as they differ by a constant, which results in an additional peak at zero frequency in the Fourier spectrum, which is outside of the interval of interest. Furthermore, the first term on the right-hand side of (\ref{2k}) is time-independent, which again leads to a peak at zero frequency. Meanwhile, the last term represents the low-frequency contribution to $A_\alpha (\omega)$ and also gives rise to peaks outside the investigated interval. Therefore, in the quench spectroscopy, we neglect these terms and only focus on the second term, whose Fourier transform is given by (\ref{Ftrao_F1F2}). In Figs. \ref{fig:reM1_om_pos_ob} and \ref{fig:reM1_om_neg_ob} we show typical Fourier spectra for $a \,\text{Re}F^{(1)}_1 (\omega) +\text{d}M_1$ together with the numerically calculated $\text{Re}M_1(\omega)$ in the oblique cases; further examples are included in Appendix \ref{Numerical_details}. The grey vertical lines mark the quasi-particle masses and resonance frequencies determined from perturbation theory. The plots demonstrate an excellent match between the analytic and numerical results, despite the presence of numerical noise in the iTEBD results.

%
%
\section{Conclusions \label{sec:conclusions}}
In this work, we have developed a comprehensive perturbative description of the dynamics of the three-state Potts quantum spin chain in the extreme ferromagnetic regime, utilising the transverse magnetic field as the expansion parameter. 

The primary achievement of this study is to provide analytical access to spectral and dynamical features beyond the reach of standard semiclassical and exact diagonalisation methods. While we recovered the known meson and bubble excitation spectra in the aligned cases, our approach proved particularly powerful in the oblique regime. In this case, switching on the longitudinal field in the direction of one vacuum leaves the remaining two degenerate, allowing for unconfined kink excitations \cite{Po25}.  

Beyond the bound state spectrum, we explored the analytic structure of the two-kink scattering amplitude. By tracking the trajectory of bound-state poles near stability thresholds, we demonstrated that the three-state Potts model follows the general scenario proposed by Fonseca and Zamolodchikov for the Ising field theory, in which stable excitations transform into resonances \cite{FZ06}. We successfully reproduced the resonances arising from the hybridisation of bound states with the unconfined continuum, matching quench-spectroscopy results with high precision.

Furthermore, we provided an explicit evaluation of post-quench magnetisation evolution. Although this perturbative method is inherently time-limited and converges more slowly in the Potts case than in the Ising model, it provides a remarkably accurate description of short- to medium-term dynamics compared with numerical iTEBD simulations.

To conclude, this work provides a robust analytical foundation for studying non-equilibrium dynamics in non-integrable spin chains. An essential next step for this research is to extend the perturbative framework to include three-kink bound states (baryons). As baryons are a distinguishing feature of the Potts model over the simpler Ising chain, capturing their contribution to the post-quench spectrum remains an intriguing challenge for future investigations.

\subsection*{Acknowledgments}
This work was partially supported by the Ministry of Culture and Innovation and the National Research, Development and Innovation Office (NKFIH) through the OTKA Grant K 138606. AK was also partially supported by the HUN-REN Hungarian Research Network through the HUN-REN-BME-BCE Quantum Technology Research Group, and by the Doctoral Excellence Fellowship Programme (DCEP), funded by the National Research Development and Innovation Fund of the Ministry of Culture and Innovation and the Budapest University of Technology and Economics, under a grant agreement with the National Research, Development and Innovation Office. GT was also supported by the Quantum Information National Laboratory of Hungary (Grant No. 2022-2.1.1-NL-2022-00004). We also acknowledge KIFÜ (Governmental Agency for IT Development, Hungary) for awarding us access to the Komondor HPC facility based in Hungary.

\appendix
\section{Details of the perturbative description of the excitation spectrum}\label{sec:pert_details}
\subsection{The pure transverse chain: kink excitations  \label{diH0_tech}}
Here we discuss the details of the perturbative analysis of the model described by the Hamiltonian $H(\mathbf{v},g)$ with $\mathbf{v}=\mathbf{0}$, which corresponds to the infinite three-state Potts spin chain with purely transverse magnetic field. The perturbative calculations describe the extreme ferromagnetic limit 
$g\ll1$ in the two-kink sector. 

As stated in the main text in Section \ref{diH0}, the first step in solving the eigenvalue problem \eqref{RS} perturbatively is the  diagonalisation of the Hamiltonian $H^{(2)}(\mathbf{0},g)$, which is given by the 'Bethe states' \eqref{eq:bethe_states}:
\begin{align}\label{Bl1_app}
&|{K}_{1,\nu}(p_1){K}_{\nu,1}(p_2)\rangle=\sum_{j_1=-\infty}^\infty\sum_{j_2=j_1+1}^\infty \sum_{\nu'=2}^3\Big[e^{i(p_1j_1+p_2 j_2)}\delta_{\nu,\nu'}+
S_{\nu,\nu'}(p_1,p_2) e^{i(p_1j_2+p_2 j_1)}\Big]\\
\nonumber
&\times| \mathbf{K}_{1,\nu'}(j_1)\mathbf{K}_{\nu',1}(j_2)\rangle,
\end{align}
where $p_1,p_2\in \mathbb{R}/2\pi \mathbb{Z}$ 
 are the quasimomenta of the two kinks, $\nu=2,3$ and $S_{\nu,\nu'}(p_1,p_2)$ are the entries of the  two-kink scattering matrix:
\begin{align}\label{scm_app}
&S_{2,2}(p_1,p_2)=S_{3,3}(p_1,p_2)=\frac{1}{2}\Big[
S_{+}(p_1,p_2)+S_{-}(p_1,p_2)
\Big],\\\nonumber
&S_{2,3}(p_1,p_2)=S_{3,2}(p_1,p_2)=\frac{1}{2}\Big[
S_{+}(p_1,p_2)-S_{-}(p_1,p_2)
\Big],\\\nonumber
&S_{\iota}(p_1,p_2)=-\frac{e^{i p_1}+e^{-i p_2}- \iota \cdot1}{e^{-i p_1}+e^{i p_2}- \iota \cdot1}\, e^{i(p_2-p_1)}, \quad \text{with }\iota=\pm.
\end{align}
The  Bethe  states $|{K}_{1,\nu}(p_1){K}_{\nu,1}(p_2)\rangle$ 
satisfy the Faddeev-Zamolodchikov commutation relation:
\begin{equation}
|{K}_{1,\nu}(p_1){K}_{\nu,1}(p_2)\rangle=\sum_{\nu'=2}^3  S_{\nu\nu'}(p_1,p_2)|{K}_{1,\nu'}(p_2){K}_{\nu',1}(p_1)\rangle,
\quad \text{with }\nu=2,3.
\nonumber
\end{equation}
Such states with $-\pi<p_2<p_1\leq\pi$ form the basis in the 
subspace $\mathcal{L}_{11}^{(2)}$, and satisfy the normalisation condition:
\begin{equation}
\langle {K}_{1,\nu}(p_2){K}_{\nu,1}(p_1)|{K}_{1,\nu'}(p_1'){K}_{\nu',1}(p_2')\rangle
=4\pi^2\, \delta_{\nu\nu'}\delta(p_1-p_1')\delta(p_2-p_2'),
\nonumber
\end{equation}
for $-\pi<p_2<p_1\leq\pi$, $-\pi<p_2'<p_1'\leq\pi$, and $\nu,\nu'=2,3$. We also use the basis \eqref{Bl2} of the  two-kink Bethe states:
\begin{align}\label{Bl2_app}
&|{K}(p_1){K}(p_2)\rangle_\iota=\frac{1}{\sqrt{2}}\Big(|{K}_{1,2}(p_1){K}_{2,1}(p_2)\rangle+\iota |{K}_{1,3}(p_1){K}_{3,1}(p_2)\rangle\Big)\\\nonumber
&=\sum_{j_1=-\infty}^\infty\sum_{j_2=j_1+1}^\infty \Big[e^{i(p_1j_1+p_2 j_2)}+
S_{\iota}(p_1,p_2) e^{i(p_1j_2+p_2 j_1)}\Big]| \mathbf{K}(j_1)\mathbf{K}(j_2)\rangle_\iota,
\end{align}
with $\iota=\pm$. From the second line of the above equation, it follows that these states diagonalise the two-kink scattering matrix:
\begin{equation}
|{K}(p_1){K}(p_2)\rangle_\iota=S_{\iota}(p_1,p_2) \, |{K}(p_2){K}(p_1)\rangle_\iota \quad \text{for }\iota=\pm.
\label{eq:scat_mx_diag_app}
\end{equation}
Most importantly, the two-kink Bethe states defined by Eqs.$\,$\eqref{Bl1_app} and \eqref{Bl2_app}, diagonalise the operators $H^{(2)}(\mathbf{0},g)$ and $\hat{T}_1$ simultanoeusly:
\begin{align}
& \hat{T}_1 \, |{K}_{1,\nu}(p_1){K}_{\nu,1}(p_2)\rangle=\exp[i(p_1+p_2)]\, |{K}_{1,\nu}(p_1){K}_{\nu,1}(p_2)\rangle,\quad 
\text{with }\nu=2,3,\label{eq:T_eig}\\
&H^{(2)}(\mathbf{0},g)|{K}_{1,\nu}(p_1){K}_{\nu,1}(p_2)\rangle=[\omega(p_1)+\omega(p_2)]|{K}_{1,\nu}(p_1){K}_{\nu,1}(p_2)\rangle, \label{eq:H_eig}\\
&  \hat{T}_1 \,  |{K}(p_1){K}(p_2)\rangle_\iota=\exp[i(p_1+p_2)]\, |{K}(p_1){K}(p_2)\rangle_\iota, \quad \text{with }\iota=\pm,\\
&H^{(2)}(\mathbf{0},g)|{K}(p_1){K}(p_2)\rangle_\iota=[\omega(p_1)+\omega(p_2)]|{K}(p_1){K}(p_2)\rangle_\iota, 
\end{align}
where
\begin{equation}
\omega(p)=1-\frac{2 g}{3}  \cos p
\end{equation}
is the kink dispersion law to the linear order in the small parameter $g$.

The scattering matrix defined by \eqref{scm_app} satisfies the unitarity relations:
\begin{align}
&\sum_{\nu''=2}^3 S_{\nu\nu''}(p_1,p_2)\,S_{\nu''\nu'}(p_2,p_1)=\delta_{\nu,\nu'},\\
&S_{\iota}(p_1,p_2)=\frac{1}{S_{\iota}(p_2,p_1)},
\end{align}
where $p_1,p_2\in \mathbb{R}$, $\nu,\nu'=2,3$, and $\iota=\pm$.

\subsection{Two-kink states in the aligned case \label{diav1_tech}}
Here we consider the perturbative analysis of the low-energy spectrum of the Hamiltonian $H(\mathbf{v},g)$ with $\mathbf{v}=(v_1,0,0)$ in the subspace $\mathcal{L}_{11}^{(2)}$. For the Hamiltonian, we use the short-cut notation $H^{(2)}(v_1,g)$ and calculate its spectrum to the leading order in $g\ll1$. We use the basis in the two-kink subspace $\mathcal{L}_{11}^{(2)}$:
\begin{equation}\label{jpm}
|j,P\rangle_\iota=\sum_{j_1=-\infty}^\infty \exp\left[i P\left(j_1+\frac{j}{2}\right)\right]|\mathbf{K}(j_1)\mathbf{K}(j_1+j)\rangle_\iota,
\end{equation}
where $\iota=\pm$, $j=1,2\ldots$, and $-\pi<P\leq\pi$ is the quasimomentum. 
It follows from \eqref{Tr1}, \eqref{Kiot}, that
\begin{equation}
\hat{T}_1 |j,P\rangle_\iota=e^{i P}|j,P\rangle_\iota.
\end{equation}
Denote by $\mathcal{L}^{(2)}(P)=\mathcal{L}^{(2)}(P,+)\oplus \mathcal{L}^{(2)}(P,-)$ the subspace of $\mathcal{L}_{11}^{(2)}$  spanned by the vectors  $|j,P\rangle_\iota$, with $j=1,2\ldots$, and $\iota=\pm$.
For all $|\psi\rangle\in \mathcal{L}^{(2)}(P)$ we have
\begin{equation}
\hat{T}_1 |\psi\rangle=e^{i P}|\psi\rangle\,.
\end{equation}
It is convenient to define a modified scalar product $\langle\ldots|\ldots\rangle$ in the subspace $\mathcal{L}^{(2)}(P)$ which satisfies: 
\begin{equation}\label{scp0}
\phantom{\sum}_{\iota}\langle j,P\Big|j',P\rangle_{\iota'}=\delta_{\iota\iota'}\delta_{jj'}\,,
\end{equation}
with $\iota,\iota'=\pm$, and $j,j'=1,2,\ldots$.

Consider now the eigenvalue problem
\begin{equation}
H^{(2)}(v_1,g) |\pi_\iota\rangle= E_\iota |\pi_\iota\rangle,
\label{eq:evaleq_app}\end{equation}
for $|\pi_\iota\rangle\in \mathcal{L}^{(2)}(P,\iota)$, 
and expand its eigenvector $|\pi_\iota\rangle$ in the basis $|j,P\rangle_\iota$:
\begin{equation}
|\pi_\iota\rangle=\sum_{j=1}^\infty \psi_\iota(j;E_\iota,P) |j,P\rangle_\iota.
\label{eq:pi_wavefunc}
\end{equation}
We simplify notations by omitting the labels $E_\iota$ and $P$ in $\psi_\iota(j;E_\iota,P)$, and use $\psi_\iota(j)$. Eq. \eqref{eq:evaleq_app} implies the following second-order linear difference equation: 
\begin{equation}\label{dE}
(
2+h_1 j -E_\iota) \psi_\iota(j)-\frac{2 g }{3 }  \cos(P/2)[\psi_\iota(j+1)+\psi_\iota(j-1)]=0\,,
\end{equation}
where $j=2,3,\ldots$.
Furthermore, the wave-function coefficients $\psi_\iota(j)$ must vanish for $j\to +\infty$, and also satisfy the boundary condition:
\begin{equation}\label{bc1}
\left(
2+h_1 - \frac{\iota\, g}{3} -E_\iota \right) \psi_\iota(1)-\frac{2 g }{3 }  \cos(P/2)\, \psi_\iota(2)=0.
\end{equation}
Note, that the difference  equation \eqref{dE} and the boundary condition \eqref{bc1} are invariant under the transformation
\begin{equation}\label{symh}
h_1\to-h_1,\quad
\iota\to-\iota,\quad
E_\iota\to 4-E_{-\iota},\quad
 \psi_\iota(j)\to (-1)^j \, \psi_{-\iota}(j)\,.
\end{equation}
Using the following identity satisfied by the Bessel function  
 of the first kind $J_{c}(Z)$:
\begin{equation}
J_{c+1}(Z)+J_{c-1}(Z)=\frac{2c}{Z} J_{c}(Z)\,,
\end{equation}
the solution for the wave-function reads
\begin{equation}\label{ps_app}
\psi_\iota(j;E_\iota,P)=(\sign h_1)^{j+1}\,J_{j+c_0}(Z)\,,
\end{equation}
where
\begin{align}\label{Zd_app}
Z=\frac{4g}{3|h_1|} \cos(P/2), \qquad
c_0=\frac{2-E_\iota}{h_1}.
\end{align}
Substituting of \eqref{ps_app} into \eqref{bc1} leads to the secular equation
\begin{align}
&J_{1+[2-E_{\iota,n}(P,h_1)]/h_1}(Z)
\left\{|h_1|+\left[2-E_{\iota,n}(P,h_1)-\frac{\iota \,g}{3}\right]\,\sign h_1\right\}\label{eq:secular_eq_of_meson_bubble_energies_app}\\\nonumber
&=
\frac{2 g}{3}\, \cos(P/2)\, J_{2+[2-E_{\iota,n}(P,h_1)]/h_1}(Z),
\end{align}
where $Z$ is given by \eqref{Zd_app}. For $h_1>0$, solutions of this equation determine the dispersion laws $E_{\iota,n}(P,h_1)$ of mesons, while for $h_1<0$ they yield the spectrum of bubbles, to the first order in the small parameter $g$. The resulting dispersion laws $E_{\iota,n}(P)$  of the few lightest mesons are depicted in Fig. \ref{fig:DL} for $g=0.2$ and $h_1=0.05$.

The energies of mesons and bubbles furthermore satisfy the following equality:
\begin{equation}\label{Epm}
E_{\iota,n}(P,h_1)+E_{-\iota,n}(P,-h_1)=4\,.
\end{equation}
which follows directly from the symmetry property \eqref{symh}, valid to the first order in the parameter $g\ll1$. 

\subsection{Two-kink states in the oblique case \label{diav2_tech}}
We now turn to the calculation of the energy spectrum of the Hamiltonian $H(\mathbf{v},g)$ in the subspace $\mathcal{L}_{11}^{(2)}$ in the oblique regime $\mathbf{v}=(0,v_2,0)$, to leading order in $g\ll1$. Once again, we introduce the shorthand notation $H^{(2)}(v_2,g)$ for the corresponding Hamiltonian.

We choose the basis
\begin{equation}\label{jP23_app}
|j,P\rangle_\nu
=\sum_{j_1=-\infty}^\infty \exp\left[i P\left(j_1+\frac{j}{2}\right)\right]|\mathbf{K}_{1,\nu}(j_1)\mathbf{K}_{\nu,1}(j_1+j)\rangle,
\end{equation}
in the subspace $\mathcal{L}_{11}^{(2)}$, with $\nu=2,3$, $j=1,2,\ldots$,  and $-\pi<P\leq\pi$. These basis vectors are simply related to the vectors $|j,P\rangle_\pm$ given in \eqref{jpm}:
\begin{equation}
|j,P\rangle_2=\frac{1}{\sqrt{2}}(|j,P\rangle_++|j,P\rangle_-),\quad 
|j,P\rangle_3=\frac{1}{\sqrt{2}}(|j,P\rangle_+-|j,P\rangle_-).
\end{equation}
At a fixed $P$, vectors $|j,P\rangle_\nu$, with $\nu=2,3$, and $j=1,2\ldots$, form the basis of the subspace $\mathcal{L}^{(2)}(P)$ introduced in the previous Section, and satisfy the normalisation condition:
\begin{equation}\label{scp}
\phantom{\sum}_\nu\langle j,P|j',P\rangle_{\nu'}=\delta_{\nu\nu'}\delta_{j,j'}.
\end{equation}
To solve the eigenvalue problem
\begin{equation}
H^{(2)}(h_2,g) |\Psi\rangle= E |\Psi\rangle,
\label{eq:eval_obliqoe_app}\end{equation}
we expand the eigenvector $|\Psi\rangle$ in the basis $|j,P\rangle_\nu$:
\begin{equation}\label{eq:Psi_def_app}
|\Psi\rangle=\sum_{j=1}^\infty \sum_{\nu=2}^3\psi_\nu(j;E,P) |j,P\rangle_\nu\,.
\end{equation}
For the sake of simplicity, we again suppress the variables $E$ and $P$ in the wave-functions $\psi_\nu(j;E,P)$ and use $\psi_\nu(j)$ instead. The eigenvalue problem \eqref{eq:eval_obliqoe_app} implies the following system of second-order linear difference equations: 
\begin{subequations}\label{dEE}
\begin{align}\label{dE2_app}
&(2-h_2 j -E) \psi_2(j)-\frac{2 g }{3 }  \cos(P/2)[\psi_2(j+1)+\psi_2(j-1)]=0,\\\label{dE3_app}
&(2 -E) \psi_3(j)-\frac{2 g }{3 }  \cos(P/2)[\psi_3(j+1)+\psi_3(j-1)]=0,
\end{align}
\end{subequations}
with $j=2,3,\ldots$.
In addition, the wave function coefficients satisfy the boundary conditions
\begin{subequations}\label{bc2_app}
\begin{align}
&\left(2-h_2 -E \right) \psi_2(1)-\frac{2 g }{3 } \cos(P/2)\, \psi_2(2)-\frac{ g }{3 } \psi_3(1) =0\,,\\
&\left(2 -E \right) \psi_3(1)-\frac{2 g }{3 } \cos(P/2)\, \psi_3(2)-\frac{ g }{3 } \psi_2(1) =0\,.
\end{align}
\end{subequations}
Furthermore, the coefficients $\psi_2(j)$ must vanish for  $j\to+\infty$, while the coefficients $\psi_3(j)$ must be at least bounded in absolute value. 

Equations \eqref{dEE} and boundary conditions \eqref{bc2_app} do not change upon the transformation:
\begin{equation}\label{sim2}
h_2\to - h_2,\quad  E\to 4-E, \quad \psi_2(j)\to (-1)^j\, \psi_2(j),\quad \psi_3(j)\to (-1)^{j+1}\, \psi_3(j)\,.
\end{equation}
The solution of equation \eqref{dE2_app} reads
\begin{equation}\label{ps2_app}
\psi_2(j;E,P)=[\sign(- h_2 )]^{j+1}\,J_{j-c_0}(Z),
\end{equation}
where $j=1,2\ldots$,
\begin{equation}
Z=\frac{4g}{3|h_2|} \cos(P/2),\qquad \text{and} \qquad
c_0=\frac{2-E}{h_2}\,,
\end{equation}
while the solution of equation \eqref{dE3_app} can be written as
\begin{equation}\label{ps3_app}
\psi_3(j;E,P)=\begin{cases}
A \sin[p(j-1)]+B\cos[p(j-1)], & \text{for }E_\text{min}(P)<E<E_\text{max}(P),\\
C \exp[-\lambda (j-1)],& \text{for } E<E_\text{min}(P),\\
C (-1)^{j+1}\exp[-\lambda (j-1)],& \text{for } E>E_\text{max}(P)\,,
\end{cases}
\end{equation}
where
\begin{align}
&p=\arccos\frac{3(2-E)}{4g \cos(P/2)}, \qquad \lambda=
\mathrm{arccosh}\,\frac{3|E-2|}{4g \cos(P/2)}.
\end{align}
The discussion of the resulting spectra is included in the main text. Here, we only note that due to the symmetry property \eqref{sim2}, the resulting meson and bubble energies in the oblique regimes satisfy the equality
\begin{equation}
E_n(P,h_2)+E_n(P,-h_2)=4.
\end{equation}
\section{True/false vacuum states for $g\ll 1$ \label{sec:vac}}

Here we apply the standard Rayleigh-Schrödinger perturbation theory to describe the properties of the vacuum states of the Potts spin chain \eqref{eq:Ham1} for $g\ll 1$. A similar analysis for the antiferromagnetic XXZ spin chain in the anisotropic limit was described by Jimbo and Miwa on Pages 14-18 of their monograph \cite{Jimbo94}.

Consider the Hamiltonian $H_N(\mathbf{v},g)$ given by \eqref{HamN} for some even number of sites $N$ with periodic boundary conditions.
The deformation $|\text{vac}(\mathbf{v},g|N)\rangle$ of the vacuum $|0\rangle^{(1)}$ satisfies the following properties:
\begin{subequations}
\begin{align}
&H_N(\mathbf{v},g) |\text{vac}(\mathbf{v},g|N)\rangle=E_\text{vac}(\mathbf{v},g|N)|\text{vac}(\mathbf{v},g|N)\rangle\,,\\
&\langle  \text{vac}(\mathbf{v},g|N)|\text{vac}(\mathbf{v},g|N)\rangle=1\,,\\
\label{vacdfi}
& |\text{vac}(\mathbf{v},g|N)\rangle=|0\rangle^{(1)}+|\delta \Phi(\mathbf{v},g|N)\rangle\,,\\
&\lim_{g\to+0}|\text{vac}(\mathbf{v},g|N)\rangle=|0\rangle^{(1)}=\bigotimes_{j=-N/2+1}^{N/2}|1\rangle_j\,.
\end{align}    
\end{subequations}
The deformed vacuum  $|\text{vac}(\mathbf{v},g|N)\rangle$ can be represented as
\begin{equation} \label{vac_expression}
|\text{vac} (\mathbf{v},g|N)\rangle=\frac{|\Omega(\mathbf{v},g|N)\rangle}{[\langle\Omega(\mathbf{v},g|N)|\Omega(\mathbf{v},g|N)\rangle]^{1/2}}\,, 
\end{equation} 
where the  vector $|\Omega(\mathbf{v},g|N)\rangle$ admits the  following asymptotic expansion:
\begin{equation}\label{dvac_app} 
|\Omega(\mathbf{v},g|N)\rangle=|0\rangle^{(1)}+g |\Omega_1(\mathbf{v}|N)\rangle+g^2 |\Omega_2(\mathbf{v}|N)\rangle+\ldots, 
\end{equation} 
such that
\[
\phantom{.}^{(1)}\langle 0| \Omega_n(\mathbf{v}|N)\rangle=0, \quad\text{for all } n=1,2,\ldots.
\]
The coefficients $|\Omega_n(\mathbf{v}|N)\rangle$ in this expansion can be perturbatively determined term by term. 
In particular:
\begin{equation}\label{eq:omega_1_app}
\lim_{N\to \infty}|\Omega_1(\mathbf{v}|N)\rangle=\frac{1}{6}\left(|1,0\rangle_2+|1,0\rangle_3\right)
\end{equation}
where we used the notation \eqref{jP23}. The norm of the vector \eqref{dvac_app}  can be expanded as
\begin{equation}
\langle \Omega(\mathbf{v},g|N)|\Omega(\mathbf{v},g|N)\rangle=1+g^2 \langle\Omega_1(\mathbf{v}|N)|\Omega_1(\mathbf{v}|N)\rangle+O(g^4) = 1+ \frac{N\,g^2}{18}+O(g^4).
\end{equation}
Denote by ${\mathfrak{N}}$ the linear operator acting on the space
\begin{equation}
\mathcal{L}_{11}=\mathcal{L}_{11}^{(0)}\oplus\mathcal{L}_{11}^{(2)}\oplus\mathcal{L}_{11}^{(3)}\oplus\ldots.
\end{equation} 
in the following way:
\begin{equation}
{\mathfrak{N}} |\Phi\rangle= n |\Phi\rangle, \quad \text{for } |\Phi\rangle\in \mathcal{L}_{11}^{(n)}.
\end{equation}
Then we can compute
\begin{align}
{\langle \text{vac}(\mathbf{v},g|N)|\mathfrak{N}|\text{vac}(\mathbf{v},g|N)\rangle}=\frac{\langle \Omega(\mathbf{v},g|N)|\mathfrak{N}|\Omega(\mathbf{v},g|N)\rangle}{\langle \Omega(\mathbf{v},g|N)|\Omega(\mathbf{v},g|N)\rangle}\\\nonumber
=
\frac{2 g^2\langle \Omega_1(\mathbf{v}|N)|\Omega_1(\mathbf{v}|N)\rangle+O(g^4)}{1+g^2\langle \Omega_1(\mathbf{v}|N)|\Omega_1(\mathbf{v}|N)\rangle+O(g^4)}
=\frac{N g^2}{9}+O(g^4).
\end{align}
This equality shows that in the state $|\text{vac}(\mathbf{v},g|N)\rangle$, the spin chain of length $N$ contains about $\frac{N g^2}{18}$ sites where the spin is oriented not in the first, but in the second or third direction. Therefore, such flipped spins are well-separated from one another for $g\ll 1$. E.g., at $g=0.2$,
the distance between the neighbouring flipped spins is about $18/0.2^2=450$ spin-chain sites. 

Similar reasoning leads to the relations
\begin{align}\label{E0vac}
&E_\text{vac}(\mathbf{0},g|N)=-\frac{2 g^2 N}{3} +O(g^3),\\\label{Avac}
&{\langle \text{vac}(\mathbf{v},g|N)|\left(P_j^1-\frac{2}{3}\right)|\text{vac}(\mathbf{v},g|N)\rangle}=-\frac{g^2}{18}+O(g^3),\\\nonumber
&\langle \text{vac}(\mathbf{v},g|N)|\left(P_j^2+\frac{1}{3}\right)|\text{vac}(\mathbf{v},g|N)\rangle\\
&=\langle \text{vac}(\mathbf{v},g|N)|\left(P_j^3+\frac{1}{3}\right)|\text{vac}(\mathbf{v},g|N)\rangle=\frac{g^2}{36}+O(g^3),
\end{align}
for $j=-N/2+1, \ldots, N/2$.
\section{Perturbative expansion for Potts quench dynamics}\label{sec:calc_potts_dyn_app}
Using the results of Appendix \ref{sec:vac}, here we determine the post-quench time evolution of the expectation value $A_\alpha(t|N)$ of the operator $\hat{A}_\alpha$, defined by (\ref{timAv},\ref{A123}), in the Potts spin chain. Note that
\begin{equation}
\hat{A}_\alpha|0\rangle^{(1)}=0, \quad \hat{A}_\alpha |\text{vac}(\mathbf{v},g|N)\rangle=O(g)\,, 
\end{equation} 
for $\alpha=1,2,3$.
Rewriting equation \eqref{vacdfi} as
\begin{equation}
 |0\rangle^{(1)}=|\text{vac}(\mathbf{v},g|N)\rangle-|\delta \Phi(\mathbf{v},g|N)\rangle\,,
\end{equation}
expanding $|\delta \Phi(\mathbf{v},g|N)\rangle$ to the second order in $g$ yields
\begin{equation}
|0\rangle^{(1)}=\left(1+\frac{N g^2}{36}\right)|\text{vac}(\mathbf{v},g|N)\rangle - g|\Omega_1(\mathbf{v}|N)\rangle- g^2|\Omega_2(\mathbf{v}|N)\rangle+O(g^3)\,.
\end{equation}
After substitution of  the right-hand side of the above equality into \eqref{timAv}, straightforward manipulations yield
\begin{align}\label{3terms}
&A_\alpha(t|N)
= \langle  \text{vac}(\mathbf{v},g|N)|\hat{A}_\alpha |\text{vac}(\mathbf{v},g|N)\rangle\\\nonumber
&-2 g^2 \,
\mathrm{Re}\, \langle  \Omega_1(\mathbf{v}|N)|
\hat{A}_\alpha \, e^{-i t[H_N(\mathbf{v},g)-E_\text{vac}(\mathbf{v},g|N)]}|\Omega_1(\mathbf{v}|N) \rangle\\\nonumber
&+g^2   \langle  \Omega_1(\mathbf{v}|N)|e^{i t[H_N(\mathbf{v},g)-E_\text{vac}(\mathbf{v},g|N)]}
\hat{A}_\alpha 
e^{-i t[H_N(\mathbf{v},g)-E_\text{vac}(\mathbf{v},g|N)]}|\Omega_1(\mathbf{v}|N) \rangle+O(g^3)\,.
\end{align}
Taking the thermodynamic limit $N\to\infty$ results in
\begin{align}\label{Aalf2}
&A_\alpha(t)
=\langle  \text{vac}(\mathbf{v},g)|\hat{A}_\alpha |\text{vac}(\mathbf{v},g)\rangle-
2 g^2 \, 
\mathrm{Re}\, \langle  \Omega_1(\mathbf{v})|
\hat{A}_\alpha\, e^{-i t\Delta H(\mathbf{v},g)}|\Omega_1(\mathbf{v}) \rangle\\\nonumber
&+g^2 \, \langle  \Omega_1(\mathbf{v})|e^{i t\Delta H(\mathbf{v},g)}
\hat{A}_\alpha e^{-i t\Delta H(\mathbf{v},g)}|\Omega_1(\mathbf{v}) \rangle+O(g^3), 
\end{align}
where following the main text we introduced the notations
\begin{subequations}\label{ThLim}
\begin{align}\label{Aal}
&A_\alpha(t)=\lim_{N\to\infty} A_\alpha(t|N),\\
&|\text{vac}(\mathbf{v},g)\rangle=\lim_{N\to\infty} |\text{vac}(\mathbf{v},g|N)\rangle,\\\label{Om}
&|\Omega_1(\mathbf{v}) \rangle=\lim_{N\to\infty}  |\Omega_1(\mathbf{v}|N) \rangle=\frac{1}{6}\left(|1,0\rangle_2+|1,0\rangle_3\right),\\\label{dHam_app}
&\Delta H(\mathbf{v},g)=\lim_{N\to\infty}\left[
H_N(\mathbf{v},g)-E_\text{vac}(\mathbf{v},g|N)
\right].
\end{align}
\end{subequations}
The first term on the right-hand side of \eqref{Aalf2} does not depend on time, and gives the expectation value of the operator $\hat{A}_\alpha$
in the deformed vacuum state $|\text{vac}(\mathbf{v},g)\rangle$. It follows from \eqref{Avac}, \eqref{A123} that
\begin{align}\label{Mvac1}
&\langle  \text{vac}(\mathbf{v},g)|\hat{A}_1 |\text{vac}(\mathbf{v},g)\rangle=-\frac{ g^2 }{18} +O(g^3),\\\label{Mvac}
&\langle  \text{vac}(\mathbf{v},g)|\hat{A}_\alpha |\text{vac}(\mathbf{v},g)\rangle=\frac{ g^2 }{36} +O(g^3), \quad \text{for  }\alpha=2,3.
\end{align}
The vector $|\Omega_1(\mathbf{v}) \rangle$ in the second and third terms on the right-hand side of \eqref{Aalf2} is given by \eqref{Om}, and corresponds to a translation invariant two-kink state:
\begin{equation} 
|\Omega_1(\mathbf{v}) \rangle\in \mathcal{L}^{(2)}(0):=\mathcal{L}^{(2)}(P)\big|_{P=0}.
\end{equation}
Operators $\hat{A}_\alpha$ act as homomorphisms in the subspace $\mathcal{L}^{(2)}(0)$, whose basis is given by vectors $\{|j,0\rangle_{2,3}\}_{j=1}^\infty$ determined by  \eqref{jP23}. The matrix elements of the operators $\hat{A}_\alpha$ in this basis are
\begin{subequations}\label{Aj}
\begin{align}\label{eq:A1_mx_element}
&\phantom{}_\nu\langle j,0|\hat{A}_1 |j',0\rangle_{\nu'}=-j\, \delta_{j,j'}\,\delta_{\nu,\nu'},\quad \text{for } \nu,\nu'=2,3,\\\label{eq:Aalpha_mx_element}
&\phantom{}_\nu\langle j,0|\hat{A}_\alpha |j,'0\rangle_{\nu'}=j\,\delta_{\alpha,\nu}\, \delta_{\alpha,\nu'}\,\delta_{j.j'}, \quad \text{for } \alpha,\nu, \nu'=2,3.
\end{align}
\end{subequations}
where no summation over repeated indices is implied on the right-hand sides. 

Note that the 
(modified) scalar product in the subspace
$\mathcal{L}^{(2)}(0)$ is defined according to \eqref{scp}:
\begin{equation}\label{scp1}
\phantom{\sum}_\nu\langle j,0|j',0\rangle_{\nu'}=\delta_{\nu\nu'}\delta_{j,j'},
\end{equation}
with $j=1,2,\ldots$, and $\nu=2,3$.
Accordingly, the completeness relation for the subspace $\mathcal{L}^{(2)}(0)$ reads
\begin{equation}\label{pro}
\mathbf{1}=\sum_{j=1}^\infty \left(| j, 0\rangle_2\,\,\phantom{.}_{2}\langle j, 0|+| j, 0\rangle_3\,\,\phantom{.}_{3}\langle j, 0|\right),
\end{equation}
where $\mathbf{1}$ is the identity operator acting on $\mathcal{L}^{(2)}(0)$. 

In general, the Hamiltonian  $\Delta H(\mathbf{v},g)$ determined by \eqref{dHam_app} does not conserve the number of kinks. Therefore at $t>0$ the state $|\Psi(t)\rangle=e^{- i t\, \Delta H(\mathbf{v},g)}|\Omega_1(\mathbf{v})\rangle $ gains 
contributions of the $n$-kink states with $n=3,4,\ldots$:
\begin{equation}\label{multi}
|\Psi(t)\rangle=|\Psi^{(2)}(t)\rangle+|\Psi^{(3)}(t)\rangle+|\Psi^{(4)}(t)\rangle+\ldots,
\end{equation}
where $|\Psi^{(n)}(t)\rangle\in \mathcal{L}_{11}^{(n)}$. This prevents the exact analytical calculation of the second and third terms on the right-hand side of \eqref{Aalf2} at $t>0$. The key simplification is provided by the two-kink approximation, which corresponds to replacing the Hamiltonian $\Delta H(\mathbf{v},g)$ by the restriction of the Hamiltonian $H(\mathbf{v},g)$ defined by  \eqref{eq:Ham1}  to the subspace $\mathcal{L}_{11}^{(2)}$:
\begin{align}\label{DHam:app}
&\Delta H(\mathbf{v},g) \to  H^{(2)}(\mathbf{v},g)={\mathcal{P}}_{11}^{(2)}\, H(\mathbf{v},g)\,  {\mathcal{P}}_{11}^{(2)}\,.
\end{align}
Note, that the Hamiltonians $H(\mathbf{v},g)$ and $\Delta H(\mathbf{v},g)$ differ 
only by the constant term  $\sim g^2$, which arises from the vacuum energy \eqref{E0vac}.
Substitution of \eqref{DHam:app} into \eqref{3terms}, \eqref{Aal}  leads to the result
\begin{align}\label{2k_app}
A_\alpha(t)\to A_\alpha^{(2)}(t)
&=g^2\,\langle   \Omega_1(\mathbf{v})|\hat{A}_\alpha | \Omega_1(\mathbf{v})\rangle-
2 g^2 \, 
\mathrm{Re}\, \langle  \Omega_1(\mathbf{v})|
\hat{A}_\alpha\, e^{-i t\,H^{(2)}(\mathbf{v},g)}|\Omega_1(\mathbf{v}) \rangle
\\\nonumber
&+g^2 \, \langle  \Omega_1(\mathbf{v})|e^{i t\,H^{(2)}(\mathbf{v},g)}
\hat{A}_\alpha e^{-i t\,H^{(2)}(\mathbf{v},g) }|\Omega_1(\mathbf{v}) \rangle+O(g^3), 
\end{align}
where the scalar product is understood in accordance with \eqref{scp1}. In the above expression, only the second and third terms exhibit time dependence, while the first (time-independent) term is determined by (\ref{vac_expression},\ref{dvac_app}) and (\ref{Mvac1},\ref{Mvac}) as
\begin{equation}
g^2\,\langle   \Omega_1(\mathbf{v})|\hat{A}_\alpha | \Omega_1(\mathbf{v})\rangle=
\begin{cases}
   \frac{-g^2}{18}+O(g^3)\,,&\text{if } \quad\alpha=1\,,\\
   \frac{ g^2 }{36} +O(g^3)\,,& \text{if  } \quad\alpha=2,3\,.
   \end{cases}  
   \label{eq:cst_term}
\end{equation}
For the remainder of this section, we focus on the oblique regime $\mathbf{v}=(0,h_2/g,0)$, and use the short-hand notation $H^{(2)}(h_2,g)$ for $H^{(2)}(\mathbf{v},g)$. We start by considering the matrix element
\begin{equation}\label{be}
\langle  \Omega_1(h_2)|
\hat{A}_\alpha\, e^{-i t\,H^{(2)}(h_2,g)}|\Omega_1(h_2) \rangle=\langle  \Omega_1(h_2)|
\hat{A}_\alpha\, e^{-i t\,H^{(2)}(h_2,g)}\, \mathbf{1}|\Omega_1(h_2) \rangle,
\end{equation}
that stands on the right-hand side of \eqref{2k_app}. 

The identity  operator \eqref{pro} acting on the subspace $\mathcal{L}^{(2)}(0)$ admits the alternative representation:
\begin{equation}\label{un2}
\mathbf{1}=\sum_{n=1}^\infty \frac{|\psi_n\rangle \langle \psi_n|}{\langle \psi_n|\psi_n\rangle}
+\int_0^\pi \frac{dp}{2\pi}\frac{|\psi(p)\rangle\langle \psi(p)| }{|\mathrm{B}(p)|^2}. 
\end{equation}
Here $|\psi_n\rangle$ and $|\psi(p)\rangle$ are the eigenstates of the Hamiltonian $H^{(2)}(h_2,g)$ corresponding
to the discrete and continuous energy spectra, respectively
\begin{align}
&H^{(2)}(h_2,g)|\psi_n\rangle=E_n |\psi_n\rangle,\\
&H^{(2)}(h_2,g)|\psi(p)\rangle=E(p)|\psi(p)\rangle,
\end{align}
where $|\psi_n\rangle, |\psi(p)\rangle\in \mathcal{L}^{(2)}(0)$, and $E(p)=2-\frac{4 g}{3}\cos p$.
These are essentially the same states as given by Eq. \eqref{eq:Psi_def_app} but reduced to the subspace $P=0$. Thus, with the notation $\psi_\nu (j;E,P=0)$ introduced in the main text for the wave-function, these states can be expressed as 
\begin{equation}\label{eq:psi_p_app}
|\psi(p)\rangle=\sum_{j=1}^\infty \sum_{\nu=2}^3\psi_\nu(j;E,0) |j,0\rangle_\nu
\end{equation}
for the continuous spectra and similarly as 
\begin{equation}\label{eq:psi_n_app}
|\psi_n\rangle=\sum_{j=1}^\infty \sum_{\nu=2}^3\psi_\nu(j;E_n,0) |j,0\rangle_\nu
\end{equation}
for the discrete case.
The Hamiltonian eigenstates corresponding to the continuous spectrum are normalised by the condition:
\begin{equation}
\langle \psi(p)|\psi(p')\rangle = 2\pi \delta(p-p') |\mathrm{B}(p)|^2,
\end{equation}
where $p,p'\in (0,\pi)$, and  $\mathrm{B}(p)=B_{in}(e^{i p}, h_2/g)$, with the function $B_{in}(z,v)$ defined by 
\eqref{Bin1}.

Combining \eqref{be} with \eqref{un2}, one finds:
\begin{align}
&\langle  \Omega_1(h_2)|\hat{A}_\alpha\, e^{-i tH^{(2)}(h_2,g)}|\Omega_1(h_2) \rangle=
\sum_{n=1}^\infty  e^{-i t E_n} \frac{\langle  \Omega_1(h_2)|\psi_n\rangle\, \langle \psi_n|\hat{A}_\alpha|\Omega_1(h_2) \rangle}{\langle \psi_n|\psi_n\rangle}\\\nonumber
&+\int_0^\pi \frac{dp}{2\pi} e^{- i t E(p)}\frac{\langle  \Omega_1(h_2)|\hat{A}_\alpha|\psi(p)\rangle\langle \psi(p)|
\Omega_1(h_2) \rangle }{|\mathrm{B}(p)|^2}.
\end{align}
All scalar products and matrix elements on the right-hand side are real.
Therefore:
\begin{align}\label{ReA}
&\mathrm{Re}\,  \langle  \Omega_1(h_2)|\hat{A}_\alpha\, e^{-i tH^{(2)}(h_2,g)}|\Omega_1(h_2) \rangle=
\sum_{n=1}^\infty  \cos( t E_n)\,  \frac{\langle  \Omega_1(h_2)|\psi_n\rangle\, \langle \psi_n|\hat{A}_\alpha|\Omega_1(h_2) \rangle}{\langle \psi_n|\psi_n\rangle}\\\nonumber
&+\int_0^\pi \frac{dp}{2\pi} \cos[E(p) t ]\,\frac{\langle  \Omega_1(h_2)|\hat{A}_\alpha|\psi(p)\rangle\langle \psi(p)|
\Omega_1(h_2) \rangle }{|\mathrm{B}(p)|^2},
\end{align}
Applying the expansions (\ref{eq:psi_p_app},\ref{eq:psi_n_app}), the expression of $|\Omega_1(h_2) \rangle$ given by  \eqref{eq:omega_1_app} and the relations \eqref{Aj}, the relevant terms can be determined as
\begin{subequations}
\label{eq:evolution_firstterm}
\begin{align}
\langle  \Omega_1(h_2)|\psi_n\rangle&=\frac{1}{6} \left( \psi_2(1,E_n,0)+\psi_3(1,E_n,0)\right), 
\\
\langle \psi_n|\hat{A}_\alpha|\Omega_1(h_2) \rangle&= \begin{cases}
   -\frac{1}{6}\left(\psi_2(1,E_n,0)+\psi_3(1,E_n,0)\right)\,,&\text{if } \quad\alpha=1\,,\\
   \frac{1}{6}\left(\psi_\alpha(1,E_n,0)\right)\,,& \text{if  } \quad\alpha=2,3\,,
   \end{cases}
\\
\langle \psi_n|\psi_n\rangle&=\sum_{j=1}^\infty \sum_{\nu=2}^3|\psi_\nu(j,E_n,0)|^2, 
\end{align}    
\end{subequations}
with similar results for the continuous part of the spectrum. The Fourier transformation of the second term on the right-hand side of \eqref{2k_app} reads:
\begin{align}
&F_\alpha^{(1)}(\omega):= \int_0^\infty dt \,e^{-i \omega t} \, \left[-2 g^2 \, 
\mathrm{Re}\,  \langle  \Omega_1(h_2)|\hat{A}_\alpha\, e^{-i tH^{(2)}(h_2,g)}|\Omega_1(h_2) \rangle\right]\label{eq:FT_analitical_def_app}\\\nonumber
&=i \,g^2 \,   \sum_{n=1}^\infty \left(\frac{1}{\omega+E_n-i 0}+\frac{1}{\omega-E_n-i 0}\right)
\frac{\langle  \Omega_1(h_2)|\psi_n\rangle\, \langle \psi_n|\hat{A}_\alpha|\Omega_1(h_2) \rangle}{\langle \psi_n|\psi_n\rangle}\\
&+i \,g^2 \,  \int_0^\pi \frac{dp}{2\pi} \left(\frac{1}{\omega+E(p)-i 0}+\frac{1}{\omega-E(p)-i 0}\right)\,\frac{\langle  \Omega_1(h_2)|\hat{A}_\alpha|\psi(p)\rangle\langle \psi(p)|
\Omega_1(h_2) \rangle }{|\mathrm{B}(p)|^2}.\nonumber
\end{align}
For the real part of $F_\alpha^{(1)}(\omega)$, this yields:
\begin{align}
&\mathrm{Re}\, F_\alpha^{(1)}(\omega)=-{\pi \,g^2  }
 \sum_{n=1}^\infty \left[\delta(\omega+E_n)+\delta(\omega-E_n)\right]
\frac{\langle  \Omega_1(h_2)|\psi_n\rangle\, \langle \psi_n|\hat{A}_\alpha|\Omega_1(h_2) \rangle}{\langle \psi_n|\psi_n\rangle}\\\nonumber
&-\frac{g^2}{2} 
 \int_0^\pi {dp}\,\left(\delta[\omega+E(p)]+\delta[\omega-E(p)]\right)\,\frac{\langle  \Omega_1(h_2)|\hat{A}_\alpha|\psi(p)\rangle\langle \psi(p)|
\Omega_1(h_2) \rangle }{|\mathrm{B}(p)|^2}. 
\end{align}
For $\omega\in (2-\frac{4g}{3},2+\frac{4g}{3}) $, we get, in particular:
\begin{equation}\label{ReF1}
\mathrm{Re}\, F_\alpha^{(1)}(\omega)=-\frac{3 g}{8 \sin p}\, \frac{\langle  \Omega_1(h_2)|\hat{A}_\alpha|\psi(p)\rangle\langle \psi(p)|
\Omega_1(h_2) \rangle}{|\mathrm{B}(p)|^2},
\end{equation}
where 
\begin{equation}
p=\arccos \frac{3( 2-\omega)}{4 g}.
\end{equation}
For $\alpha=1,2$, the explicit form of formula \eqref{ReF1} reads:
\begin{align}\label{eq:ReF_12_om}
&\mathrm{Re}\, F_1^{(1)}(\omega)=\frac{3 g}{8 \sin p}\,\frac{[\psi_2(1;\omega, 0)+\psi_3(1;\omega, 0)]^2}{36|\mathrm{B}(p)|^2 },\\
&\mathrm{Re}\, F_2^{(1)}(\omega)=-\frac{3 g}{8 \sin p}\,\frac{\psi_2(1;\omega, 0)\, [\psi_2(1;\omega, 0)+\psi_3(1;\omega, 0)]}{36|\mathrm{B}(p)|^2 }.
\end{align}
Here we recall that
\begin{subequations}\label{psiB}
\begin{align}
&\psi_2(1;\omega, 0)=J_{1-\frac{4 g \cos p}{3 h_2}}\left(\frac{4 g}{3|h_2|}\right),\\
&\psi_3(1;\omega, 0)=\mathrm{B}(p)\,e^{-i p}+\mathrm{B}(-p)\,e^{i p},\\
&\mathrm{B}(p)=B_{in} (e^{ip},h_2/g),
\end{align}
\end{subequations}
where  the function $B_{in}(z,v_2)$ is given by \eqref{Bin1}.

Let us turn now to the matrix element
\begin{align}\label{be2}
\begin{aligned}
&\langle  \Omega_1(h_2)|e^{i t\,H^{(2)}(h_2,g)}
\hat{A}_\alpha e^{-i t\,H^{(2)}(h_2,g) }|\Omega_1(h_2) \rangle\\
=&\langle  \Omega_1(h_2)|e^{i t\,H^{(2)}(h_2,g)}
\hat{A}_\alpha \, \mathbf{1}\, e^{-i t\,H^{(2)}(h_2,g) }|\Omega_1(h_2) \rangle,    
\end{aligned}
\end{align}
which appears in the second line on the right-hand side of \eqref{2k_app}. Exploiting the completeness identity in the form \eqref{pro}, together with formulas \eqref{Aj}, the right-hand side
of \eqref{be2} can be evaluated as:
\begin{align}
&\langle  \Omega_1(h_2)|e^{i t\,H^{(2)}(h_2,g)}
\hat{A}_1 e^{-i t\,H^{(2)}(h_2,g) }|\Omega_1(h_2) \rangle=
-\sum_{j=1}^\infty j\, \left[|\varphi_2(j,t)|^2+|\varphi_3(j,t)|^2\right],\\\label{Anu}
&\langle  \Omega_1(h_2)|e^{i t\,H^{(2)}(h_2,g)}
\hat{A}_\nu e^{-i t\,H^{(2)}(h_2,g) }|\Omega_1(h_2) \rangle=
\sum_{j=1}^\infty j\, |\varphi_\nu(j,t)|^2,\quad \text{with }\nu=2,3,
\end{align}
where
\begin{equation}
\varphi_\nu(j,t):=\phantom{.}_{\nu}\langle j,0| e^{-i t\,H^{(2)}(h_2,g) }|\Omega_1(h_2) \rangle.
\end{equation}
Using the completeness identity in the form \eqref{un2}, together with formula \eqref{Om}, the latter function can be represented in the explicit form: 
\begin{align}\label{phinu}
&\varphi_\nu(j,t)=\phantom{.}_{\nu}\langle j,0| e^{-i t\,H^{(2)}(h_2,g) }
\, \mathbf{1}\,|\Omega_1(h_2) \rangle\\\nonumber
&=\frac{1}{6}\sum_{n=1}^\infty e^{-i t E_n}\frac{\phantom{.}_{\nu}\langle j,0|\psi_n\rangle\,  \left(\langle\psi_n|1,0\rangle_2+
\langle\psi_n|1,0\rangle_3\right)}{\langle\psi_n|\psi_n\rangle}\\\nonumber
&+\frac{1}{6}\int_0^\pi \frac{dp}{2\pi }\,e^{-i t E(p)}
\frac{\psi_\nu(j;E,0)\,  \left[\psi_2(1;E,0)+\psi_3(1;E,0)\right]}{|\mathrm{B}(p)|^2},
\end{align}
where 
\begin{equation}
   \phantom{.}_{\nu}\langle j,0|\psi_n\rangle =\psi_\nu(1;E_n,0), \quad \langle \psi_n|1,0\rangle_2=\psi_2(1;E_n,0), \quad \langle \psi_n|1,0\rangle_3=\psi_3(1;E_n,0)
\end{equation}
and of course, once again $E(p)=2-\frac{4 g}{3}\cos p$.

The post-quench time evolution of the magnetisations can be then calculated by the means of equations \eqref{A123} and (\ref{2k_app},\ref{eq:cst_term},\ref{ReA},\ref{eq:evolution_firstterm},\ref{Anu},\ref{phinu}). The resulting time evolution of the magnetisations $M_2(t)$, and $M_3(t)$ are displayed in Fig. \ref{fig:M} at $g=0.2$ in the positive (a),
and negative (b) oblique regimes with $h_2=0.1$, and $h_2=-0.1$, respectively.

Our perturbative approach predicts
\begin{equation}
M_3(t)+\frac{1}{3}= c\,  t+O(1), \quad \text{at }t\to\infty.
\label{eq:inf_time_behav_app}
\end{equation}
that is also reflected in Fig. \ref{fig:M} for the first stage of the time evolution.
The explicit form of $c$ 
can be found by means of the asymptotical analysis of 
equations \eqref{Anu}, \eqref{phinu} at $t\to\infty$. The final result reads:
\begin{equation}\label{ch2g_app}
c=\frac{g^3}{54 \pi}\int_0^\pi dp\, \sin p \, \frac{ \left[\psi_2(1,p)+\psi_3(1,p)\right]^2}{|\mathrm{B}(p)|^2}.
\end{equation}
\section{Quantum quench in the Ising spin chain at a
weak transverse magnetic field}\label{ApIs}

Here, we test our perturbative approach using quantum quenches in the transverse-field Ising  chain  (TFIC), defined by the Hamiltonian
\begin{equation}
\label{HamIN}
     H_{N} (h_x)=
        -\frac{1}{2}\sum_{j=-N/2+1}^{N/2} \left((\sigma_j^z \,\sigma_{j+1}^z-1)+ h_x \sigma_j^x  \right) \,,
\end{equation}
where $N$ is even, and the periodic boundary condition $\sigma_{N/2+1}^z=\sigma_{-N/2+1}^z$ is imposed. This Hamiltonian acts in the $2^N$-dimensional Hilbert space
\begin{equation}
W_N=\bigotimes_{j=-N/2+1}^{N/2} [\mathbb{C}^2]_j.
\end{equation}
We denote the basis of the two-dimensional local vector space $[\mathbb{C}^2]_j$ associated with the site $j$ as $|\mu\rangle_j$, with $\mu=1,2$. 

As mentioned in the Introduction, this model is integrable for any value of the transverse magnetic field $h_x$ and admits a representation in terms of non-interacting fermions via the Jordan-Wigner transformation. The free-fermionic representation allows one to describe the dynamics in the TFIC in a very simple way \cite{2011PhRvL.106v7203C,2012JSMTE..07..016C,2012JSMTE..07..022C}, making it a paradigmatic model for quantum quenches. 

The small parameter of our perturbation theory is 
$h_x\ll1$. We consider the quench protocol where the initial state is the pure ferromagnetic state along direction 1:
\begin{equation}
|0\rangle^{(1)}=\bigotimes_{j=-N/2+1}^{N/2}|1\rangle_j.
\end{equation}
This state represents one of two degenerate  ground states of the Hamiltonian \eqref{HamIN} at $h_x=0$:
\begin{equation}
H_N(0)|0\rangle^{(1)}=0.
\end{equation}
Let us tune the transverse magnetic field $h_x$ to some small positive value, and consider the eigenvalue problem:
\begin{equation}
H_N(h_x) |\Omega(h_x|N)\rangle=E_\text{vac}(h_x|N)|\Omega(h_x|N)\rangle,
\end{equation}
in which $E_\text{vac}(h_x|N)$ and $|\Omega(h_x|N)\rangle$ are understood as formal power series in $h_x$. 
In particular: 
\begin{equation}\label{dvac1}
|\Omega(h_x|N)\rangle=|\Omega_0(N)\rangle+h_x |\Omega_1(N)\rangle+h_x^2 |\Omega_2(N)\rangle+\ldots. 
\end{equation}
We require $|\Omega_0(N)\rangle=|0\rangle^{(1)}$, and 
$\phantom{.}^{(1)}\langle 0| \Omega_n(N)\rangle=0$, for all  $n=1,2,\ldots$.
The normalised deformed vacuum state associated with \eqref{dvac1} reads:
\begin{equation}
|\text{vac}(h_x|N)\rangle=\frac{|\Omega(h_x|N)\rangle}{[\langle\Omega(h_x|N))|\Omega(h_x|N))\rangle]^{1/2}}.
\end{equation}
Let us consider the operator $\hat{A}$, which differs by the constant shift $\frac{1}{2}$ from the magnetisation operator
$\hat{M}$:
\begin{align}
&\hat{A}=\hat{M}-\frac{1}{2},\\
&\hat{M}=\frac{1}{ 2N} \sum_{j=-N/2+1}^{N/2}\sigma_j^z.
\end{align}
The time evolution of its expectation value is given by
\begin{equation}\label{timAvis}
A(t|N)=\phantom{.}^{(1)}\langle 0|\exp[i \,H_N(h_x)t]\, \hat{A}\,\exp[-i\, H_N(h_x)t]|0\rangle^{(1)}.
\end{equation}
Following the procedure described in Section  \ref{sec:Pottsquench}, one obtains  in the thermodynamic limit $N\to\infty$
the small-$h_x$ asymptotic formula for the magnetisation $M(t)$:
\begin{align}\label{At}
M(t)
=&\frac{1}{2}-\frac{h_x^2}{16}-
2 h_x^2 \, \,
\mathrm{Re}\, \langle  \Omega_1|
\hat{A}\, e^{-i t\,\Delta H(h_x)}|\Omega_1 \rangle\\\nonumber
&+h_x^2 \, \langle  \Omega_1|e^{i t\,\Delta H(h_x)}
\hat{A}\, e^{-i t\,\Delta H(h_x)}|\Omega_1 \rangle+O(h_x^3), 
\end{align}
which is analogous to \eqref{Aalf2}, where we introduced the notations
\begin{subequations}\label{ThLim1}
\begin{align}
&M(t)=\frac{1}{2}+\lim_{N\to\infty}A(t|N),\\
&|\text{vac}(h_x)\rangle=\lim_{N\to\infty} |\text{vac}(h_x|N)\rangle,\\\label{Om1}
&|\Omega_n \rangle=\lim_{N\to\infty}  |\Omega_n(N) \rangle,\\\label{dHamis}
&\Delta H(h_x)=\lim_{N\to\infty}\left[
H_N(h_x)-E_\text{vac}(h_x|N)
\right].
\end{align}
\end{subequations}
The next crucial step of the calculations is provided by the two-kink approximation, which replaces the 
Hamiltonian  $\Delta H(h_x)$ in the asymptotic formula \eqref{At} by its restriction $\Delta H^{(2)}(h_x)$ to the  subspace
$\mathcal{L}_{11}^{(2)}$ spanned by the translation invariant two-kink states:
\begin{equation}\label{j}
|j\rangle=\sum_{j_1=-\infty}^\infty |\mathbf{K}_{12}(j_1)\mathbf{K}_{21}(j_1+j)\rangle, \quad \text{with }j=1,2,\ldots,
\end{equation}
which yields
\begin{align}\label{M2}
&M(t)\to M^{(2)}(t)=
\frac{1}{2}-\frac{h_x^2}{16}-
2 h_x^2 \,\, 
\mathrm{Re}\, \langle  \Omega_1|
\hat{A}\, e^{-i t\, \Delta H^{(2)}(h_x)}|\Omega_1 \rangle\\\nonumber
&+h_x^2 \, \langle  \Omega_1|e^{i t\Delta H^{(2)}(h_x)}
\hat{A}\, e^{-i t\Delta H^{(2)}(h_x)}|\Omega_1 \rangle,
\end{align}
where
\begin{equation}
 \Delta H^{(2)}(h_x)=\mathcal{P}_{11}^{(2)} \,
\Delta H(h_x)\,\mathcal{P}_{11}^{(2)}\,,
\end{equation}
and $\mathcal{P}_{11}^{(2)}$ is the projection operator onto the subspace $\mathcal{L}_{11}^{(2)}$. Taking into account the evident equality
\begin{equation}
|\Omega_1 \rangle=\frac{1}{4}\, |1\rangle,
\end{equation}
the result \eqref{M2} can be rewritten as
\begin{equation}\label{MIs2}
M^{(2)}(t)=\frac{1}{2}-\frac{h_x^2}{16}+A_1^{(2)}(t)+A_2^{(2)}(t),
\end{equation}
where
\begin{align}\label{A12}
&A_1^{(2)}(t)=- \frac{h_x^2}{8} \, 
\mathrm{Re}\, \langle 1|
\hat{A}\, e^{-i t \Delta H^{(2)}(h_x)}|1 \rangle,\\\label{A22}
&A_2^{(2)}(t)=\frac{h_x^2}{16} \, \langle  1|e^{i t\Delta H^{(2)}(h_x)}
\hat{A}\, e^{-i t\Delta H^{(2)}(h_x)}|1 \rangle.
\end{align}
The function $A_1^{(2)}(t)$ determined by \eqref{A12} describes high-frequency oscillations of the magnetisation $M^{(2)}(t)$. 
The explicit representation for  this function reads:
\begin{align}\label{A1Is}
A_1^{(2)}(t)=\frac{h_x^2}{4\pi}\int_0^\pi dp\,  \cos[(2-2 h_x \cos p )t ]\, \sin^2 p
=h_x^2\, \frac{J_1(\tau)}{4 \tau}\,\cos(2 t),
\end{align}
where  $J_j(\tau)$ is the Bessel function, and $\tau=2 h_x t$.

In turn, the function $A_2^{(2)}(t)$ determined by \eqref{A22} contains the slowly varying contribution to the magnetisation, and its explicit form reads:
\begin{align}\label{A2Is}
&A_2^{(2)}(t)=-\frac{h_x^2}{16}\sum_{j=1}^\infty j \,|\varphi(j,t)|^2\\\nonumber
&=\frac{h_x^2}{48} [
-(3+4 \tau^2)J_0^2(\tau)+4\tau J_0(\tau)J_1(\tau)-(1+4 \tau^2)J_1^2(\tau)],
\end{align}
where 
\begin{align}
\varphi(j,t):=\langle j|e^{-i tH^{(2)}(h_x)}|1 \rangle
= i^{j-1}\, e^{-2 i t}\,\frac{ 2\, j\,  J_j(\tau)}{\tau}.
\end{align}
The asymptotic behaviour of $A_2^{(2)}(t)|_{t=\tau/(2 h_x)}$ at large $\tau$ is given by:
\begin{equation}
A_2^{(2)}(t)|_{t=\tau/(2 h_x)}=h_x^2\left[-\frac{\tau}{6\pi}-\frac{1}{16\pi \tau}+O(\tau^{-3})\right].
\label{A22_asymp}\end{equation}
This result can be matched against the analytic results of Ref. \cite{2012JSMTE..07..016C}, where the large-time asymptotic of the magnetisation is obtained as\footnote{Note that our convention for the Hamiltonian and the magnetisation both differ by a factor of $1/2$, which accounts for the difference of Eq. \eqref{eq:CEF_result} from Eq. (15) of Ref. \cite{2012JSMTE..07..016C}.} 
\begin{align}
M(t)=&\frac{1}{2}\left(1-h_x^2\right)^{1/8}\exp
\left[
-t\int\limits_0^\pi \frac{dk}{\pi}\epsilon'(k)2K^2(k)
\right] \label{eq:CEF_result}, \\
&\epsilon(k)=\sqrt{1+h_x^2-2h_x\cos k}\nonumber\\
&K^2(k)=\frac
{\sqrt{1+h_x^2-2h_x\cos k}-1+h_x\cos k}
{\sqrt{1+h_x^2-2h_x\cos k}+1-h_x\cos k}.\nonumber
\end{align}
Expanding this expression to order $h_x^3$ gives
\begin{equation}
M(t)=\frac{1}{2}-\frac{h_x^2}{16}-\frac{h_x^3}{3\pi}t+O(h^4)
\label{CEF_expanded}
\end{equation}
which correctly reproduces the linear term in \eqref{A22_asymp}. The full exponential relaxation of the magnetisation, as manifest in \eqref{eq:CEF_result}, can be obtained by a form-factor resummation \cite{2012JSMTE..07..016C}; the perturbative method we use only gives its lowest-order expansion, which is linear in time. However, as noted in \cite{2012JSMTE..07..016C}, the form factor resummation cannot reproduce the oscillating contributions which are clearly absent from (\ref{eq:CEF_result}, \ref{CEF_expanded}). These oscillating terms were computed \cite{2012JSMTE..04..017S} and numerically verified \cite{2016NuPhB.911..805R} in the scaling field theory limit, which cannot be compared to the perturbative result valid for $h_x\ll 1$, since the scaling limit is obtained in the vicinity of the critical point $h_x=1$.  

\begin{figure}[t]
\centering
\includegraphics[width=.55
\linewidth, angle=00]{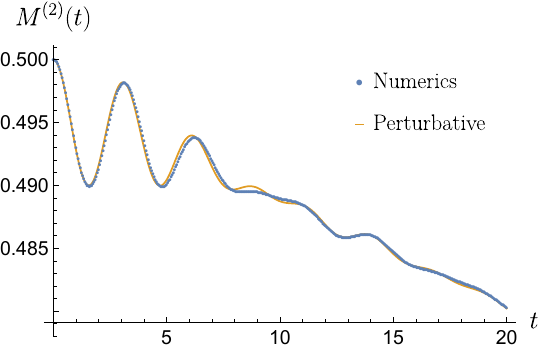}
\caption{Validation Case: Transverse-Field Ising Chain (TFIC).\\
Post-quench magnetisation $M^{(2)}(t)$ for the Ising spin chain at $h_{x}=0.2$. The perturbative result (orange curve) derived from equations  \eqref{MIs2}, \eqref{A1Is}, and \eqref{A2Is} is compared against iTEBD numerics (blue dots). This benchmark confirms the accuracy of the perturbative method in capturing the initial high-frequency oscillations and the linear initial relaxation phase.
 \label{fig:MIs} } 
\end{figure}

Fig. \ref{fig:MIs} illustrates the post-quench time evolution of the magnetisation determined by 
equations \eqref{MIs2}, \eqref{A1Is}, 
and \eqref{A2Is} at $h_x=0.2$, compared to the results of the numerical simulation of the time evolution. The simulation was performed using the infinite volume Time-Evolving Block Decimation (iTEBD) method with second-order Trotterisation, time step $\delta t = 0.005$ and maximal bond dimension $\chi_\text{max} = 300$. The Fourier images $F_\alpha(\omega)$ of the functions $A_\alpha^{(2)}(t)$ are defined according  to:
\begin{equation}
F_\alpha(\omega):=\int_0^\infty dt\,e^{-i \omega t} A_\alpha^{(2)}(t), \quad \text{for }\alpha=1,2.
\end{equation}
Here we present only the remarkably simple formula for the real part of the function $F_1(\omega)$:
\begin{equation}
\mathrm{Re}\, F_1(\omega)=\begin{cases}
0,& \text{if }  |2-\omega|\ge 2h_x \\
\frac{1}{32}\sqrt{4 h_x^2-(2-\omega)^2}, & \text{if }  |2-\omega|< 2h_x,
\end{cases}
\end{equation}
which holds at $\omega>0$.

\section{Details of the numerical simulations
\label{Numerical_details}}
The numerical simulations of the time evolution after quantum quenches were performed using the infinite volume Time-Evolving Block Decimation (iTEBD) method. In the simulations, we used second-order Trotterisation with time step $\delta t=0.005$, evaluating the magnetisation(s) after every 20th step. The maximal bond dimension was set at $\chi_\text{max}=300$ for simulations running to longer times ($t=1000$), and at $\chi_\text{max}=800$ for simulations running up to shorter times (approximately $t\approx 200$). Larger bond dimensions ensure higher precision; however, they consume much more computer resources and run much more slowly. Nevertheless, to achieve sufficient resolution for quench spectroscopy, long-time runs are required. The resulting Fourier spectra of the long-time simulations with $\chi_\text{max}=300$ and the short-time ones with $\chi_\text{max}=800$ essentially agree, as shown in Ref. \cite{Po25}.

\begin{figure}
\centering
\begin{subfigure}{0.495\textwidth}
    \includegraphics[width=\textwidth]{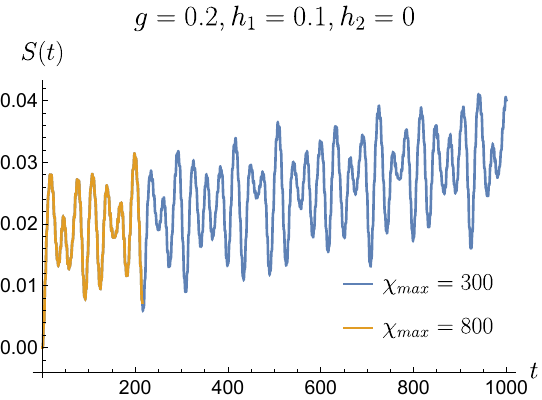}
   \caption{Entanglement entropy after the positively aligned quench.}
\end{subfigure}
\hfill
\begin{subfigure}{0.495\textwidth}
    \includegraphics[width=\textwidth]{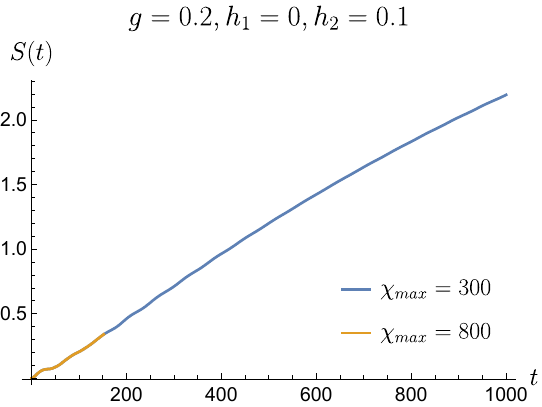}
    \caption{Entanglement entropy after the positive oblique quench.}
\end{subfigure}
\caption{Time evolution of entanglement entropy $S(t)$ for different bond dimensions $\chi_{max} = 300$ (blue) and $800$ (orange). In the positively aligned case (a), entropy exhibits slow drift due to kink confinement. In the oblique case (b), the entropy grows rapidly, signalling the presence of unconfined kink excitations unique to the Potts model.}
\label{fig:entropy}
\end{figure}

After the quenches, we observe persistent oscillations in the magnetisation(s), as illustrated in the main text in Fig. \ref{fig:magnetisation_t_num}. In the following, we supplement these results by showing the entanglement entropy $S$ for two qualitatively distinct cases in Fig. \ref{fig:entropy} for both settings of the maximum bond dimension. 
In Fig. \ref{fig:entropy} (a), it is demonstrated that for the positively aligned quench, the entropy exhibits an apparent saturation up to short times but manifests a slow drift in the long-time simulations. In the negatively aligned case, the time evolution of the entropy shows qualitatively the same features. This behaviour results from the confining/anticonfining nature of the quenches. However, in the oblique cases, demonstrated by the positive oblique case in Fig. \ref{fig:entropy} (b), the entropy grows fast due to the contribution from the unconfined kink degrees of freedom  \cite{Po25}. 

To compare the numerical and the analytical Fourier spectra, it is necessary to take into account the different definitions of the Fourier transform (\ref{eq:FT_analitical_def_app}) and (\ref{eq:FT_numerical_def}). As stated in the main text, these can differ by a multiplicative factor and a constant shift. By comparing the definitions, the multiplicative factor $a$ turns out to be given by 
\begin{equation}
    a=\frac{1}{\sqrt{N_{\text{FT}}}\,   \delta t_{\text{FT}}}.
    \label{eq:FT_cst_match}
\end{equation}
In the long-time simulations, we time-evolved the initial state until $t=1000$ with $\delta t=0.005$, as already stated, yielding $ N_s=200001$ steps, including the $t=0$ point. However, as we noted previously, we only calculated the magnetisation(s) at every 20th time step, which means that the number of steps relevant for the Fourier transformation is $N_{FT}=10001$, while the relevant time step is $\delta t_{\text{FT}}=0.1$, which can be substituted into Eq. (\ref{eq:FT_cst_match}) yielding $a\approx0.10$. 

\begin{figure}
\centering
\begin{subfigure}{0.495\textwidth}
    \includegraphics[width=\textwidth]{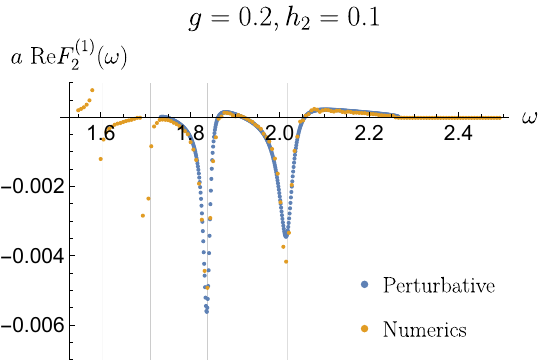}
   \caption{}
\end{subfigure}
\hfill
\begin{subfigure}{0.495\textwidth}
    \includegraphics[width=\textwidth]{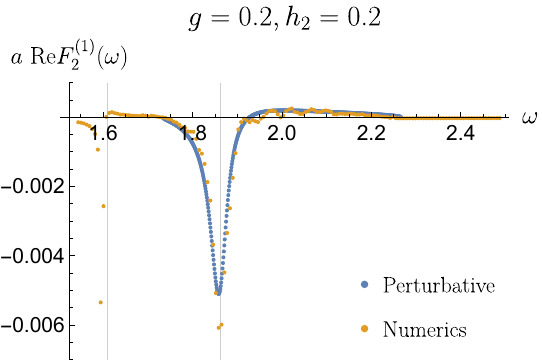}
    \caption{}
\end{subfigure}
\caption{Comparison of the analytical $a\,\mathrm{Re}\,F_{2}^{(1)}(\omega)$ with numerical $ReM_{2}(\omega)$ for positive  oblique quenches with parameters $g=0.2$ and $h_2=0.1$ (a) and $h_2=0.2$ (b). The grey lines correspond to the collisionless bubble masses and the locations of the resonances, determined from perturbation theory. These plots confirm that the perturbative framework successfully predicts the dynamical response for magnetisation component $M_{2}$ with a vanishing constant shift $dM_{2} = 0$.}
\label{fig:reM2_om_pos_ob}
\end{figure}

\begin{figure}
\centering
\begin{subfigure}{0.495\textwidth}
    \includegraphics[width=\textwidth]{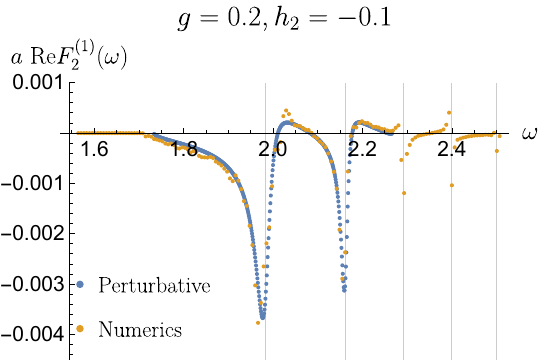}
   \caption{}
\end{subfigure}
\hfill
\begin{subfigure}{0.495\textwidth}
    \includegraphics[width=\textwidth]{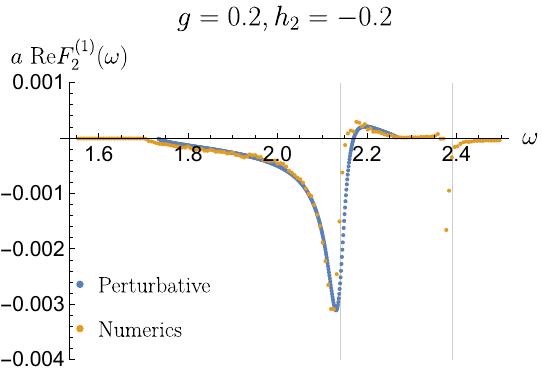}
    \caption{}
\end{subfigure}
\caption{Comparison of the analytical $a\,\mathrm{Re}\,F_{2}^{(1)}(\omega)$ with numerical $\mathrm{Re}M_{2}(\omega)$ for the negative oblique quenches with parameters $g=0.2$ and $h_2=-0.1$ (a) and $h_2=-0.2$ (b). The grey lines correspond to the collisionless meson masses and the locations of the resonances, determined from perturbation theory. These plots confirm that the perturbative framework successfully predicts the dynamical response for magnetisation component $M_{2}$ with a vanishing constant shift $dM_{2} = 0$.}
\label{fig:reM2_om_neg_ob}
\end{figure}

To supplement the main text, we also show typical Fourier spectra of the analytically calculated $a \,\text{Re}\,F^{(1)}_2 (\omega) +\text{d}M_2$ together with  $\text{Re}\,M_2(\omega)$ determined in the iTEBD simulations for oblique quenches in Figs. \ref{fig:reM2_om_pos_ob} and \ref{fig:reM2_om_neg_ob}. The numerical and analytical data show excellent agreement, with the shift $\text{d} M_2$ vanishing. 

\clearpage

\bibliography{Bank_Potts_chain} 

\begin{thebibliography}{10}
\providecommand{\url}[1]{\texttt{#1}}
\providecommand{\urlprefix}{URL }
\expandafter\ifx\csname urlstyle\endcsname\relax
  \providecommand{\doi}[1]{doi:\discretionary{}{}{}#1}\else
  \providecommand{\doi}{doi:\discretionary{}{}{}\begingroup \urlstyle{rm}\Url}\fi
\providecommand{\eprint}[2][]{\url{#2}}

\bibitem{Wilson74}
K.~G. Wilson,
\newblock \emph{Confinement of quarks},
\newblock Phys. Rev. D \textbf{10}, 2445 (1974),
\newblock \doi{10.1103/PhysRevD.10.2445}.

\bibitem{Nar04}
S.~Narison,
\newblock \emph{QCD as a Theory of Hadrons},
\newblock Cambridge University Press, Cambridge (2004).

\bibitem{McCoy78}
B.~M. McCoy and T.~T. Wu,
\newblock \emph{Two dimensional {I}sing field theory in a magnetic field: Breakup of the cut in the two-point function},
\newblock Phys. Rev. D \textbf{18}, 1259 (1978),
\newblock \doi{10.1103/PhysRevD.18.1259}.

\bibitem{1996NuPhB.473..469D}
G.~{Delfino}, G.~{Mussardo} and P.~{Simonetti},
\newblock \emph{{Non-integrable quantum field theories as perturbations of certain integrable models}},
\newblock Nucl. Phys. B \textbf{473}, 469 (1996),
\newblock \doi{10.1016/0550-3213(96)00265-9},
\newblock \eprint{hep-th/9603011}.

\bibitem{1998NuPhB.516..675D}
G.~{Delfino} and G.~{Mussardo},
\newblock \emph{{Non-integrable aspects of the multi-frequency sine-Gordon model}},
\newblock Nucl. Phys. B \textbf{516}, 675 (1998),
\newblock \doi{10.1016/S0550-3213(98)00063-7},
\newblock \eprint{hep-th/9709028}.

\bibitem{FonZam2003}
P.~Fonseca and A.~B. Zamolodchikov,
\newblock \emph{Ising field theory in a magnetic field: Analytic properties of the free energy},
\newblock J. Stat. Phys. \textbf{110}, 527 (2003),
\newblock \doi{https://doi.org/10.1023/A:1022147532606}.

\bibitem{FZ06}
P.~Fonseca and A.~B. Zamolodchikov,
\newblock \emph{Ising spectroscopy ${{\rm I}} $: Mesons at ${T}<{T}_c$} (2006), \eprint{arXiv:hep-th/0612304}.

\bibitem{Rut08a}
S.~B. {Rutkevich},
\newblock \emph{{Energy Spectrum of Bound-Spinons in the Quantum Ising Spin-Chain Ferromagnet}},
\newblock J. Stat. Phys. \textbf{131}, 917 (2008),
\newblock \doi{10.1007/s10955-008-9495-1},
\newblock \eprint{0712.3189}.

\bibitem{RutP09}
S.~B. {Rutkevich},
\newblock \emph{{Two-kink bound states in the magnetically perturbed Potts field theory at $T<T_{c}$}},
\newblock J. Phys. A Math. Gen. \textbf{43}, 235004 (2010),
\newblock \doi{10.1088/1751-8113/43/23/235004},
\newblock \eprint{0907.3697}.

\bibitem{Rut18}
S.~B. {Rutkevich},
\newblock \emph{{Kink confinement in the antiferromagnetic XXZ spin-(1/2) chain in a weak staggered magnetic field}},
\newblock EPL (Europhys. Lett.) \textbf{121}, 37001 (2018),
\newblock \doi{10.1209/0295-5075/121/37001},
\newblock \eprint{1710.11605}.

\bibitem{Lagnese_2020}
G.~Lagnese, F.~M. Surace, M.~Kormos and P.~Calabrese,
\newblock \emph{Confinement in the spectrum of a {H}eisenberg{\textendash}{I}sing spin ladder},
\newblock J. Stat. Mech. Theor. Exp. \textbf{2020}, 093106 (2020),
\newblock \doi{10.1088/1742-5468/abb368}.

\bibitem{2020PhRvB.102a4426R}
F.~B. {Ramos}, M.~{Lencs{\'e}s}, J.~C. {Xavier} and R.~G. {Pereira},
\newblock \emph{{Confinement and bound states of bound states in a transverse-field two-leg Ising ladder}},
\newblock Phys. Rev. B \textbf{102}, 014426 (2020),
\newblock \doi{10.1103/PhysRevB.102.014426},
\newblock \eprint{2005.03145}.

\bibitem{Mus22}
M.~Lencs\'es, G.~Mussardo and G.~Tak\'acs,
\newblock \emph{Confinement in the tricritical {I}sing model},
\newblock Phys. Lett. B \textbf{828}, 137008 (2022),
\newblock \doi{10.1016/j.physletb.2022.137008}.

\bibitem{Rut23s}
S.~Rutkevich,
\newblock \emph{{Soliton confinement in the double sine-{G}ordon model}},
\newblock SciPost Phys. \textbf{16}, 042 (2024),
\newblock \doi{10.21468/SciPostPhys.16.2.042},
\newblock \eprint{2311.07303}.

\bibitem{Kor16}
M.~Kormos, M.~Collura, G.~Tak{\'a}cs and P.~Calabrese,
\newblock \emph{Real-time confinement following a quantum quench to a non-integrable model},
\newblock Nat. Phys. \textbf{13}, 246 (2017),
\newblock \doi{10.1038/nphys3934}.

\bibitem{2019PhRvB..99r0302M}
P.~P. {Mazza}, G.~{Perfetto}, A.~{Lerose}, M.~{Collura} and A.~{Gambassi},
\newblock \emph{{Suppression of transport in nondisordered quantum spin chains due to confined excitations}},
\newblock Phys. Rev. B \textbf{99}, 180302 (2019),
\newblock \doi{10.1103/PhysRevB.99.180302},
\newblock \eprint{1806.09674}.

\bibitem{2020PhRvB.102d1118L}
A.~{Lerose}, F.~M. {Surace}, P.~P. {Mazza}, G.~{Perfetto}, M.~{Collura} and A.~{Gambassi},
\newblock \emph{{Quasilocalized dynamics from confinement of quantum excitations}},
\newblock Phys. Rev. B \textbf{102}, 041118 (2020),
\newblock \doi{10.1103/PhysRevB.102.041118},
\newblock \eprint{1911.07877}.

\bibitem{Po22}
O.~Pomponio, M.~A. Werner, G.~Zar\'and and G.~Tak\'acs,
\newblock \emph{{Bloch oscillations and the lack of the decay of the false vacuum in a one-dimensional quantum spin chain}},
\newblock SciPost Phys. \textbf{12}, 061 (2022),
\newblock \doi{10.21468/SciPostPhys.12.2.061}.

\bibitem{DG08}
G.~{Delfino} and P.~{Grinza},
\newblock \emph{{Confinement in the q-state Potts field theory}},
\newblock Nucl. Phys. B \textbf{791}, 265 (2008),
\newblock \doi{10.1016/j.nuclphysb.2007.09.003},
\newblock \eprint{0706.1020}.

\bibitem{Rut15B}
S.~B. {Rutkevich},
\newblock \emph{{Baryon masses in the three-state Potts field theory in a weak magnetic field}},
\newblock J. Stat. Mech. Theor. Exp. \textbf{2015}, 01010 (2015),
\newblock \doi{10.1088/1742-5468/2015/01/P01010},
\newblock \eprint{1408.1818}.

\bibitem{LT2015}
M.~{Lencs{\'e}s} and G.~{Tak{\'a}cs},
\newblock \emph{{Confinement in the q-state Potts model: an RG-TCSA study}},
\newblock J. High Energ. Phys. \textbf{2015}, 146 (2015),
\newblock \doi{10.1007/JHEP09(2015)146},
\newblock \eprint{1506.06477}.

\bibitem{Roy26}
A.~{Roy}, R.~M. {Konik} and D.~{Rogerson},
\newblock \emph{{Universal Euler-Cartan Circuits for Quantum Field Theories}} arXiv:2407.21278 (2024),
\newblock \doi{10.48550/arXiv.2407.21278},
\newblock \eprint{2407.21278}.

\bibitem{Po25}
O.~Pomponio, A.~Krasznai and G.~Tak\'acs,
\newblock \emph{{Confinement and false vacuum decay on the {P}otts quantum spin chain}},
\newblock SciPost Phys. \textbf{18}, 082 (2025),
\newblock \doi{10.21468/SciPostPhys.18.3.082}.

\bibitem{Ising_ak_gt}
A.~{Krasznai} and G.~{Tak{\'a}cs},
\newblock \emph{{Escaping fronts in local quenches of a confining spin chain}},
\newblock Scipost Phys. \textbf{16}, 138 (2024),
\newblock \doi{10.21468/SciPostPhys.16.5.138},
\newblock \eprint{2401.04193}.

\bibitem{2016NuPhB.911..805R}
T.~{Rakovszky}, M.~{Mesty{\'a}n}, M.~{Collura}, M.~{Kormos} and G.~{Tak{\'a}cs},
\newblock \emph{{Hamiltonian truncation approach to quenches in the Ising field theory}},
\newblock Nucl. Phys. B \textbf{911}, 805 (2016),
\newblock \doi{10.1016/j.nuclphysb.2016.08.024},
\newblock \eprint{1607.01068}.

\bibitem{Rut05}
S.~B. {Rutkevich},
\newblock \emph{{Large-n Excitations in the Ferromagnetic Ising Field Theory in a Weak Magnetic Field: Mass Spectrum and Decay Widths}},
\newblock Phys. Rev. Lett. \textbf{95}, 250601 (2005),
\newblock \doi{10.1103/PhysRevLett.95.250601},
\newblock \eprint{hep-th/0509149}.

\bibitem{Rut09}
S.~B. {Rutkevich},
\newblock \emph{{Formfactor perturbation expansions and confinement in the Ising field theory}},
\newblock J. of Phys. A Math. Gen. \textbf{42}, 304025 (2009),
\newblock \doi{10.1088/1751-8113/42/30/304025},
\newblock \eprint{0901.1571}.

\bibitem{Rap06}
{\'A}.~{Rapp} and G.~{Zar{\'a}nd},
\newblock \emph{{Dynamical correlations and quantum phase transition in the quantum Potts model}},
\newblock Phys. Rev. B \textbf{74}, 014433 (2006),
\newblock \doi{10.1103/PhysRevB.74.014433},
\newblock \eprint{cond-mat/0507390}.

\bibitem{Rut10C}
S.~B. {Rutkevich},
\newblock \emph{{On the weak confinement of kinks in the one-dimensional quantum ferromagnet CoNb$_{2}$O$_{6}$}},
\newblock J. Stat. Mech. Theor. Exp. \textbf{2010}, 07015 (2010),
\newblock \doi{10.1088/1742-5468/2010/07/P07015},
\newblock \eprint{1003.5654}.

\bibitem{Shiba_80}
N.~{Ishimura} and H.~{Shiba},
\newblock \emph{{Dynamical Correlation Functions of One-Dimensional Anisotropic Heisenberg Model with Spin 1/2. I: Ising-Like Antiferromagnets}},
\newblock Progress of Theoretical Physics \textbf{63}, 743 (1980),
\newblock \doi{10.1143/PTP.63.743}.

\bibitem{Bera17}
A.~K. {Bera}, B.~{Lake}, F.~H.~L. {Essler}, L.~{Vanderstraeten}, C.~{Hubig}, U.~{Schollw{\"o}ck}, A.~T.~M.~N. {Islam}, A.~{Schneidewind} and D.~L. {Quintero-Castro},
\newblock \emph{{Spinon confinement in a quasi-one-dimensional anisotropic Heisenberg magnet}},
\newblock Phys. Rev. B \textbf{96}, 054423 (2017),
\newblock \doi{10.1103/PhysRevB.96.054423},
\newblock \eprint{1705.01259}.

\bibitem{Rut22}
S.~B. {Rutkevich},
\newblock \emph{{Spinon confinement in the gapped antiferromagnetic XXZ spin-1/2 chain}},
\newblock Phys. Rev. B \textbf{106}, 134405 (2022),
\newblock \doi{10.1103/PhysRevB.106.134405},
\newblock \eprint{2207.12215}.

\bibitem{Rut24}
S.~B. {Rutkevich},
\newblock \emph{{Confinement of spinons in the XXZ spin-1/2 chain in presence of a transverse magnetic field}},
\newblock Phys. Rev. B \textbf{109}, 014411 (2024),
\newblock \doi{10.1103/PhysRevB.109.014411},
\newblock \eprint{2307.08328}.

\bibitem{CZ92}
L.~{Chim} and A.~{Zamolodchikov},
\newblock \emph{{Integrable Field Theory of the q-State Potts Model with $0<q<4$}},
\newblock Int. J. Mod. Phys. A \textbf{7}, 5317 (1992),
\newblock \doi{10.1142/S0217751X9200243X}.

\bibitem{KS88}
A.~N. Kirillov and F.~A. Smirnov,
\newblock \emph{Local fields in scaling field theory associated with $3$-state {P}otts model},
\newblock {\it ITF Kiev preprint} ITF-88-73R (in Russian) (1988).

\bibitem{2007PhRvL..98p0405R}
{\'A}.~{Rapp}, G.~{Zar{\'a}nd}, C.~{Honerkamp} and W.~{Hofstetter},
\newblock \emph{{Color Superfluidity and ``Baryon'' Formation in Ultracold Fermions}},
\newblock Phys. Rev. Lett. \textbf{98}, 160405 (2007),
\newblock \doi{10.1103/PhysRevLett.98.160405},
\newblock \eprint{cond-mat/0607138}.

\bibitem{2020arXiv200707258L}
F.~{Liu}, S.~{Whitsitt}, P.~{Bienias}, R.~{Lundgren} and A.~V. {Gorshkov},
\newblock \emph{{Realizing and Probing Baryonic Excitations in Rydberg Atom Arrays}} arXiv:2007.07258 (2020),
\newblock \doi{10.48550/arXiv.2007.07258},
\newblock \eprint{2007.07258}.

\bibitem{2023PhRvR...5d3020W}
M.~A. {Werner}, C.~P. {Moca}, M.~{Kormos}, {\"O}.~{Legeza}, B.~{D{\'o}ra} and G.~{Zar{\'a}nd},
\newblock \emph{{Spectroscopic evidence for engineered hadronic bound state formation in repulsive fermionic SU(N ) Hubbard systems}},
\newblock Phys. Rev. Research 043020 (2023),
\newblock \doi{10.1103/PhysRevResearch.5.043020},
\newblock \eprint{2207.00994}.

\bibitem{1977PhRvD..15.2929C}
S.~{Coleman},
\newblock \emph{{Fate of the false vacuum: Semiclassical theory}},
\newblock Phys. Rev. D \textbf{15}, 2929 (1977),
\newblock \doi{10.1103/PhysRevD.15.2929}.

\bibitem{2024PhRvL.133x0402L}
G.~{Lagnese}, F.~M. {Surace}, S.~{Morampudi} and F.~{Wilczek},
\newblock \emph{{Detecting a Long-Lived False Vacuum with Quantum Quenches}},
\newblock Phys. Rev. Lett. \textbf{133}, 240402 (2024),
\newblock \doi{10.1103/PhysRevLett.133.240402},
\newblock \eprint{2308.08340}.

\bibitem{Jimbo94}
M.~Jimbo and T.~Miwa,
\newblock \emph{Algebraic {A}nalysis of {S}olvable {L}attice {M}odels},
\newblock Conference Board of the Mathematical Sciences. American Mathematical Soc. (1995).

\bibitem{LL3}
L.~D. Landau and E.~M. Lifshitz,
\newblock \emph{Quantum Mechanics: Non-Relativistic Theory},
\newblock Course of Theoretical Physics. Elsevier Science,
\newblock ISBN 9780080503486 (1981).

\bibitem{Gr18}
D.~J. Griffiths and D.~F. Schroeter,
\newblock \emph{Introduction to Quantum Mechanics},
\newblock Cambridge University Press,
\newblock ISBN 9781107189638 (2018).

\bibitem{taylor2012scattering}
J.~R. Taylor,
\newblock \emph{Scattering Theory: The Quantum Theory of Nonrelativistic Collisions},
\newblock Dover Books on Engineering. Dover Publications,
\newblock ISBN 9780486142074 (2012).

\bibitem{2014JPhA...47N2001D}
G.~{Delfino},
\newblock \emph{{Quantum quenches with integrable pre-quench dynamics}},
\newblock J. Phys. A Math. Gen. \textbf{47}, 402001 (2014),
\newblock \doi{10.1088/1751-8113/47/40/402001},
\newblock \eprint{1405.6553}.

\bibitem{2017JPhA...50h4004D}
G.~{Delfino} and J.~{Viti},
\newblock \emph{{On the theory of quantum quenches in near-critical systems}},
\newblock J. of Phys. A Math. Gen. \textbf{50}, 084004 (2017),
\newblock \doi{10.1088/1751-8121/aa5660},
\newblock \eprint{1608.07612}.

\bibitem{2018ScPP....5...27H}
K.~{H{\'o}ds{\'a}gi}, M.~{Kormos} and G.~{Tak{\'a}cs},
\newblock \emph{{Quench dynamics of the Ising field theory in a magnetic field}},
\newblock SciPost Phys. \textbf{5}, 027 (2018),
\newblock \doi{10.21468/SciPostPhys.5.3.027},
\newblock \eprint{1803.01158}.

\bibitem{2019JHEP...08..047H}
K.~{H{\'o}ds{\'a}gi}, M.~{Kormos} and G.~{Tak{\'a}cs},
\newblock \emph{{Perturbative post-quench overlaps in quantum field theory}},
\newblock J. High Energ. Phys. \textbf{2019}, 47 (2019),
\newblock \doi{10.1007/JHEP08(2019)047},
\newblock \eprint{1905.05623}.

\bibitem{Rut80}
S.~B. Rutkevich,
\newblock \emph{Relaxation dynamics of a quantum chain of harmonic oscillators},
\newblock Ukr. Phys. J. \textbf{25}, 1135 (1980).

\bibitem{Rut12}
S.~B. Rutkevich,
\newblock \emph{Partial thermalization in the quantum chain of harmonic oscillators} (2012), \eprint{arXiv:1201.0578v2}.

\bibitem{2012JSMTE..04..017S}
D.~{Schuricht} and F.~H.~L. {Essler},
\newblock \emph{{Dynamics in the Ising field theory after a quantum quench}},
\newblock J. Stat. Mech. Theor. Exp. \textbf{2012}, 04017 (2012),
\newblock \doi{10.1088/1742-5468/2012/04/P04017},
\newblock \eprint{1203.5080}.

\bibitem{2014JSMTE..10..035B}
B.~{Bertini}, D.~{Schuricht} and F.~H.~L. {Essler},
\newblock \emph{{Quantum quench in the sine-Gordon model}},
\newblock J. Stat. Mech. Theor. Exp. \textbf{2014}, 10035 (2014),
\newblock \doi{10.1088/1742-5468/2014/10/P10035},
\newblock \eprint{1405.4813}.

\bibitem{2017NuPhB.925..362P}
L.~{Piroli}, B.~{Pozsgay} and E.~{Vernier},
\newblock \emph{{What is an integrable quench?}},
\newblock Nucl. Phys. B \textbf{925}, 362 (2017),
\newblock \doi{10.1016/j.nuclphysb.2017.10.012},
\newblock \eprint{1709.04796}.

\bibitem{2019ScPP....6...62P}
B.~{Pozsgay}, L.~{Piroli} and E.~{Vernier},
\newblock \emph{{Integrable Matrix Product States from boundary integrability}},
\newblock SciPost Phys. \textbf{6}, 062 (2019),
\newblock \doi{10.21468/SciPostPhys.6.5.062},
\newblock \eprint{1812.11094}.

\bibitem{2011PhRvL.106v7203C}
P.~{Calabrese}, F.~H.~L. {Essler} and M.~{Fagotti},
\newblock \emph{{Quantum Quench in the Transverse-Field Ising Chain}},
\newblock Phys. Rev. Lett. \textbf{106}, 227203 (2011),
\newblock \doi{10.1103/PhysRevLett.106.227203},
\newblock \eprint{1104.0154}.

\bibitem{2012JSMTE..07..016C}
P.~{Calabrese}, F.~H.~L. {Essler} and M.~{Fagotti},
\newblock \emph{{Quantum quench in the transverse field Ising chain: I. Time evolution of order parameter correlators}},
\newblock J. Stat. Mech. Theor. Exp. \textbf{2012}, 07016 (2012),
\newblock \doi{10.1088/1742-5468/2012/07/P07016},
\newblock \eprint{1204.3911}.

\bibitem{2012JSMTE..07..022C}
P.~{Calabrese}, F.~H.~L. {Essler} and M.~{Fagotti},
\newblock \emph{{Quantum quenches in the transverse field Ising chain: II. Stationary state properties}},
\newblock J. Stat. Mech. Theor. Exp. \textbf{2012}, 07022 (2012),
\newblock \doi{10.1088/1742-5468/2012/07/P07022},
\newblock \eprint{1205.2211}.

\end{thebibliography}
\end{document}